\documentclass[ejsv2]{imsart}
\newcommand{\blind}{1}

\RequirePackage[numbers]{natbib}
\RequirePackage[colorlinks,citecolor=blue,urlcolor=blue]{hyperref}
\RequirePackage{graphicx}
\usepackage{appendix}

\usepackage{amsmath}
\usepackage{mathtools}
\usepackage{algorithm}
\usepackage{algpseudocode}
\usepackage{textcomp,enumerate,bbm,latexsym, bm}
\usepackage{enumitem} 
\usepackage{graphicx} 
\usepackage{url,algorithm,verbatim}
\usepackage[normalem]{ulem}
\usepackage{algpseudocode}
\usepackage{natbib}

\usepackage{multirow}
\usepackage{booktabs}
\usepackage{chngcntr, subcaption}

\usepackage{dsfont}

\usepackage{subcaption}
\usepackage{amsthm}
\newcommand{\argmin}{\text{argmin}}
\newcommand{\tr}{\text{tr}}

\renewcommand{\argmin}{\mathop{\rm argmin}}
\newcommand{\argmax}{\mathop{\rm argmax}}

\newcommand{\wh}{\widehat}

\newcommand{\supp}{{\rm supp}}

\newcommand{\R}{\mathbb{R}}
\renewcommand{\tr}{\operatorname{tr}}

\makeatletter
\newcommand{\doublewidetilde}[1]{{%
  \mathpalette\double@widetilde{#1}%
}}
\newcommand{\double@widetilde}[2]{%
  \sbox\z@{$\m@th#1\widetilde{#2}$}%
  \ht\z@=.9\ht\z@
  \widetilde{\box\z@}%
}

\usepackage{xr}
\usepackage{hyperref}
\usepackage{cleveref}
\usepackage{wrapfig}
\arxiv{2507.11160}
\startlocaldefs
\theoremstyle{plain}

\newtheorem{theorem}{Theorem}[section]
\newtheorem{lemma}[theorem]{Lemma}
\newtheorem{corollary}{Corollary}[theorem]
\theoremstyle{definition}

\newtheorem{remark}[theorem]{Remark}

\theoremstyle{remark}

\endlocaldefs

\begin{document}
\begin{frontmatter}
\title{Efficient Canonical Correlation Analysis with Sparsity}
\runtitle{Efficient Canonical Correlation Analysis with Sparsity}

\begin{aug}
\author[A]{\fnms{Zixuan}~\snm{Wu}\ead[label=e1]{zixuanwu@uchicago.edu}},
\author[A]{\fnms{Coralie}~\snm{Rousseau}\ead[label=e4]{crousseau@uchicago.edu}},
\author[B]{\fnms{Elena}~\snm{Tuzhilina}\ead[label=e2]{elena.tuzhilina@utoronto.ca}}
\and
\author[A]{\fnms{Claire}~\snm{Donnat}\ead[label=e3]{cdonnat@uchicago.edu}}
\address[A]{Department of Statistics,
University of Chicago\printead[presep={,\ }]{e1,e4,e3}}

\address[B]{Department of Statistics,
University of Toronto\printead[presep={,\ }]{e2}}
\runauthor{Z. Wu et al.}
\end{aug}

\begin{abstract}
In high-dimensional settings, Canonical Correlation Analysis (CCA) often fails, and existing sparse methods force an untenable choice between computational speed and statistical rigor. This work introduces a fast and provably consistent sparse CCA algorithm (ECCAR) that resolves this trade-off. We formulate CCA as a high-dimensional reduced-rank regression problem, which allows us to derive consistent estimators with high-probability error bounds without relying on computationally expensive techniques like Fantope projections. The resulting algorithm is scalable, projection-free, and significantly faster than its competitors. We validate our method through extensive simulations and demonstrate its power to uncover reliable and interpretable associations in three complex biological datasets, as well as in an ML interpretability task. Our work makes sparse CCA a practical and trustworthy tool for large-scale multimodal data analysis. A companion R package has been made available.\if1\blind
{\footnote{The package can be found at the following link: \url{https://cran.r-project.org/web/packages/ccar3/index.html}}}
\end{abstract}

\begin{keyword}[class=MSC]
\kwd[Primary ]{62H20}
\kwd{62J07}
\kwd[; secondary ]{62P10}
\end{keyword}

\begin{keyword}
\kwd{Multi 'omics analysis}
\kwd{Multivariate Analysis}
\kwd{Sparsity}
\kwd{High-Dimensional Statistics}
\kwd{Unsupervised Learning}
\end{keyword}

\end{frontmatter}

\section{Introduction}
\label{sec:intro}

In an era of increasingly prevalent multimodal datasets, from genomics to imaging and beyond, methods that can effectively uncover shared information across different data views are indispensable. Canonical correlation analysis (CCA), a classical technique introduced by \citet{hotelling1936relations}, remains a cornerstone for identifying linear associations between two sets of variables. Given two datasets $ X \in \mathbb{R}^{n \times p} $ and $ Y \in \mathbb{R}^{n \times q} $ assumed to be jointly sampled from a centered multivariate normal distribution {with  covariance $ \Sigma $, CCA finds $r$ linear projections of $ X $ and $ Y $ which are maximally correlated in expectation}. Mathematically, this objective can be expressed as:
\begin{equation}\label{eq:cca}
\begin{split}
 \forall j \leq r, \qquad   &   u^\star_j, v^\star_j= \argmax_{u \in \R^{p},~v\in \R^q} u^\top{\Sigma}_{XY}v \\
   \text{subject to}      \quad u^\top {\Sigma}_{X}u =1, \quad v^\top{\Sigma}_{Y}v =1  \quad &\text{and }   \quad \forall l<j,  \quad  u_l^\top{\Sigma}_{X}u =0, \quad v_l^\top{\Sigma}_{Y}v=0.
\end{split}
\end{equation}

In practice, the population covariance matrices $ \Sigma_X, \Sigma_Y, \Sigma_{XY} $ are replaced by their empirical counterparts ($ \widehat{\Sigma}_X = \frac{1}{n}X^\top X,$ $\widehat{\Sigma}_Y = \frac{1}{n}Y^\top Y
$ and $\widehat{\Sigma}_{XY} = \frac{1}{n}X^\top Y$, respectively). For each pair of resulting canonical directions, $u^\star_j$ and $v^\star_j$, the corresponding variates,  $X u^\star_j$ and $Y v^\star_j$, represent projections of the two datasets onto a subspace that captures shared information. The resulting diagonal matrix $\Lambda^\star = \operatorname{diag}(\lambda^\star_1,\ldots,\lambda^\star_r)$, where, for each $j$, $\lambda^\star_j = u_j^{^\star\top} \Sigma_{XY}v^\star_j$, encodes correlations between pairs of canonical variates.

Canonical correlation analysis has found widespread applications across diverse disciplines, including sociology, economics \citep{fan2018canonical, thorndike2000canonical, dos2014canonical, mazuruse2014canonical}, and extensively in biology \citep{le2009sparse, revilla2021multi, rodosthenous2020integrating, safo2018integrative, jiang2023canonical, lin2013group}, where the rise of multimodal 'omics' data has renewed interest in CCA as an integrative analysis tool.
{We refer the reader to Section~\ref{sec:real} and Appendix~C for examples of the use of CCA in biological settings.}

Although canonical correlation analysis is a popular tool for scientific discovery, its reliability in increasingly common high-dimensional settings  (i.e. when $\max(p, q )\geq n$) is a growing concern. In fact, this method is known to often produce unstable results and inconsistent canonical directions in such scenarios \citep{gao2017sparse}. This issue, long discussed in statistical circles, is now being highlighted in applied research \citep{helmer2024stability,mcintosh2021comparison,nakua2024comparing}. For example, recent work by \citet{helmer2024stability} studying the reliability of  CCA in neuroscience settings highlights that empirically, the  ``stability of CCA [...]  in high-dimensional datasets is questionable'':  the correlations found on training data do not match those on held-out data, indicating overfitting and inconsistent canonical directions. Such failures to generalize critically undermine CCA's role in generating reliable insights and discovering associations.

  \paragraph{Towards a theoretical understanding of CCA in high dimensions.}   Although these limitations have recently attracted renewed attention (especially in applied communities), the instability of CCA in high-dimensional settings is a well studied phenomenon.  A large body of work has already characterized the behavior of the sample canonical correlations themselves. In proportional-dimensional regimes, where $p$, $q$, and $n$ grow on the same order, \citet{yang2012convergence} establish a limiting empirical distribution of sample canonical correlation coefficients, while \citet{yang2022sample} provides local laws and
Tracy--Widom limits for the largest sample canonical correlations. 
Others have assumed that only finitely many population canonical correlations are nonzero, and studied the resulting outlier sample canonical correlations, phase transitions, and limiting fluctuations  \citep{bao2019canonical,yang2022limiting, ma2023sample}. 
All of these works are primarily spectral: they characterize the canonical-correlation spectrum and its fluctuations, rather than aiming to recover canonical directions.

Another line of work focuses more directly on the canonical variables themselves. In proportional high-dimensional regimes, \citet{bykhovskaya2023high}
show that classical CCA direction estimates need not be consistent, even when signal
canonical correlations separate from the noise bulk, and quantify the resulting angles
between sample and population canonical variables. High-dimensional CCA has also been
studied in the context of independence testing based on regularized
sample canonical correlations \citep{yang2015independence}, post-selection inference \citep{mckeague2022significance}, and
de-biased inference for leading sparse canonical directions and strengths
\citep{laha2023statistical}. These results clarify the behavior of classical canonical
correlations and variables estimates, along with associated tests, in high dimensions. However, when the goal is to directly improve inference and estimate stable and interpretable canonical directions, additional
structure on the canonical direction matrices is needed.

 To address this challenge, several sparse CCA approaches have been developed. These methods impose structural constraints directly on the canonical direction matrices,
assuming row-wise, or coordinate-level, sparsity: only a subset of variables contributes to
the canonical variates. This is distinct from finite-rank or spiked CCA models, where only a
fixed number of population canonical correlations are nonzero but the associated canonical
directions may still be dense. Sparse variants of CCA thus simultaneously perform variable selection while estimating the {CCA subspaces} $U^\star=\big[u^\star_1|\cdots|u^\star_r\big]$ and $V^\star=\big[v^\star_1|\cdots|v^\star_r\big]$, where the $\star$ notation indicates the underlying population quantities. Existing solutions to sparse CCA  can be broadly classified into two categories: heuristic methods vs. theory-based methods \footnote{A more comprehensive and detailed review of sparse CCA methods is provided in Appendix~A.}. Heuristic approaches typically rely on alternating algorithms to impose sparsity (usually in the form of an $\ell_1$ penalty), holding one direction fixed to estimate the other. These algorithms tend to be fast but sensitive to initialization. Among these, the approach of \citet{witten2009penalized} has become particularly popular due to its simplicity and computational tractability. Yet this method relies on a crucial and often overlooked simplification: it assumes that the covariance matrices $ \Sigma_X $ and $ \Sigma_Y $ are diagonal. While this assumption simplifies the optimization problem, it ignores the true geometry of the data and can lead to spurious or misleading associations—particularly when the variables exhibit strong dependencies (see Figure~\ref{fig:stategra_res}).

In contrast, recent theoretical work has established consistent procedures for sparse CCA that respect the full covariance structure \citep{gao2015minimax, gao2017sparse, gao2023sparse}, though these methods often rely on computationally intensive steps --- such as a Fantope-based initialization --- that scale poorly with the dataset size. More recently, \citet{donnat2024canonical} proposed a computationally efficient alternative by reframing CCA as a reduced-rank regression problem, offering consistent and efficient CCA estimators when one of the datasets is low-dimensional. However, their method does not support sparsity in both $ X $ and $ Y $.
To provide intuition for why the problem becomes substantially harder when both dimensions are high,
note that CCA seeks to estimate canonical directions under unknown covariance structures
$\Sigma_X$ and $\Sigma_Y$.
\citet{donnat2024canonical} exploit a regression interpretation of CCA: regressing $Y$ onto $X$
yields a consistent  estimator of the product $U^\star \Lambda^\star V^{\star\top}\Sigma_Y$.
When $Y$ is low-dimensional, the precision matrix $\Sigma_Y^{-1}$ of $Y$ can be estimated accurately, and a penalized regression of $Y \wh \Sigma_Y^{-1/2}$ onto $X$ allows one to
 recover the canonical product $U^\star \Lambda^\star V^{\star\top}$.
However, this approach breaks down when both $X$ and $Y$ are high-dimensional, since the sample covariance matrices
are singular and cannot be inverted reliably.

Fundamentally, under the regression interpretation of CCA as regressing $Y$ onto 
$X$, the double high-dimensional regime turns sparse CCA into a high-dimensional estimation problem in which consistent estimation relies on exploiting sparsity in both response and predictor variables. Our formulation addresses this difficulty by optimizing over the product  $B = U^\star \Lambda^\star V^{\star\top}$ through the quantity $XBY^\top$.
The sparsity constraint on $B$ effectively restricts the problem to subsets of variables from both views,
which enables stable estimation on these lower-dimensional supports and permits localized matrix inversion.
In this way, the method circumvents the need for global covariance estimation while addressing the intrinsic
coupling between the two datasets.

\paragraph{Contributions.}In this paper, we propose a novel two-step procedure for estimating CCA directions in high dimensions. This approach, similar to most theoretical and provably consistent CCA methods, initially transforms the problem into a convex estimation problem by estimating $B = U^\star \Lambda^\star V^{\star\top}$ before estimating the components. Unlike \cite{gao2015minimax, gao2017sparse, gao2023sparse}, our convex estimation procedure is scalable. Unlike \cite{donnat2024canonical}, our method does not rely on the low dimensionality of one of the datasets.

(1) We show in Section~\ref{subsec:low_dim} that the matrix product $B = U^{\star} \Lambda^{\star} V^{\star \top}$ 
can be estimated by solving the optimization problem:
\begin{equation}\label{eq:1}
\widehat{B} = \argmin_{B \in \mathbb{R}^{p \times q}} \Big\| \frac{1}{n} X B Y^\top - I_n \Big\|_F^2, \qquad \text{ such that } \operatorname{rank}(B) = r.
\end{equation} Here $I_n$ denotes the identity matrix of size $n$.
To extend applicability to high-dimensional contexts (Section~\ref{subsec:high_dim}), we propose a penalize-then-truncate solution to Equation~\ref{eq:1}, and incorporate an $\ell_1$ penalty (or $\ell_1$-group penalty, when applicable) on the matrix $B$.

(2) {We establish the statistical consistency of our estimates in Section~\ref{sec:theory} and show that our method controls false-positive variable selection with high probability. Under an additional minimum-signal condition, this further implies exact recovery of the row and column supports of the true canonical subspaces,  validating the practical utility of sparse CCA as a hypothesis-generating tool.}
While our method presents a minor loss in statistical efficiency compared to established methods such as \cite{gao2017sparse,gao2023sparse}, it offers significantly improved computational efficiency (Table~3 in the Appendix~A).

(3) We illustrate the effectiveness of our approach through simulation studies (Section~\ref{sec:sim}) and applications to four real-world datasets (Section~\ref{sec:real}), demonstrating its accuracy and capability to identify scientifically relevant associations across a variety of fields. Notably, our method was run on a transcriptomics dataset (section 5.4) within a few minutes, whereas alternative theory-based approaches fail to complete even after 12 hours of computation.

\paragraph{Notations.} For any \( t \in \mathbb{Z}^+ \), let \( [t] \) denote the set \( \{1, 2, \dots, t\} \). For a set \( S \), we let \( |S| \) denote its cardinality and \( S^c \) its complement. For any \( a, b \in \mathbb{R} \), define \( a \vee b = \max(a, b) \) and \( a \wedge b = \min(a, b) \).
For a vector \( u \in \mathbb{R}^d \) (denoted by a lowercase letter), we define the Euclidean norm \( \|u\| = \sqrt{\sum_i u_i^2} \), the \( \ell_0 \)-norm \( \|u\|_0 = \sum_i \boldsymbol{1}_{\{u_i \neq 0\}} \), and the \( \ell_1 \)-norm \( \|u\|_1 = \sum_i |u_i| \).
For a matrix \( A = (a_{ij}) \in \mathbb{R}^{p \times k} \) (denoted by an uppercase letter), let \( A_{i \cdot} \) denote its \( i \)-th row {and \( A_{ \cdot j} \) denote its \( j \)-th column.} {For two index subsets \( S_1 \subseteq [p] \) and \( S_2 \subseteq [k] \), we denote by \( A_{S_1 S_2} \) the submatrix of \( A \) consisting of the rows and columns indexed by \( S_1 \) and \( S_2 \), respectively.
} The support of \( A \), denoted \( \mathrm{supp}(A) \), is the index set of its nonzero rows, i.e.
\(
\mathrm{supp}(A) = \{ i \in [p] : \| A_{i \cdot} \| > 0 \}.
\)
We use the notation $\mathrm{supp}_e(A) $ to denote the entry-wise support of a matrix $A\in \mathbb{R}^{p_1\times p_2}$: \(
\mathrm{supp}_e(A) = \{ i \in [p_1],  j \in [p_2]: | A_{ij} | > 0 \}.
\)
The \( i \)-th largest singular value of \( A \) is denoted \( \sigma_i(A) \), with \( \sigma_{\max}(A) = \sigma_1(A) \) and \( \sigma_{\min}(A) = \sigma_{\operatorname{rank}(A)}(A) \). The Frobenius norm and the operator norm of \( A \) are defined as \( \|A\|_F = \sqrt{\sum_{i,j} a_{ij}^2} \) and \( \|A\|_{\mathrm{op}} = \sigma_1(A) \), respectively. The $\ell_{1,1}$ norm of the matrix $A$ is denoted as \( \|A\|_{11} = {\sum_{i,j} |A_{ij}|} \),  and its $\ell_{2,1}$ norm  is defined as  the sum of its row norms: \( \|A\|_{21} = {\sum_{i=1}  \|A_{i\cdot}\|} \) .  The $\ell_0$ norm of the matrix $A$ is denoted as $A_{0} = |\mathrm{supp}_e(A) |$.
The pseudo-inverse of a matrix \( A \) is denoted by \( A^\dagger \).
If \( A \) is positive semi-definite, \( A^{1/2} \) denotes its principal square root. For two matrices \( A, B \in \mathbb{R}^{p \times k} \), the trace inner product is given by \( \langle A, B \rangle = \operatorname{Tr}(A^\top B) \).

\paragraph{Parameter Space.} In this work, we assume the \textit{canonical pair model} described by \cite{gao2017sparse}. Under this formalism, we observe \(n\) independently and identically distributed (i.i.d.) pairs of vectors \((X_i, Y_i)_{i=1}^n\) drawn from a joint multivariate Gaussian distribution \(\mathcal{N}_{p+q}(0, \Sigma)\). Here, \(\Sigma = \left(\begin{smallmatrix} 
\Sigma_X & \Sigma_{XY} \\ \Sigma_{XY}^\top & \Sigma_Y 
\end{smallmatrix}
\right)\) and the cross-covariance matrix \(\Sigma_{XY}\) admits the reparametrization
\begin{equation}\label{reparam}
\Sigma_{XY} = \Sigma_X U^\star \Lambda^{\star} {V^\star}^\top \Sigma_Y,
\end{equation}
and the matrices \( U^\star = [u^\star_1\,|\, \dots\,|\, u^\star_r] \in \mathbb{R}^{p \times r} \) and \( V^\star = [v^\star_1\,|\, \dots\,|\, v^\star_r] \in \mathbb{R}^{q \times r} \) store the \( r \) canonical directions column-wise. The diagonal matrix \( \Lambda^\star = \operatorname{diag}(\lambda^\star_1, \dots, \lambda^\star_r) \) contains the corresponding canonical correlations, ordered in decreasing order: \( \lambda^\star_1 \geq \lambda^\star_2 \geq \cdots \geq \lambda^\star_r \).  

Throughout this paper, we consider scenarios where the leading canonical direction vectors are row-wise sparse, { so that only a small subset of covariates has nonzero loadings across the leading canonical directions.} More precisely, we define by \(\mathcal{F}(s_u, s_v, p, q, r ; M)\) the set of all covariance matrices \(\Sigma\) that satisfy Equation \ref{reparam} and such that:

\begin{enumerate}
    \item {\it the support of $U^\star$ and $V^\star$ is small}, i.e. 
    $|\text{supp}(U^\star)| \leq s_u \text{ and } |\text{supp}(V^\star)| \leq s_v,$  where the support  of a matrix is defined as the set of its non-zero rows;
    \item {\it the covariances are well-conditioned,} i.e.
    $$\sigma_{\min}(\Sigma_X) \wedge \sigma_{\min}(\Sigma_Y) \geq \frac{1}{M} \qquad \text{ and  } \qquad  \sigma_{\max}(\Sigma_X) \vee \sigma_{\max}(\Sigma_Y) \leq M,$$
    where $M>0$ is a constant independent of the dimension. 
\end{enumerate}
The probability space considered throughout this paper, therefore, is given by
\begin{equation}\label{eq:param_space}
\big\{ (X_i, Y_i) \overset{\text{i.i.d.}}{\sim} \mathcal{N}_{p+q}(0, \Sigma) \mid \Sigma \in \mathcal{F}(s_u, s_v, p, q, r; M) \big\}.
\end{equation}

\section{CCA through regression}
\label{sec:meth}

\subsection{The low-dimensional case: high $n$, low $p$ and $q$}\label{subsec:low_dim}
We begin by analyzing the classical setting where the sample size 
$n$ is large relative to the covariate dimensions, i.e.,  $n \gg  p \vee q$. This regime provides the basis for the derivation of our algorithm in the high-dimensional setting. We consider the convex objective
\begin{equation}\label{eq:regression1}
        \mathcal{L}(B) =  \frac{1}{2}\Big\|\frac1nXBY^\top - I_n\Big\|_F^2, 
    \end{equation}
and show that its minimizer is a consistent estimator of the population matrix $B^\star = U^\star \Lambda^\star V^{\star\top}$, as established in the following theorem.

\begin{theorem}\label{theorem_consistency_product}
    Assume that $p$ and $q$ are fixed as $n$ is allowed to grow. Let $\wh B$ denote the minimizer of $\mathcal{L}(B)$. 
    Under the canonical pair model, 
    if $n$ grows to infinity, then
        $ \wh{B}  \overset{a.s.}{\to} U^\star\Lambda^\star {V^\star}^\top.$
    \end{theorem}

We therefore adopt a denoise‑then‑truncate strategy— we first solve the convex problem in Equation \ref{eq:regression1} to estimate $B$, then project onto rank 
$r$ by SVD—which avoids the non‑convexity and tuning instability of explicit rank penalties while preserving convex optimization and enabling our high‑probability guarantees (Section \ref{sec:theory}).
Having established the initial value \( \widehat{B} \), estimates of \( U^{\star} \) and \( V^{\star} \) are thus obtained by performing a {rank-$r$} singular value decomposition (SVD) of the matrix product
\(
\wh \Sigma_X^{\frac12} \wh B \wh \Sigma_Y^{\frac12}
\)
.
The following theorem establishes the consistency of $\wh U $ and $\wh V$. Its proof can be found in Appendix~E.

\begin{theorem}\label{theorem:consistency_low_dim}
Let $\wh B$ denote the minimizer of $\mathcal{L}(B)$. Denote by $\widehat{U}_0$ and $\widehat{V}_0$ the left and right singular vector matrices of $\wh \Sigma_X^{\frac12} \wh B \wh \Sigma_Y^{\frac12}$. Additionally, denote by $U_0$ and $V_0$ the left and right singular vector matrices of $ \Sigma^{\frac12}_X  B^\star   \Sigma^{\frac12}_Y$.
Under the canonical pair model,  
        \begin{equation}\label{eq:consistency_low_dim0}
            \lim_{n \to \infty} \max\left\{ \min_{O \in \mathcal{O}_{r} }  \| \widehat{U}_0-U_0 O \|_F ,\min_{\tilde{O} \in \mathcal{O}_{r}}  \| \widehat{V}_0-V_0 \tilde{O} \|_F  \right\} =0, 
        \end{equation} 
        where $\mathcal{O}_{r}$ denotes the set of orthogonal matrices in $\R^{r \times r}.$ Consequently, the estimates  $\wh U = \wh \Sigma_X^{-\frac12} \wh U_0 $ and $\wh V = \wh \Sigma^{-\frac12}_Y \wh V_0$ are consistent estimators of $U^{\star}$ and $V^{\star}$ in the sense that:
    \begin{equation}\label{eq:consistency_low_dim}
        \lim_{n \to \infty}\max\left\{\min_{O \in \mathcal{O}_{r}} \| \widehat{U}-U^{\star} O \|_F , \min_{ \tilde{O} \in \mathcal{O}_{r}} \| \widehat{V}-V^{\star} \tilde{O} \|_F  \right\} =0.
    \end{equation}
    \end{theorem}

\subsection{The high-dimensional case: low $n$, high $p$ or $q$}\label{subsec:high_dim}
In the high-dimensional case, the sample estimates $\wh \Sigma_X$,  $\wh \Sigma_Y$  and $\wh \Sigma_{XY}$ are no longer guaranteed to provide consistent estimators of the covariance matrices $\Sigma_X$, $\Sigma_Y$ and $\Sigma_{XY}$. As a result, the estimators of the CCA directions $U^\star$ and $V^\star$ described in the previous subsection are no longer guaranteed to be consistent without assuming further structure. 

In this paper, we consider the case where $U^\star$ and $V^\star$ can be assumed to be row-sparse, i.e.
$ | \supp({U^\star})| \leq s_u,\quad | \supp({V^\star})| \leq s_v. $
As in \citet{gao2017sparse}, we note in this case that the row-sparsity of $U^{\star}$ and $V^{\star}$ naturally translates into element-wise sparsity of the product $B^\star = U^\star \Lambda^\star V^{\star\top}$.
 Therefore, to accommodate the structure of the problem, a natural extension of the previous estimator to the high-dimensional setting is to transform Equation~\ref{eq:regression1} into the following loss function:
    \begin{equation}
        \mathcal{L}(B) = \frac{1}{2}\Big\|\frac1nXBY^\top - I_n\Big\|_F^2 + \rho \|B\|_{11},
        \label{eqn:lasso}
    \end{equation}
where $\rho$ is a tuning parameter controlling the solution sparsity. The sparsity constraint on $B^\star$ is here relaxed to  an $\ell_{1,1}$ penalty, thus turning the estimation of $B^\star$ into a convex optimization problem. { Estimates of $U^{\star}$ and $V^{\star}$ can then be obtained by normalizing the rank-$r$ singular value decomposition of $\wh{B}$. }
This design is also the key to scalability: the first stage solves the convex, projection‑free problem in Equation \ref{eqn:lasso} via simple ADMM updates (matrix multiplications + soft‑thresholding), and we compute only a single rank‑$r$ SVD at the end—yielding the quadratic‑in‑dimension cost detailed below in Section~\ref{subsec:algo} rather than the cubic cost of Fantope‑based methods.

The full procedure is outlined in Algorithm~\ref{alg:eccar}. Moreover, the formulation in Equation~\ref{eqn:lasso} provides a flexible framework that can be readily extended to incorporate other types of structural regularization by changing the type of penalty from an $\ell_{11}$ penalty to a more general function. For instance, one may replace the element-wise penalty with group sparsity constraints or structured norms to reflect known groupings among variables (see Section~\ref{subsec:high_dim_group}).

\begin{algorithm}

\caption{Efficient Sparse CCA via Reduced Rank Regression (ECCAR)}
\label{alg:eccar}
\begin{algorithmic}[1]
\Require $X\in\mathbb{R}^{n\times p}$,\; $Y\in\mathbb{R}^{n\times q}$,\; regularization $\rho\ge 0$,\; rank $r\in\mathbb{N}$
\Ensure $\wh U\in\mathbb{R}^{p\times r}$,\; $\wh V\in\mathbb{R}^{q\times r}, \; \wh \Lambda \in \mathbb{R}^{r\times r}$
\State \textbf{Sparse regression:}
\[
\wh B \gets \argmin_{B\in\mathbb{R}^{p\times q}}
\frac{1}{2}\,\Big\|\tfrac{1}{n}\,X B Y^{\top}-I_n\Big\|_{F}^{2}
\;+\;\rho\,\|B\|_{11}
\]
\State \textbf{Correlation Coefficients: } 
\[
{ \wh \Lambda = \mathrm{diag} \left(d_1\left(\wh \Sigma_X^{1/2} \wh B \wh \Sigma_{Y}^{1/2}\right), \cdots, d_r\left(\wh \Sigma_X^{1/2} \wh B \wh \Sigma_{Y}^{1/2}\right) \right) }
\]
\State \textbf{Truncated SVD:}
\[
\wh B_r = \wh U_{B}\,\wh\Lambda_{B}\,\wh V_{B}^{\top}
\;\gets\;
\operatorname{svd}_{r}\!\big( \wh B \big)
\]
\State \textbf{Normalization:}
\[
\wh U \;\gets\; \wh U_B \, \big( \wh U_B^\top\, \wh \Sigma_X \, \wh U_B \big)^{-1/2},
\qquad
\wh V \;\gets\; \wh V_B \, \big( \wh V_B^\top\, \wh \Sigma_Y \, \wh V_B \big)^{-1/2},
\]
\Statex \textit{where } $\wh\Sigma_{X}=X^{\top}X/n$,
$\wh\Sigma_{Y}=Y^{\top}Y/n$,\;
$\|B\|_{11}=\sum_{i,j}|B_{ij}|$, $d_k(\cdot)$ returns the k-th singular value of a matrix
and $\operatorname{svd}_{r}(\cdot)$ returns the rank-$r$ SVD.
\end{algorithmic}
\end{algorithm}

\subsection{Characterizing the computational complexity of our approach}\label{subsec:algo}
\label{subsec:implementation}
The efficiency of our method is tied to that of the  solver of the penalized regression problem  of Equation~\ref{eqn:lasso}. Here, we propose solving it using the Alternating Direction Method of Multipliers (ADMM; \cite{boyd2011}). 
 Appendix~B provides a detailed breakdown of the updates and steps of the algorithm.
    
Let $T$ denote the number of iterations, and assume $n \leq p, q$. The total computational complexity of our algorithm is then 
$O\big(p^2n + q^2n + T( (p \wedge q)n^2 + pqn)\big)$,
which scales quadratically with the larger dimension. For comparison, the Fantope-based initialization in \citet{gao2023sparse} also employs ADMM steps. However, this approach requires computing at each step the SVD of a $(p + q) \times (p + q) $ matrix, resulting in a total complexity of 
$O\big(T(p+q)^3\big)$ that scales cubically with the larger dimension \citep{gao2023sparse}.

\subsection{CCA with group sparsity}\label{subsec:high_dim_group}

In many datasets, variables are naturally organized into groups. For example, genes may be grouped by biological function, brain voxels by anatomical region, and geographical measurements by spatial location. In such settings, a common objective is to identify a sparse subset of these covariate groups that captures the key relationships in the data. To address this, we extend the sparse CCA method introduced in the previous subsection to incorporate group structure. Specifically, we replace the sparsity penalty in Equation~\ref{eqn:lasso} by a group-sparse modification leading to:
\begin{equation}  \mathcal{L}(B) = \frac{1}{2} \Big\|\frac 1nX B Y^\top - I_n\Big\|_F^2 + \rho \sum_{g \in G} \sqrt{T_g}\|B_g\|_{F}.
\label{eq:gsparse}
\end{equation}
Here $G$ denotes the set of groups, $B_g$ denotes the restriction of the matrix $B$ to the $g$th group, while $T_g$ denotes the size of group $g$. 
As in Section \ref{subsec:implementation}, the minimizer of Equation~\ref{eq:gsparse} can be efficiently found via an ADMM procedure (Appendix~B).

\section{Theoretical Guarantees}\label{sec:theory}

In this section, we first demonstrate that the constrained optimization setting of Equations~\ref{eqn:lasso} and~\ref{eq:gsparse} allows accurate estimation of the product \(B^\star = U^{\star} \Lambda^{\star} V^{\star\top}\). Combined with steps 2 and 3 of Algorithm~\ref{alg:eccar}, our procedure can thus be shown to produce consistent estimates of the canonical directions $U^\star$ and $V^\star$.

\subsection{Consistency in the sparse high-dimensional setting}
The following theorem  provides an upper bound on the distance between \(\wh{B}\) and \(B^\star\).

\begin{theorem}
     Consider the parameter space $\mathcal{F}(s_u, s_v, p, q, r; M )$ for the covariance matrix $\Sigma$ (Equation~\ref{reparam} and conditions therein), and let $\Delta = \widehat{B}-B^\star$, where $\widehat{B}$ is the estimate obtained in step 1 of Algorithm~\ref{alg:eccar} and $B^\star$ is the underlying population quantity: $B^\star = U^\star \Lambda^\star V^{\star \top}$.
     Assume \(
         n \geq c s_u s_v \log(p + q) \)  for some sufficiently large constant~$c$.          
         There exist constants $a, b, C$ depending on $M$ and $c$ such that  if ${\rho \geq a\sqrt{\log(p + q)/n}} $,  then with probability at least 
         $1 - \exp(-b s_u\log(ep/s_u)) - \exp(- b s_v \log(eq/s_v )) - (p + q)^{-b}$, we have: 
            $$ \|\Delta\|_F \le  C  \rho \sqrt{s_u s_v}.$$  
In particular, if $\rho$ is chosen on the order of
$\sqrt{\log(p + q)/n}$, then
\begin{align}\label{eq:bound_D}
\|\Delta\|_F \lesssim
\sqrt{\frac{s_u s_v \log(p + q)}{n}},
\end{align}
and the estimator $\widehat{B}$ is sparse in the sense that
\[
\|\widehat{B}\|_0 \lesssim s_u s_v .
\]
    \label{theorem1}
\end{theorem}

Our approach's error bound compares favorably to that of the initialization procedure proposed by \citet{gao2017sparse}. Their two-step method first estimates the product  \(A^\star = U^\star V^{\star\top}\) using a Fantope-based approach (Theorem 4.1), resulting in an error bound that scales with the inverse of the signal strength $\lambda^\star_r$: $\|\widehat{A} - A\|_F \lesssim \frac{1}{\lambda^{\star}_r}\sqrt{\frac{s_us_v \log(p+q)}{n}}$. We achieve the same bound, but without the $\lambda^{\star}_r$-term in the denominator. This difference arises because our method directly targets the complete product  \(U^\star \Lambda^\star V^{\star\top}\).

From a computational standpoint, our method offers a distinct advantage. The approach in \citet{gao2017sparse} involves a costly projection onto a Fantope set at each iteration. In contrast, our algorithm is projection-free and relies only on efficient matrix multiplications, resulting in significantly improved speed. Building on this, we now demonstrate how our initial estimate of $B$ facilitates the final estimation of $U^\star$ and $V^\star$.

{\begin{theorem}
     Suppose that the assumptions of Theorem \ref{theorem1} are satisfied. Assume $n \geq c s_u s_v \log(p + q)/\lambda^{\star 2}_r $ for some sufficiently large constant $c$.  
     There exist constants $a_1, a_2, b, C_1, C_2$ depending on $M$ and $c$ such that if $\rho\in \Big[a_1 \sqrt{\frac{\log(p + q)}{n}}, a_2 \sqrt{\frac{\log(p + q)}{n}}\Big]$,   then with probability at least $1 - \exp(-br) - \exp(-b (s_u + \log(ep/s_u)) )- \exp(- b (s_v + \log(eq/s_v ))) - (p + q)^{-b}$, we have 
     $$\|\wh \Lambda - \Lambda^\star\|_F \le C_1\sqrt{ \frac{s_u s_v \log(p + q)}{n} }.$$
     $$ {\max\Big\{\min_{W \in \mathcal{O}_{r}} \| \widehat{U} - U^\star W \|_F, \min_{\tilde W \in  \mathcal{O}_{r}} \| \widehat{V} - V^\star \tilde{W} \|_F \Big\}   \le  C_2 \frac{1}{\lambda^{\star}_r } \sqrt{ \frac{s_u s_v \log(p + q)}{n} }}$$      
    \label{theorem2}
\end{theorem}

}

Theorem~\ref{theorem2} provides error bounds for $\wh{U}$ and $\wh{V}$. We note that our bound scales with the square root of the product $s_u s_v$, thus, is less efficient (from a statistical perspective) than the one in \citet{gao2017sparse}, for which the authors showed that:
\[
 \min_{W \in \mathcal{O}_r} \left\|  \wh{U} - U^\star W \right\|_F \leq C \frac{1}{\lambda^{\star}_r M^{\frac12}} \sqrt{\frac{s_u (r + \log(p))}{n}},
 \]
which is independent of $q$ and $s_v$ and scales with the inherent dimension of the problem $s_u r$ (shown by the authors to be minimax optimal). However, their procedure has practical limitations. It requires splitting the data into three folds, which reduces statistical power, and relies on a computationally intensive Fantope projection --- an approach rarely used in practice. While a more recent method by \citet{gao2023sparse} avoids sample splitting and achieves a bound of order $s_u + s_v$, it still depends on this computationally expensive initialization and requires tuning multiple hyperparameters. Therefore, our method offers a valuable trade-off: it sacrifices some statistical efficiency in exchange for substantial computational gains, making it a practical and scalable solution for sparse CCA.

\subsection{Guarantees on the support recovery}

{Canonical Correlation Analysis (CCA) is frequently employed to identify associated subsets of variables between two datasets, a process central to scientific discovery. In this context, an important statistical objective is to avoid selecting variables that do not belong to the true canonical subspaces. This result establishes control of false positive discoveries of $\wh B$.}

\begin{theorem}\label{theorem:supp_recov}
Consider the parameter space $\mathcal{F}(s_u, s_v, p, q, r; M )$ for the covariance matrix $\Sigma$ (Equation~\ref{reparam} and conditions therein). Assume 
$n > \max\{s_u, s_v\}$   and 
     \begin{equation}
        \frac{\| \wh{\Sigma}_{XY} -  \wh{\Sigma}_X B^\star \wh{\Sigma}_Y\|_\infty}{\rho} +  \frac{2(\| \wh{\Sigma}_{XY} -  \wh{\Sigma}_{X} B^\star \wh{\Sigma}_{Y}\|_\infty + \rho )}{\rho}\frac{\wh{\tau}(X, Y) \sqrt{s_u s_v} }{\sigma_{\min}( (\wh{\Sigma}_{X})_{S_uS_u}) \sigma_{\min}( (\wh{\Sigma}_{Y})_{S_vS_v}) }  < 1,
     \end{equation}
     where 
     $S_u = \operatorname{supp}(U^\star) \mbox{, } S_v = \operatorname{supp}(V^\star)$ and
     \begin{align*}\wh{\tau}(X, Y ) = \max\Big\{ &\big\|(\wh{\Sigma}_X)_{S_u^cS_u}    \big\|_{2, \infty} \big \|(\wh{\Sigma}_Y)_{S_v^cS_v}    \big\|_{2, \infty} ,\big\|(\wh{\Sigma}_X)_{S_u^cS_u}    \big\|_{2, \infty}\big\| (\wh{\Sigma}_Y)_{S_vS_v}\big\|_{op}, \\
     &\big\|(\wh{\Sigma}_X)_{S_uS_u} \big\|_{op} \big\|(\wh{\Sigma}_Y)_{S_v^cS_v}   \big\|_{2, \infty}\Big\}.
     \end{align*}  Then the minimizer of  Equation \ref{eqn:lasso} is unique and satisfies
        $\text{supp}_e(\wh{B}) \subset S_u \times S_v.$
    \label{thm:support}
\end{theorem}

\begin{remark}[Intuition for $\tau$] The population counterpart of $\hat{\tau}(X,Y)$ (denoted here 
$\tau(X,Y)$) is an incoherence (cross‑talk) index: it aggregates the worst‑case correlation between off‑support rows and the on‑support rows of $\Sigma_X$	and $\Sigma_Y$, adjusted by the conditioning of the on‑support blocks. Small 
$\tau$ means variables outside the true supports cannot be well represented by those inside, so the $\ell_{11}$ penalty can exclude them; large 
$\tau$ indicates strong leakage and makes support recovery harder. In the block‑diagonal ideal where $ (\Sigma_X)_{S_u^cS_u} = 0, (\Sigma_Y)_{S_v^cS_v} = 0$, we have 
$\tau=0$ and the support condition is trivially satisfied.    
\end{remark}
To make the assumptions of the previous theorem more transparent (and deterministic), we refine Theorem~\ref{theorem:supp_recov} by the following corollary providing guarantees on the support of $\widehat{B}$ with high probability.
\begin{corollary}
      Consider the parameter space $\mathcal{F}(s_u, s_v, p, q, r; M )$ for the covariance matrix $\Sigma$ (Equation~\ref{reparam} and conditions therein), and assume
       \begin{equation}
        \frac{2 {\tau}(X, Y)\sqrt{s_u s_v} }{\sigma_{\min}\big( ({\Sigma}_{X})_{S_uS_u}\big) \sigma_{\min}\big(({\Sigma}_{Y})_{S_vS_v}\big) }  < 1 - \alpha
     \end{equation}
     for some $\alpha \in (0, 1]$,
     where $\tau(X,Y)$ is the population version of $\wh\tau(X, Y)$.
There exist some constants $a, b, c$ depending on $M$ and $\alpha$ such that  if 
   $n \geq c (s_u + s_v ) s_us_v\log(p + q)$
  and $\rho \geq a\sqrt{\log(p + q)/n} $, then the minimizer of  \ref{eqn:lasso} is unique and satisfies 
      $\text{supp}_e(\wh{B}) \subset S_u \times S_v
     $ with probability at least $1 - \exp(-b s_u\log(ep/s_u)) - \exp(- b s_v \log(eq/s_v )) - (p + q)^{-b}$.
\label{corollary}
\end{corollary}

{Corollary~\ref{corollary} provides false-positive selection control for $\widehat B$.} The conditions of Theorem~\ref{theorem:supp_recov} and Corollary~\ref{corollary} ensure that if the covariance between variables within the true support ($S_u$) and those outside of it ($S_u^c$)  is sufficiently small, then the non-zero entries of the estimator  $\widehat{B}$
  will be contained within the true support of $B^\star$. This requirement is intuitive: when two variables are highly correlated, a robust method should not assign a large canonical loading to one while excluding the other. 
{
  If in addition we assume that \begin{equation}
        \min \left \{ \min_{i \in S_u}\| B^\star_{i \cdot }\|, \min_{j \in S_v}\| B^\star_{\cdot j}\| \right\} \gtrsim \sqrt{\frac{s_us_v \log(p + q)}{n}},
     \end{equation} then the estimation error bound in Theorem~\ref{theorem1},  with $\rho$  chosen on the order of
$\sqrt{\log(p + q)/n}$, immediately implies that\begin{equation}
    S_u \subseteq\mathrm{supp} (\widehat B),
\qquad
S_v \subseteq \mathrm{supp}(\widehat B\,^\top),  
     \end{equation}  provided that $n/ \sqrt{s_us_v\log(p+q)} $ is sufficiently large.
Combining this with the support-containment guarantee in Corollary~\ref{corollary} yields exact recovery of the row and column supports.} Consequently, since by the design of our algorithm, sparsity in  $\widehat{B}$ implies sparsity in $\wh{U}$ and $\wh{V}$, our theorem's guarantees on support recovery extend directly to  
$\wh{U}$ and $\wh{V}$. This implies that the set of non-zero elements in the estimated vectors is a subset of the true support.

 \subsection{Rank Selection}

Rank selection is a practical challenge in existing sparse CCA procedures \citep{gao2017sparse, gao2023sparse}, particularly for methods based on
Fantope projections, which require projection onto an $r$-dimensional space
at every iteration. In contrast, the first stage of
our approach is rank-agnostic and does not require prior knowledge of $r$.
It directly estimates the product $U^\star \Lambda^\star V^{\star\top}$,
a matrix of rank $r$.
The following corollary characterizes the singular values of { $\wh B$ and }
$\widehat{\Sigma}_X^{1/2}\,\widehat{B}\,\widehat{\Sigma}_Y^{1/2}$.

\begin{corollary}
     Suppose that the assumptions of Theorem \ref{theorem1} are satisfied. Assume $n \geq c s_u s_v \log(p + q)/\lambda^{\star 2}_r $ for some sufficiently large constant $c$. {Denote the k-th singular value of a matrix $A$ by  $d_k(A)$.}
     There exist constants $a_1, a_2, b, C$ depending on $M$ and $c$ such that if $\rho\in \Big[a_1 \sqrt{\frac{\log(p + q)}{n}}, a_2 \sqrt{\frac{\log(p + q)}{n}}\Big]$,   then with probability at least $1  - \exp(-b (s_u + \log(ep/s_u)) )- \exp(- b (s_v + \log(eq/s_v ))) - (p + q)^{-b} $, we have 
{
$$d_r(\wh B) \geq \frac{\lambda_r^\star}{M} - C \sqrt{ \frac{s_u s_v \log(p + q)}{n} }, \quad  d_r(\widehat{\Sigma}_X^{1/2}\,\widehat{B}\,\widehat{\Sigma}_Y^{1/2}) \geq \lambda_r^\star - C  \sqrt{ \frac{s_u s_v \log(p + q)}{n} } $$

$$d_{r + 1}(\wh B) \le C  \sqrt{ \frac{s_u s_v \log(p + q)}{n} }, \quad d_{r + 1}(\widehat{\Sigma}_X^{1/2}\,\widehat{B}\,\widehat{\Sigma}_Y^{1/2}) \le C \sqrt{ \frac{s_u s_v \log(p + q)}{n} }$$}
     
    \label{corollary:rank}
\end{corollary}

{The proof of Corollary~\ref{corollary:rank} is provided in  Appendix~F.1.5.  Corollary \ref{corollary:rank} shows that the leading $r$ singular values of  $\widehat{\Sigma}_X^{1/2}\,\widehat{B}\,\widehat{\Sigma}_Y^{1/2}$  concentrate around
the population canonical correlations $\Lambda^\star$, while the remaining singular values 
are dominated by estimation noise. Consequently, instead of using the
singular values of $\widehat{\Sigma}_{XY}$, which can be highly unstable in
high-dimensional settings, one can examine the spectrum of 
$\widehat{\Sigma}_X^{1/2}\,\widehat{B}\,\widehat{\Sigma}_Y^{1/2}$ as a
denoised surrogate. Alternatively, we also show that the spectrum of $\widehat{B}$ provides an alternative way to determine the rank, as its singular values also exhibit a gap around the true rank whenever the smallest canonical correlation (divided by $M$) dominates the estimation noise. Contrary to $\widehat{\Sigma}_X^{1/2}\,\widehat{B}\,\widehat{\Sigma}_Y^{1/2}$, though,  its singular values are not directly equal to the canonical correlations and are thus less directly interpretable. 

\begin{wrapfigure}{r}{0.4\textwidth}
  \centering
  \includegraphics[width=\linewidth]{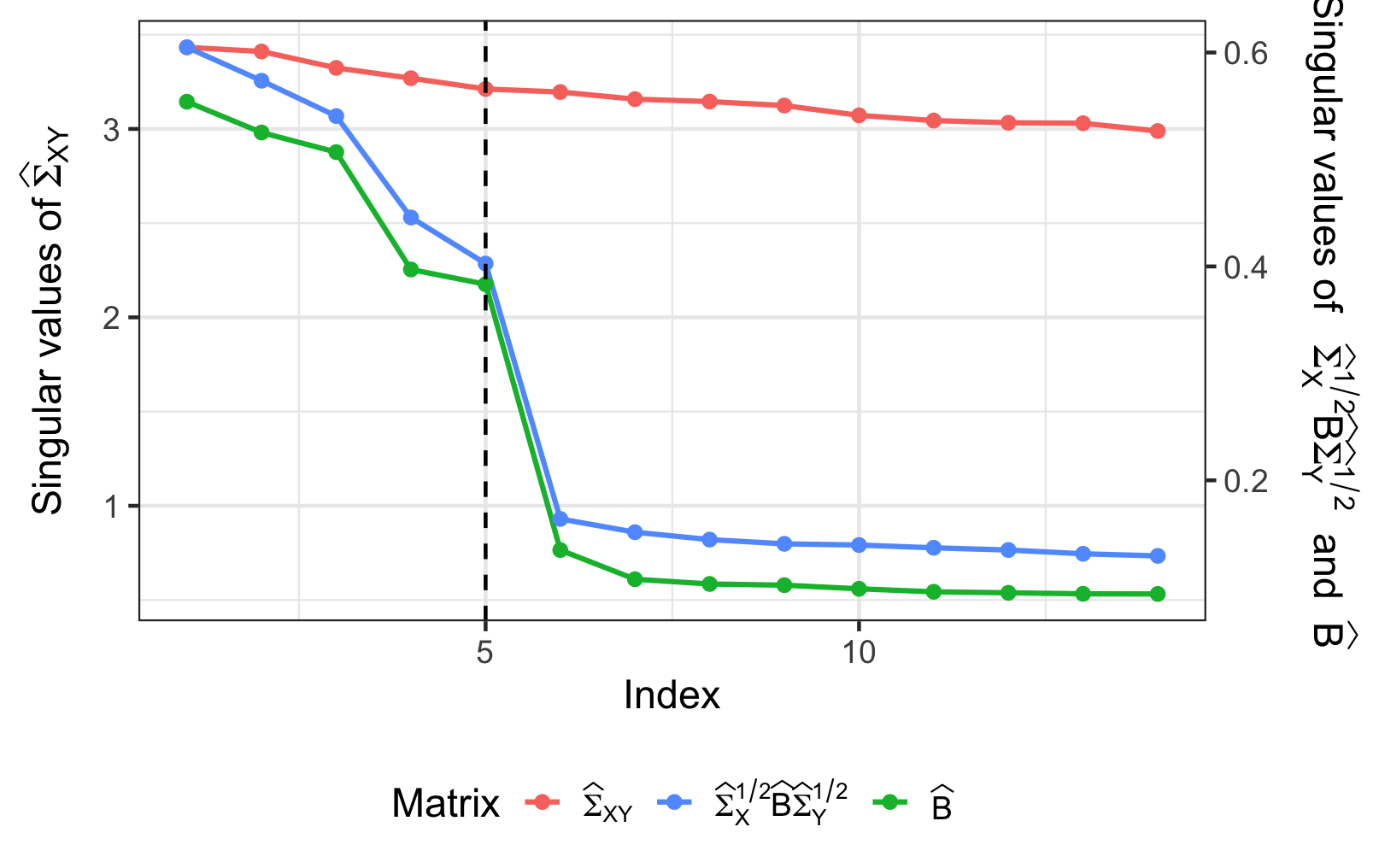}
  \caption{Comparison of the singular values of $\widehat{B}$, $\widehat{\Sigma}_X^{1/2}\,\widehat{B}\,\widehat{\Sigma}_Y^{1/2}$ and $\widehat{\Sigma}_{XY}$ in a simulation setting with $n=400$, $p=q=600$, $r=5$, and $\lambda_i=0.7$ for $i=1,\ldots,5$ (see Section \ref{sec:sim} for details). The black dashed line indicates the true rank. While the singular values of $\widehat{\Sigma}_{XY}$ decay smoothly and do not clearly reveal the rank, those of $\widehat{\Sigma}_X^{1/2}\,\widehat{B}\,\widehat{\Sigma}_Y^{1/2}$  and $\widehat{B}$ exhibit a clear separation that allows accurate identification of the true rank.}
  \label{fig:eigens}
\end{wrapfigure}

In practice, the rank $r$ may thus be selected by identifying a spectral gap
in these matrices or by thresholding singular values below a noise level (e.g., using, for instance, the Kneedle algorithm \cite{satopaa2011finding}). For example, Figure~\ref{fig:eigens} illustrates a simulation setting in which the rank can be clearly identified from the spectra of $\widehat{\Sigma}_X^{1/2}\,\widehat{B}\,\widehat{\Sigma}_Y^{1/2}$ or $\widehat{B}$. In the real-world experiments presented in the next section, we provide visualizations of the screeplots for all three matrices (  $\widehat{\Sigma}_X^{1/2}\,\widehat{B}\,\widehat{\Sigma}_Y^{1/2}$, $\widehat{B}$ and $\widehat{\Sigma}_{XY}$) to show how these can be used to select an appropriate rank.

Alternatively, the rank may be selected using a permutation-based procedure, in the spirit of the parallel analysis proposed in \citet{horn1965rationale}. Let $\widehat{\Lambda}_0$ denote the singular values obtained
from the SVD of
$\widehat{\Sigma}_X^{1/2}\,\widehat{B}\,\widehat{\Sigma}_Y^{1/2}$.
To approximate the null distribution, one may randomly permute the rows
of $X$ (or $Y$), thereby preserving the marginal distributions of the two
datasets while destroying any cross-view dependence, and then recompute
the estimator. Repeating this procedure yields a collection of singular
values under the null hypothesis of no association. The rank $r$ can then
be chosen as the number of leading singular values of $\widehat{\Lambda}_0$
that exceed a reference level derived from the permutation replicates,
for example, their mean or 95\% quantile.

}

{
\subsection{Handling Collinearity in High Dimensions}
 
In high-dimensional applications, the sample covariance matrices of $X$
and $Y$ are typically ill-conditioned. In such regimes, these matrices
can be nearly singular, exhibit unstable eigenstructures, and provide
unreliable estimates of the underlying population covariances.  To cope
with such collinearity, Lasso-type methods typically rely on sparsity,
effectively restricting estimation to a subset of the support so that the
relevant design matrix only needs to be estimated locally. A similar
principle applies here: the solution depends on a small proportion of the
coordinates of $X$ and $Y$, which prevents the algorithm from breaking down
even when the full design matrices are ill-conditioned. Our proposed ADMM algorithm also remains numerically stable
even under strong collinearity; in fact, when the empirical covariance matrices are ill-conditioned (e.g. when $n\ll p$), the algorithm explicitly leverages the low rank of the covariance matrices to improve its efficiency by performing most of its updates in the image space of $\widehat{\Sigma}_X$ ($\widehat{\Sigma}_Y$, respectively).  A detailed discussion is provided in
Appendix~B.

However, as in standard Lasso problems, the estimator tends to select a
single variable from a group of highly correlated variables, typically the
one most strongly associated with the opposite view, while excluding the
others, even when they all belong to the true support. This behavior favors
maximizing cross-view correlation with minimal sparsity.

To mitigate this issue, additional preprocessing  (e.g., choosing one representative amongst a cluster of correlated variables) or regularization may be
introduced. For example, one can incorporate an elastic net penalty into the
optimization objective to encourage the joint selection of correlated variables.
Alternatively, shrinkage estimators for the covariance matrices may be used
\citep{ledoit2004well}. When prior knowledge about grouped covariates is
available, a group-Lasso penalty can further promote the joint selection of
correlated variables rather than choosing a single representative, as
discussed in Section~\ref{subsec:high_dim_group}.}

\section{Simulations}
\label{sec:sim}
We conduct a series of synthetic experiments to evaluate the performance of different sparse CCA approaches in high-dimensional settings. 
We begin by constructing a block-diagonal covariance matrix \(\Sigma_{X}\), defined as $\Sigma_{X} = \left(\begin{smallmatrix}    
       D_X +  U_X \Lambda_X U_X^\top & 0\\
        0 &  I
    \end{smallmatrix}\right),$
where \(U_X \in \mathbb{R}^{p_1 \times r_{\mathrm{pca}}}\) is a randomly generated matrix with orthonormal columns. The matrices  \( D_X \in \mathbb{R}^{p_1 \times p_1}, \Lambda_X \in \mathbb{R}^{r_{\mathrm{pca}}\times r_{\mathrm{pca}}}\)  are diagonal and chosen so that $\Sigma_X$ has unit diagonal entries. 
This construction induces a non-trivial covariance structure for \(X\), in which the leading \(r_{\mathrm{pca}}\) principal components are sparse and span only \(p_1\) out of the total \(p\) rows. In our experiments, we set \(p_1 = 20\) and $r_{\mathrm{pca}} = 5$. The covariance matrix \(\Sigma_Y\) is constructed analogously. 

Next, we construct the cross-covariance matrix \(\Sigma_{XY}\) as
$\Sigma_{XY} = \Sigma_X U^{\star} \Lambda^{\star}  V^{\star \top} \Sigma_Y,$
where \(U^{\star} \in \mathbb{R}^{p \times r}\) and \(V^{\star} \in \mathbb{R}^{q \times r}\) are random matrices with uniformly sampled entries. Both \(U^{\star}\) and \(V^{\star}\) are row-sparse, with exactly $s_u = s_v$ non-zero rows. The matrices are then normalized, ensuring that \(U^{\star \top} \Sigma_X U^\star = I_r\) and \(V^{\star \top} \Sigma_Y V^\star = I_r\). 
{We define the canonical correlation matrix as \(\Lambda^\star = \lambda I_r\)  with $\lambda = 0.9$ indicating ``high'' signal strength, $0.7$ for ``medium'', and $0.5$ for ``weak''.}
Finally, we generate the observations \((X_i, Y_i)_{i=1}^n\) by sampling from the joint multivariate normal distribution
$\mathcal{N}_{p+q}\left(0, \left(\begin{smallmatrix}
    \Sigma_X & \Sigma_{XY} \\
    \Sigma_{XY}^\top & \Sigma_Y
\end{smallmatrix}\right) \right).$  To test the robustness of our method under different distributions with potentially heavier tails, we also consider sampling from a t-distribution.

We use this synthetic data to compare the performance of our sparse CCA method with the following four versions of CCA with sparsity:  the SAR algorithm of \citet{wilms2015sparse}, the sparse CCA approach of \citet{witten2009penalized}, as well as that of \citet{waaijenborg2009} and \citet{parkhomenko2009sparse}. We also include a comparison of our method against the theory-based approaches, including the Fantope initialization, as well as the full CCA procedure (CoLAR) of \citet{gao2017sparse}, and SGCA by \citet{gao2023sparse}. 
In all versions of our method, we select an optimal regularization parameter $\rho$ by either plugging in the theoretical value of $\rho$ from Theorem~\ref{theorem1} or by 10-fold cross-validation.  Similarly, we select the regularization parameters in all heuristic methods using either 5-fold cross-validation or a BIC criterion to avoid the need for refitting the method multiple times. 
Due to the computational complexity of the theory-based methods, we use the theoretical value of $\rho$ in the Fantope initialization procedure instead of relying on cross-validation. We cross-validate on the value of the regularization in the second step of CoLAR, and use the default values in SGCA. 
To evaluate the resulting CCA reconstructions, we measure the distances between subspaces that span the canonical variates by computing the principal angle  between the subspaces associated with the matrices \citep{jordan1875essai}.

\paragraph{Estimation error with increasing $p$.}
 We set $n=400$ and $p = q = 200 , 400 , 600 , 800 , 1000 $ with $r=2$. For the group-sparse ECCAR, we divide the cross covariances into 5 by 5 blocks and treat each block as a group. We consider three sparsity levels, $s_u  =s_v  =5,10,15$. We estimate the $\sin \Theta$ distance between the stacked $\left(\begin{smallmatrix}
     U^\star\\
     V^\star
 \end{smallmatrix}\right) \in \R^{ (p+q) \times r}$ and the estimates, and the running time based on 25 independent experiments. The results for our method and best performing benchmarks are displayed in Figure \ref{fig:sim(p)} (a more complete figure with the full set of benchmarks is provided in Appendix~C.1).  Our method is able to handle very high-dimensional cases such as $p = q = 1000$, even with weak signals when the canonical variates are sparse enough. In contrast, all other methods fail to achieve reasonable accuracy when $p \geq 400$. Similar results are observed for t-distributed data, as shown in Appendix~C.1.
 
 Strikingly, as shown in Figure~\ref{fig:sim_time}, our method (using a regularization parameter $\rho$ computed through theory) is 100 times faster than the Fantope and 1,000 times faster than both CoLAR \citep{gao2017sparse} and  SGCA \citep{gao2023sparse}. Even accounting for cross-validation, our method remains more than 10 times faster than these state-of-the-art methods (which, technically, ought to be cross-validated to ensure an optimal choice of hyperparameters), and is faster than popular heuristic methods such as that of \citet{witten2009penalized} and \citet{wilms2015sparse}.

\paragraph{Estimation error with increasing $r$.}
We set \(n = 400\), \(p = q = 200\), and \(r = 2, 3, 4, 5\) to examine how the performance of our method changes with respect to \(r\). We consider the same sparsity levels \(s_v=s_u\) and signal strength levels as in the previous simulation setting. The results are displayed in Figure~13  of Appendix~C.1. Notably, the performance of our method does not deteriorate with increasing \(r\).

\begin{figure}[p]
     \centering
     \includegraphics[width= 0.9\textwidth]{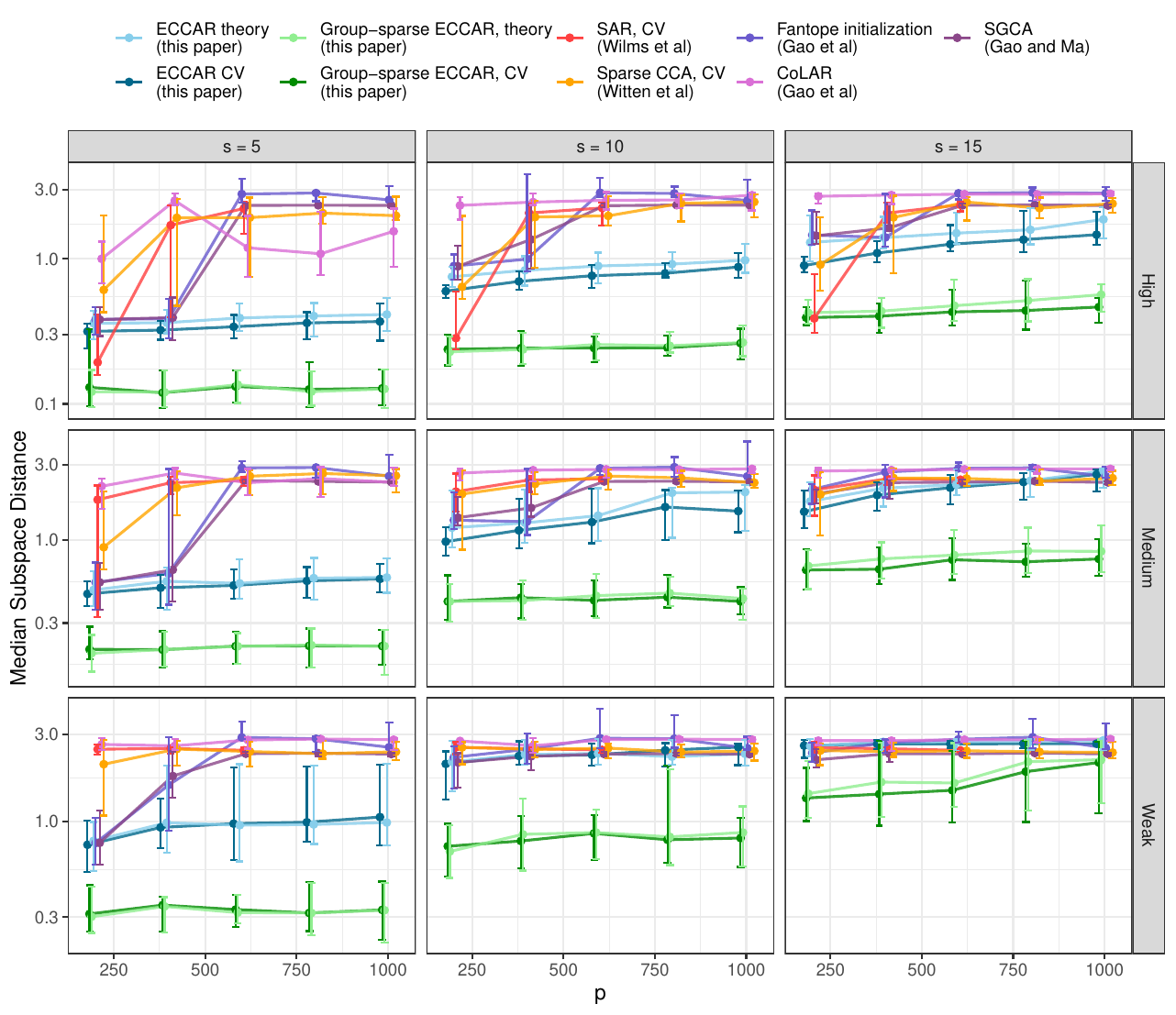}
\caption{Estimation error as a function of $p(=q)$, the support sizes $s_u=s_v = s$  (columns) and the strength of the signal (value of $\lambda^\star$, rows) for $n=400$. Points indicate the median subspace distance over 25 independent experiments  with their interquartile range.}
     \label{fig:sim(p)}
 \end{figure}

 \begin{figure}[p]
     \centering
     \includegraphics[ width= 0.9\textwidth]{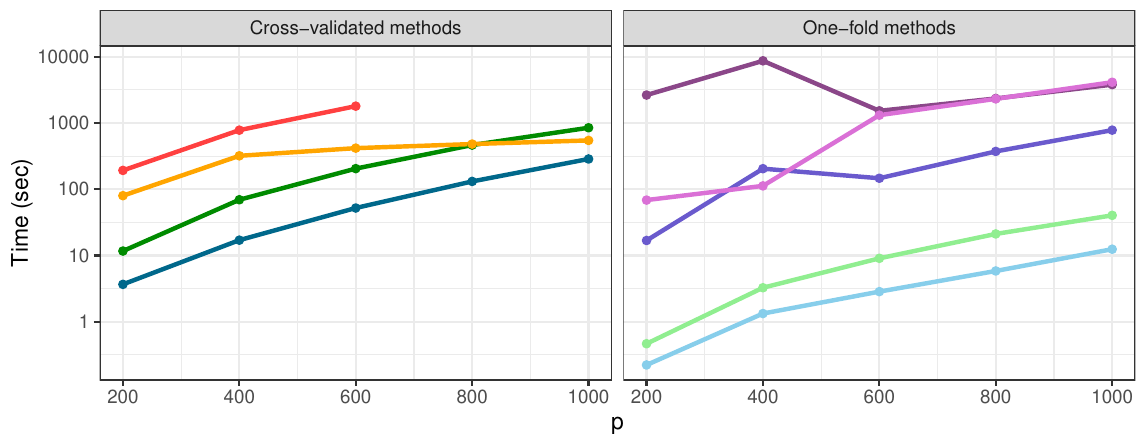}
\caption{Time as a function of $p$ and $q$ for $n=400$, $\lambda^\star=0.9$ and $s_u=5$. Points indicate median time over 25 independent experiments. The missing values in SAR are caused by computational issues that occurred during model fitting.}
     \label{fig:sim_time}
 \end{figure}

\section{Real-Data Analysis}
\label{sec:real}

In this section, we illustrate the benefits of our approach on four real-life examples stemming from different fields: genetics (Subsection~\ref{subsec:aud}), neuroscience (Subsection~\ref{subsec:abide}), natural language processing (Subsection~\ref{subsec:llm}) and transcriptomics (Subsection~\ref{subsec:stategra}). We further showcase our method on the Nutrimouse dataset (Appendix~C.2), often employed as a benchmark for CCA  \citep{wilms2015sparse, donnat2024canonical, mixomics}.

\subsection{Alcohol Use Disorder}\label{subsec:aud}

Alcohol-use disorders (AUDs) --- encompassing both abuse and dependence --- are major drivers of morbidity and premature death. Their etiologies are highly heterogeneous: risk is believed to arise from variation in many genes, gene-gene interactions, and extensive transcriptomic reprogramming triggered by long-term alcohol exposure. To investigate these risk factors, \citet{zhang2014differentially} profiled genome-wide DNA methylation levels and genome-wide gene expression levels in the postmortem prefrontal cortex of 23 European Australian AUD cases and 23 matched controls. The same dataset was later revisited in \citet{luo2016canonical} using supervised joint-modeling techniques.  Here we re-examine this dataset to assess the validity and efficiency of our proposed CCA approach (ECCAR). As in \citet{witten2009penalized} and \citet{luo2016canonical}, a marginal test was performed on all genes and CpG sites, and the top $p = 300$ genes and $q = 500$ CpG sites most strongly associated with AUD were selected for further analysis.

\begin{table}[t]
\centering
\begin{tabular}{lccc}
\toprule
\textbf{Method} & \textbf{Test} & \textbf{Test} & \textbf{Accuracy} \\
 & \textbf{MSE} & \textbf{ Correlation} & \textbf{ (SVM)} \\
\midrule
ECCAR, CV  (this paper)      & \textbf{0.604} & 0.400 &\textbf{0.958} \\ 
\hline
SAR, CV  (\citet{wilms2015sparse})     &  0.857  & 0.274 &  0.869   \\
\hline
SAR,  BIC  (\citet{wilms2015sparse})     & 0.905 &0.311  &  0.879 \\
\hline
Sparse CCA, CV  (\citet{witten2009penalized})        & 1.070  & 0.258 &  0.823 \\
\hline
Sparse CCA, permuted  (\citet{witten2009penalized})     & 0.731  & \textbf{0.483} & 0.933  \\
\hline
Sparse CCA, CV  (\citet{waaijenborg2009})   &  1.825 &-0.351 &  0.955  \\
\hline
Sparse CCA, BIC (\citet{waaijenborg2009}) & 1.865  & -0.383  & 0.955 \\
\hline
Fantope Initialization (\citet{gao2017sparse})     & 40.081  & 0.106 & 0.570  \\
\hline
SGCA (\citet{gao2023sparse})    & 1.617  &  0.193 & 0.736 \\
\bottomrule
\end{tabular}
\caption{Performance comparison of sparse CCA methods on the AUD data.}
\label{tab:AUD}
\end{table}

 { Following \citet{luo2016canonical}, we apply ECCAR with $r = 2$ to estimate the first two pairs of canonical directions. This choice is further supported by the screeplots of the empirical cross covariance matrix as well as of the matrix $\widehat{B}$ (see Figure 17 in Appendix C.3)}.  
The regularization parameter was tuned with an 8-fold cross-validation scheme, where six folds were used for training, one for validation (and selection of the hyperparameter) and one for testing. The optimal penalty parameter $\rho$ was selected by minimizing the average validation mean-squared error (MSE) of the difference
$XU - YV$. Table \ref{tab:AUD} compares the test MSE and correlation of ECCAR with those provided by the benchmarks.  To assess the downstream predictive utility of ECCAR, we projected the genetic data onto the estimated canonical directions and classified the resulting scores $XU$ as either AUD cases or controls using support vector machine (SVM). In this case, ECCAR achieved the lowest test MSE and the highest classification accuracy.

\begin{figure*}[p]
    \centering
    \begin{subfigure}[b]{0.31\textwidth}
        \centering
        \includegraphics[width = \textwidth]{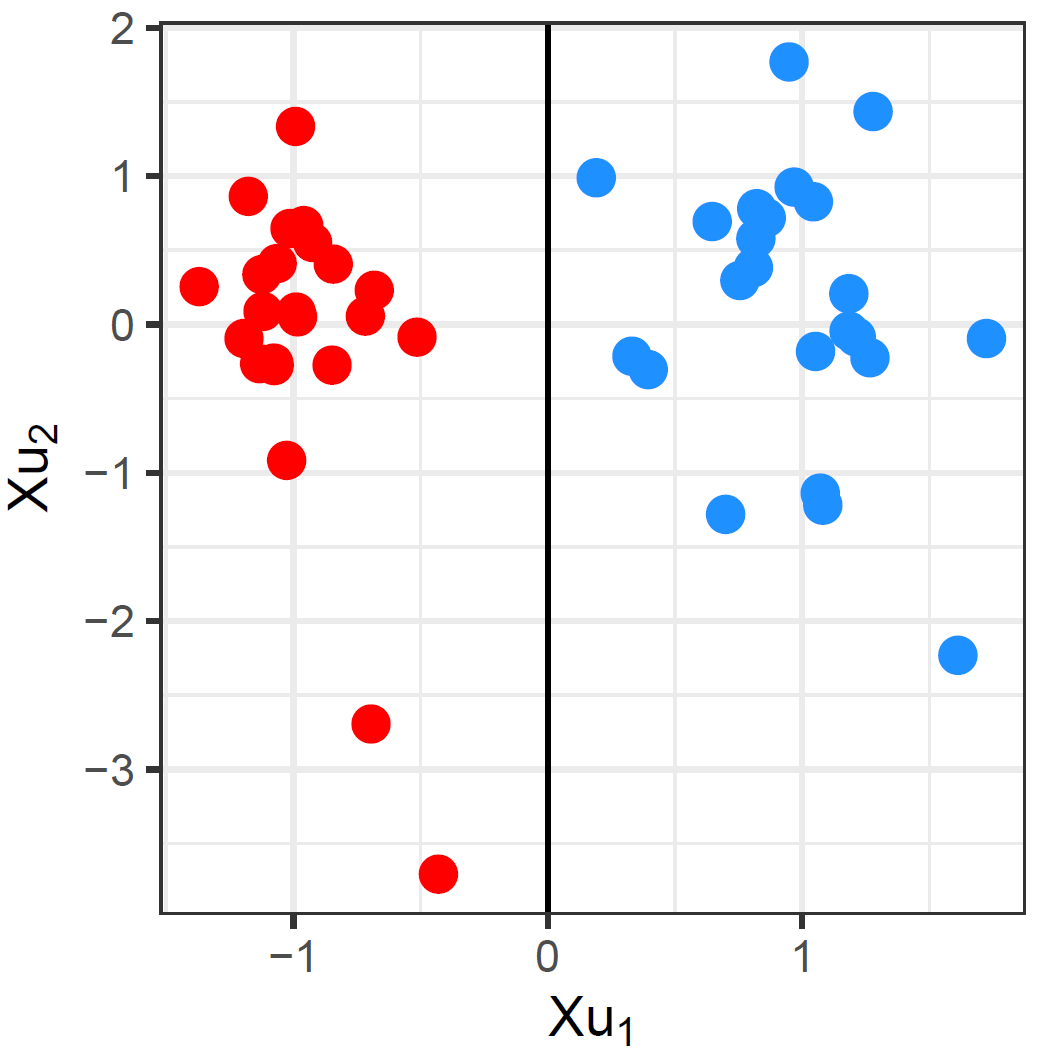}
        \caption{ECCAR}
    \end{subfigure}
    ~ 
    \begin{subfigure}[b]{0.4\textwidth}
        \centering
        \includegraphics[width = \textwidth]{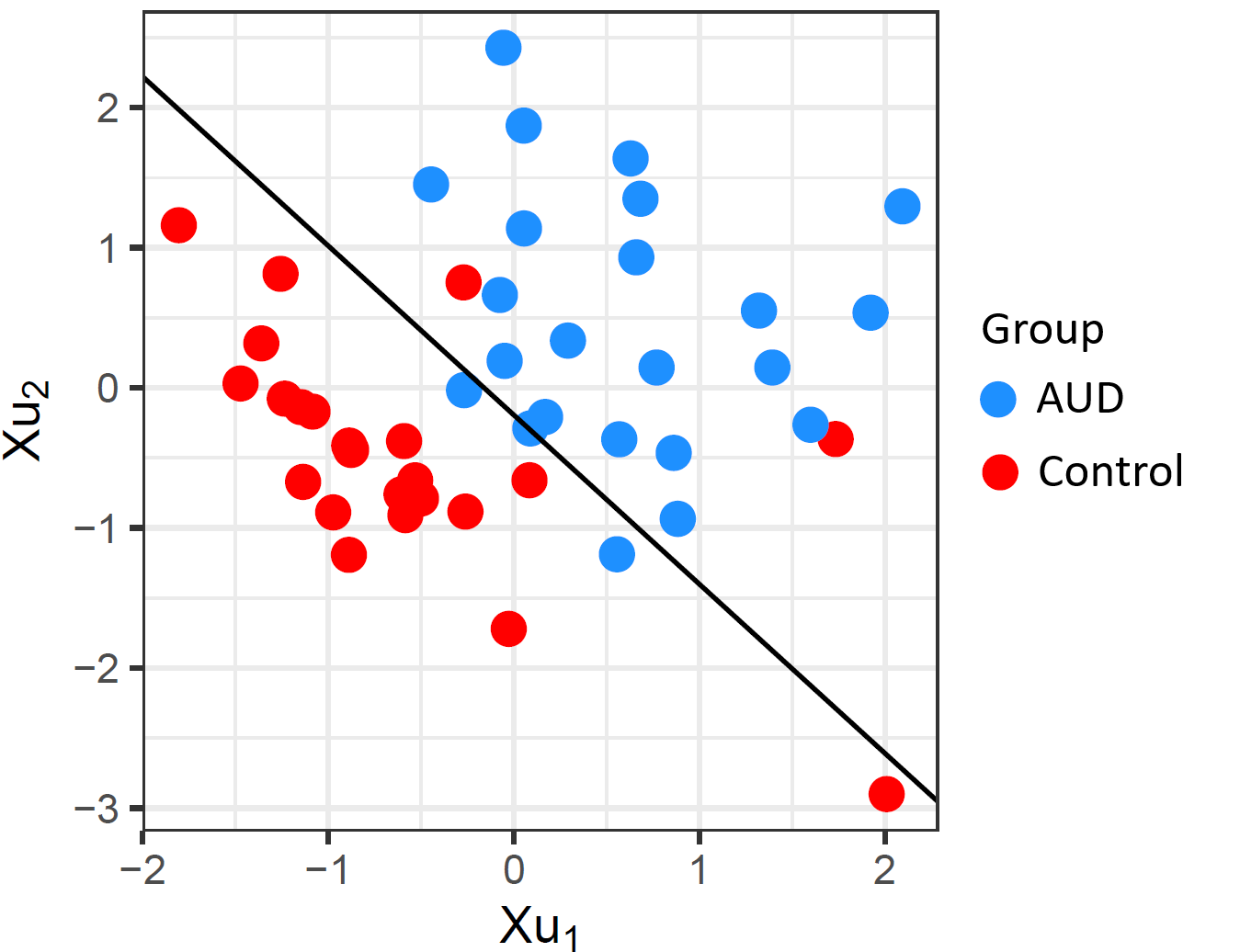}
        \caption{Sparse CCA by \citet{witten2009penalized}}
    \end{subfigure}
    \caption{The scatter plot for the
first two canonical variates produced for the AUD dataset. (a) The black line $Xu_1 = 0$ perfectly separates two classes. (b) The black line represents the logistic regression decision boundary.}
    \label{fig:aud_sep}
\end{figure*}

  \begin{figure}[p]
     \centering
     \includegraphics[width= 0.6 \textwidth]{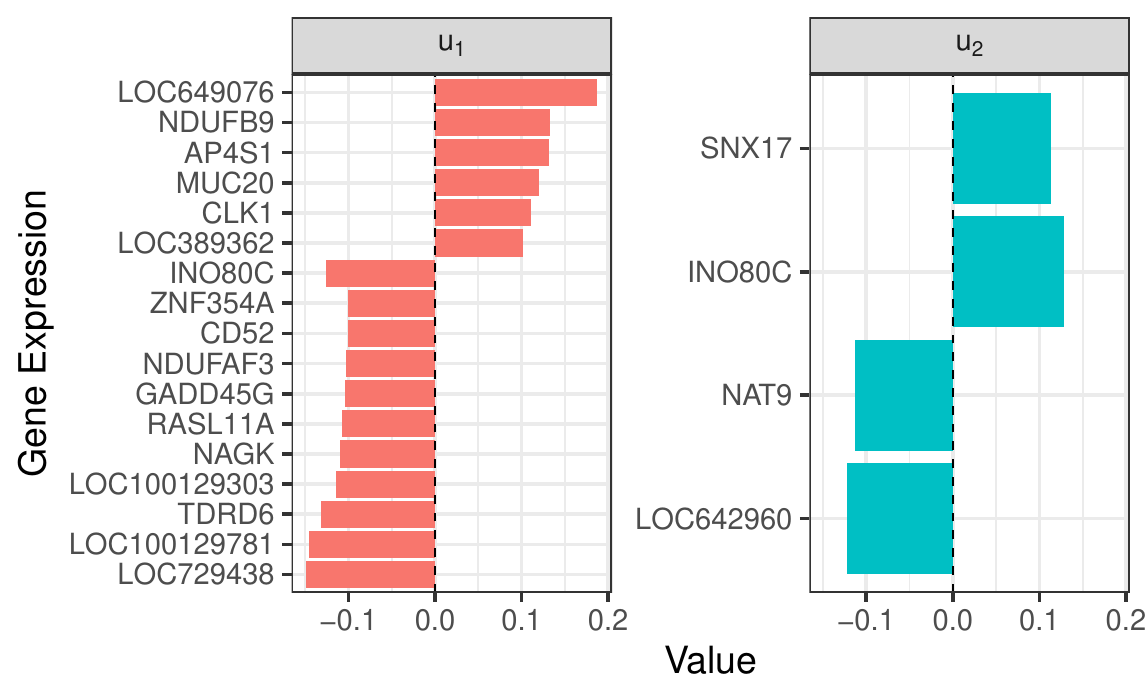}
\caption{The barplot for the first and second canonical direction vectors $u_1$ and $u_2$ corresponding to the genes obtained by ECCAR.}
     \label{fig:aud_gene}
 \end{figure}

  \begin{figure}[p]
     \centering
     \includegraphics[width= 0.6 \textwidth]{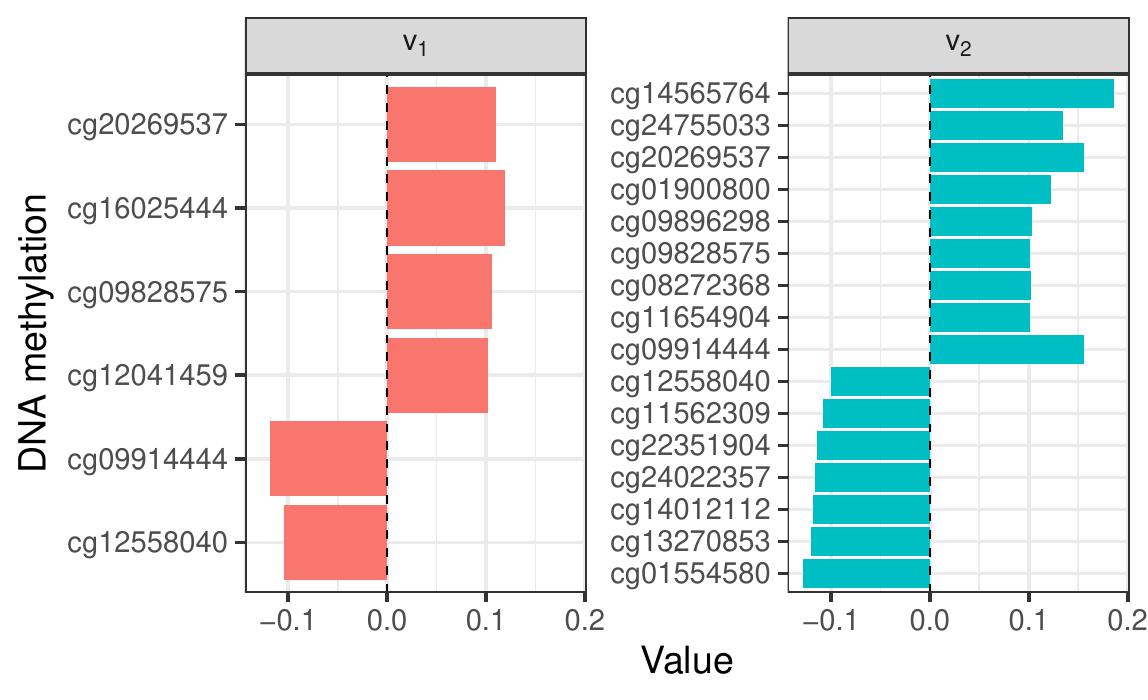}
\caption{The barplot for the first and second canonical direction vectors $v_1$ and $v_2$ corresponding to the CpG sites obtained by ECCAR.}
     \label{fig:aud_dna}
 \end{figure}

Figure \ref{fig:aud_sep} visualizes the separation achieved by ECCAR and by the sparse CCA variant of \citet{witten2009penalized}. Remarkably, ECCAR's first canonical variate yields a perfect separation between AUD cases and controls, while the approach of \citet{witten2009penalized} exhibits more mixing between AUD cases and controls.

Turning to the investigation of the first canonical variates, ECCAR identifies 20 genes  (Figure \ref{fig:aud_gene}) and 18 CpG sites (Figure \ref{fig:aud_dna}) associated with DNA methylation. Since this direction cleanly separates AUD from control cases, we subsequently compare our loadings to existing scientific literature. Notably, ECCAR's findings echo those of \citet{zhang2014differentially}, who reported the differential expressions of ZNF354A and RASL11A as significantly different between AUD and control cases. Additionally, NDUFAF3 was found to be significantly differentially expressed between alcohol-fed  and control pair-fed  mice, as noted by \citet{zhang2018label}. \citet{gavin2016role} suggested that lower levels of the DNA demethylation protein GADD45G may influence Bdnf expression, potentially altering alcohol consumption behavior. Moreover, GADD45G has been identified as a potential target for the anti-cancer effects of 4MOD in liver cancer \citep{zeng2023identify}, which is closely related to alcohol consumption. NDUFB9 was found to be differentially expressed in the medial prefrontal cortex of binge drinkers \citep{mcclintick2018gene}. These results provide more scientific validation for the effectiveness of our method. There is limited evidence regarding the identified CpG sites due to the sparse existing literature on DNA methylation functions. Nevertheless, we found that cg20269537 and cg11562309 are associated with AUD cases, as previously reported by \citet{hu2018ancogeneDB}.

\subsection{The Autism Brain Imaging Data Exchange}\label{subsec:abide}

To further evaluate our method, we use preprocessed resting-state fMRI data from the Autism Brain Imaging Data Exchange (ABIDE; \citep{di2014autism}) to assess ECCAR's performance. ABIDE is a multi-site consortium that provides data from subjects with Autism Spectrum Disorder (ASD) and controls. 
For the purpose of this example, we consider cases for which both fMRI time series and Vineland Adaptive Behavior tests \citep{sparrow2005vineland} were available, resulting in a set of 106 participants (65 with an autism diagnosis and 41 controls). The Vineland Adaptive Behavior tests consist of 14 behavioral scores measuring skills such as communication, coping, daily living skills, and socialization. The resulting scores were stored row-wise in matrix \(Y\), with \(q = 14\).

For each subject, mean fMRI time series were extracted for a set of 110 regions of interest (ROIs) parcellated using the Harvard-Oxford (HO) atlas \citep{jenkinson2012fsl,desikan2006automated}. A \(110 \times 110\) functional connectivity matrix was then computed by correlating the mean time series between all ROI pairs. The upper-triangular elements of each subject’s connectivity matrix were vectorized and stored row-wise in \(X\), yielding \(p = 5,995\) features per subject.
 We subsequently mapped the 110 ROIs to Yeo's 7-network parcellation \citep{yeo2011organization}, supplemented with the Subcortical network. This resulted in 8 functional networks: Subcortical (Scor), Frontoparietal Control Network (FPN), Default Mode Network (DMN), Dorsal Attention Network (DAN), Limbic Network (LIN), Salience/Ventral Attention Network (SAN), Somatomotor Network (SMN), and Visual Network (VIN). Based on these assignments, the features in \(X\) were grouped into 36 sets, each corresponding to interactions between pairs of networks (including within-network interactions).

\begin{table}[h]
\centering

\begin{tabular}[t]{lcc}
\toprule
\textbf{Method} & \textbf{Test MSE} &  \textbf{Compute time (s)} \\
\midrule
ECCAR, CV  (this paper) & 2.19 (1.84; 2.63) & 412s (411; 416)\\
\hline
Group ECCAR, CV (this paper) & 1.96 (1.63; 2.19)  & 411s (410; 419)\\
\hline
Row-sparse ECCAR, CV (this paper) & \textbf{1.76} (1.39; 2.16) & 619s (617; 624)\\
\hline
SAR, CV (\citet{wilms2015sparse}) & 3.17 (2.16; 3.43) &  11372s (10985; 12495)\\
\hline
SAR, BIC (\citet{wilms2015sparse}) & 3.75 (3.49; 3.77) &  10769s (10579; 10839)\\
\hline
Sparse CCA, CV (\citet{witten2009penalized}) & 2.48 (2.23; 2.92) &  250s (249; 250)\\
\hline
Sparse CCA, permuted (\citet{witten2009penalized}) & 2.81 (2.22; 2.97) &  57s (57; 57)\\
\hline
SCCA, CV (\citet{parkhomenko2009sparse}) & 2.33 (1.81; 2.62)  & 26s (26; 26)\\
\bottomrule
\end{tabular}
\caption{\label{tab:pred_summary} Summary of the performance of the different methods on the ABIDE dataset. Results are reported as the median test MSE and computation time across 10 folds, with interquartile ranges shown in brackets.}
\end{table}

 \begin{figure*}[h]
    \centering
    \begin{subfigure}[b]{0.43\textwidth}
        \centering
        \includegraphics[width = \textwidth]{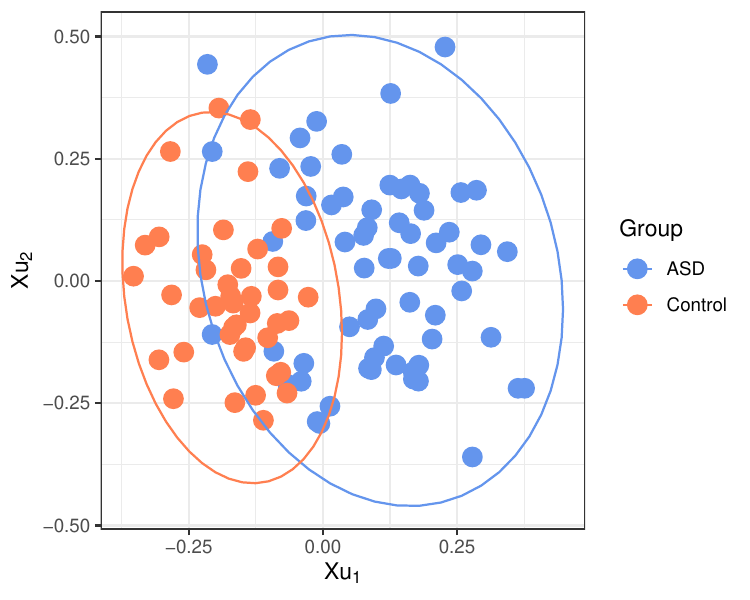}
    \end{subfigure}
    ~ 
    \begin{subfigure}[b]{0.54\textwidth}
        \centering
       \includegraphics[width = \textwidth,]{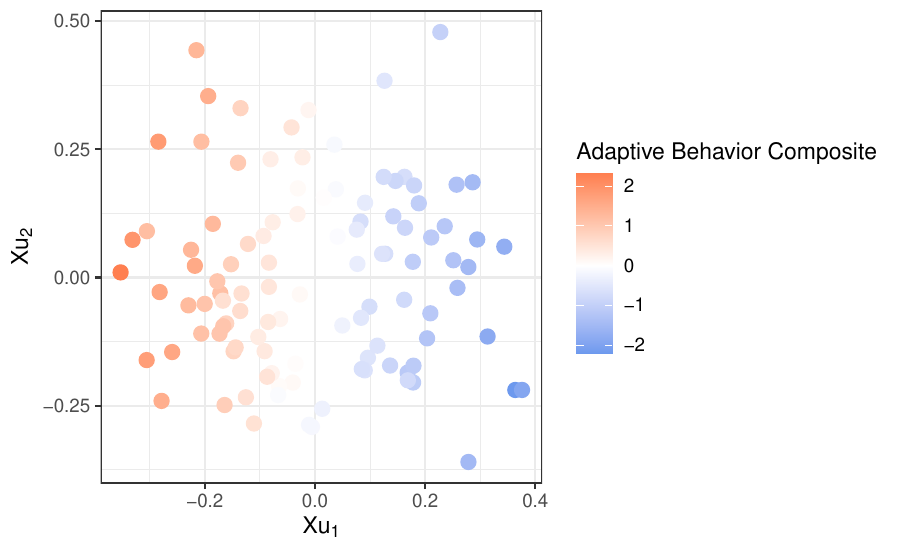}
    \end{subfigure}
    \caption{Scatter plots for the first two canonical variates produced by row-sparse ECCAR. Left: The points are colored by diagnosis, the plot includes 95\% confidence ellipses for each diagnosis group. Right: the points are colored by Adaptive Behavior Composite.}
\label{fig:abide:xu}
\end{figure*}

We run various sparse CCA methods with $r=2$ on the ABIDE dataset and evaluate the performance using nested 10-fold cross-validation (see Appendix~C.4, Figure~18 for a justification of the rank). For our method, we consider three settings: (i) ECCAR, which imposes element-wise sparsity on \( B \); (ii) row-sparse ECCAR, which encourages sparsity at the row level by grouping elements of \( B \) according to their row indices; and (iii) group ECCAR, which groups the rows of \( B \) based on predefined network interaction groups (36 in total).
We report the median MSE across the folds in Table~\ref{tab:pred_summary}. As shown in the table, the row-sparse and group ECCAR methods achieve a substantially lower MSE compared to all other approaches. 

By analyzing the scatter plot of the canonical variates (left panel of Figure~\ref{fig:abide:xu}) obtained from the row-sparse ECCAR method, we observe a clear separation between ASD patients and controls. This suggests that the first canonical direction effectively captures diagnostic differences between the two groups. The corresponding canonical direction vector $u_1$, associated with brain connectivities, is shown in Figure~21 in Appendix~C.4. 
As expected, many connectivity features receive zero weights.

To investigate which network-level interactions are most relevant to disease status, we compute the average absolute loading of $u_1$ within each of 36 predefined brain network interaction groups and visualize the results as a heatmap. As shown in the left panel of Figure 19,
the interaction between the dorsal attention network (DAN) and the somatomotor network (SMN) receives the highest loading, although the difference from other interactions is not pronounced.
To more explicitly identify disease-relevant network interactions, we also analyze the same heatmap based on the group-sparse ECCAR method. As shown in Figure 22, 
the direction $u_1$ exhibits a block-sparse structure, as expected from the grouping penalty. The corresponding network-level summary, presented in the right panel of Figure 19(b),
indicates that the limbic network (LIN) shows the strongest involvement, particularly in its interactions with the dorsal attention network (DAN) and the frontoparietal network (FPN). Given that the limbic network is primarily involved in emotion regulation, memory processing, and motivational behavior, its engagement aligns with established neurobiological characteristics of autism spectrum disorder \citep{haznedar2000limbic}.
We also analyze the canonical vector $v_1$, associated with the Vineland Adaptive Behavior Scales, as estimated by the row-sparse ECCAR. As shown in Figure 20, 
the Adaptive Behavior Composite (ABC) receives the highest loading, indicating its primary contribution to the canonical correlation. This observation is supported by the right panel of Figure~\ref{fig:abide:xu} that shows very clear separation between patients with positive and negative ABC values.

We illustrate the results of the sparse CCA approaches of \citet{parkhomenko2009sparse} and \citet{witten2009penalized} in Appendix~C.4. The corresponding canonical variate scatter plots (top row of Figure 25) 
show weaker separation between ASD and control groups compared to our method. The direction vectors $u_1$, as well as their network-level summaries are shown in Figures 23, 24, and  Figure 25.
We notice that the method of \citet{parkhomenko2009sparse} yields a dense $u_1$, assigning similar weights across many network interactions and failing to pinpoint specific networks (Figure 25). 
In contrast, the method of \citet{witten2009penalized} results in an overly sparse $u_1$, highlighting only the interaction between the default mode network (DMN) and the somatomotor network (SMN), potentially overlooking the critical role of the limbic system identified by our method (Figure 25).

\subsection{Interpretability of LLM Embeddings}\label{subsec:llm}

To show that the application of our method extends beyond biology, we illustrate its utility in the context of algorithm interpretability. While modern foundation models excel at generating powerful, high-dimensional embeddings from data, the structure of these embeddings is not inherently interpretable. We thus consider the following question: \textit{Given a point cloud of document embeddings, can we explain its structure?} We propose a principled framework to probe and gain insight into this setting: first, a set of interpretable features is constructed from the data (e.g., document topics or linguistic features from text); second, canonical correlation analysis is employed to identify and quantify the linear relationships between these features and the model's embeddings. Our approach adapts a paradigm previously used in computer vision to compare internal neural network activations \citep{raghu2017svcca,morcos2018insights,andrew2013deep}, and repurposes it to directly explain a model's final output embeddings in terms of external, interpretable variables.

We analyze embeddings of the 20 Newsgroups dataset, a corpus of articles belonging to one of 20 topics, which we further coarsen into 7 main categories (see Table~5 in Appendix~C.5). Each document is encoded into a vector of dimensionality $p=750$ using the \texttt{all-mpnet-base-v2} SentenceTransformer model available on Hugging Face. This model is based on the MPNet encoder and has been fine-tuned with contrastive and cosine-similarity objectives to produce semantically rich sentence representations that work well in downstream similarity and clustering tasks. To interpret the resulting embedding space, we align these embeddings with a corresponding Term Frequency-Inverse Document Frequency (TF-IDF) representation, constructed using the top $q$ words with the highest TF-IDF frequencies. We then apply CCA to these two data views using ranks $r=4$ and $10$ {(see Figure 26 of the Appendix for a justification of these two levels of resolution)}, to uncover structure at various levels of granularity. The performance of different CCA methods is evaluated using MSE on held-out data and running time, as a function of both the number of documents ($n$) used to train the CCA model and the number of TF-IDF features ($q$) retained for interpreting the transformer embeddings.

\begin{figure}
    \centering
    \includegraphics[width=0.9\linewidth]{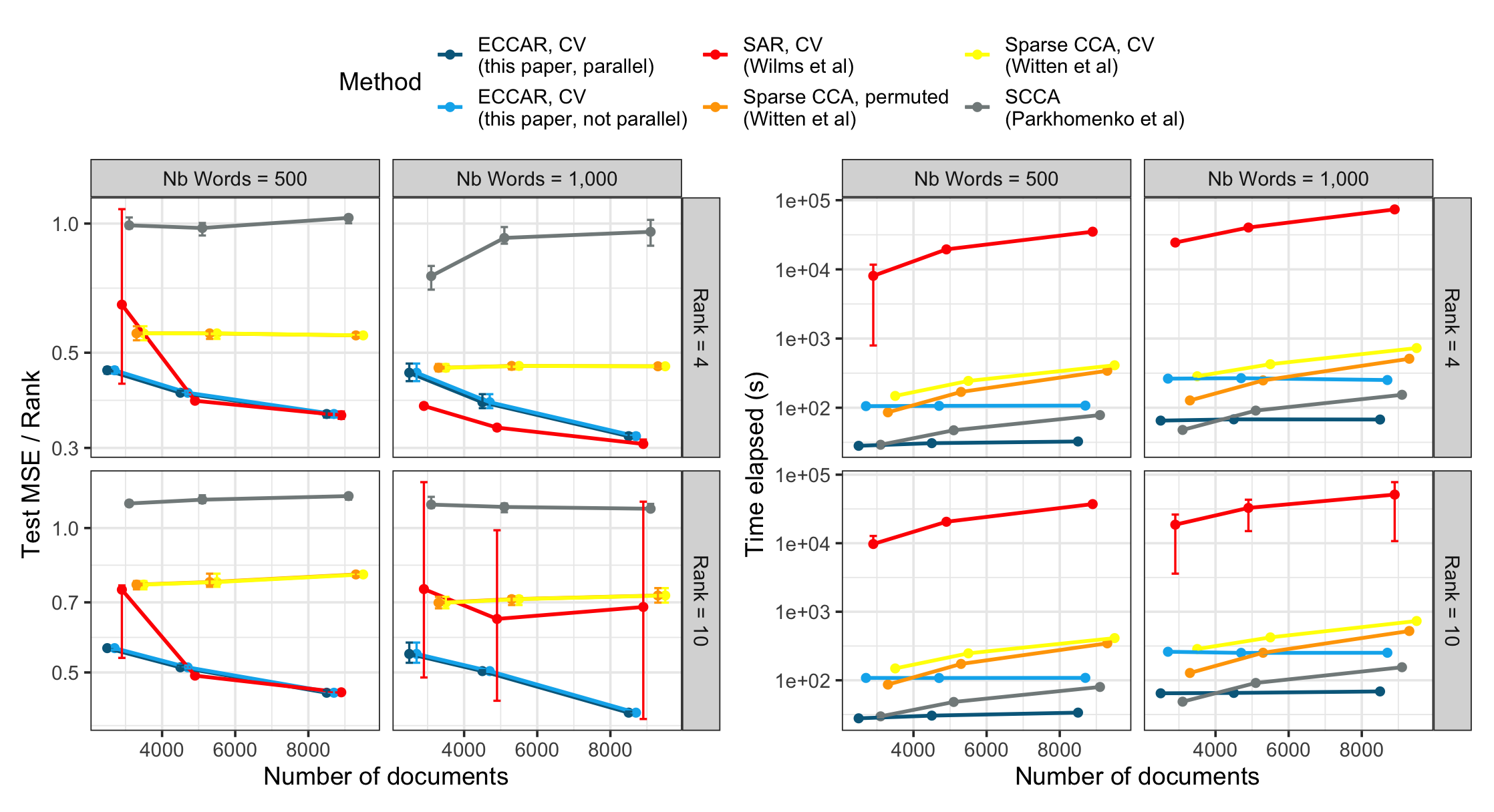}
    \caption{Results of the different sparse CCA methods on the LLM interpretability task. The performance is evaluated using the mean-squared error on a  held-out test set (left), and by reporting the compute time (right).}
    \label{fig:results_llm}
\end{figure}

The results presented in Figure~\ref{fig:results_llm} show that our proposed estimator ECCAR and SAR \citep{wilms2015sparse} achieve comparable performances, with SAR having on occasion a slightly smaller test MSE, but displaying higher variability, particularly when the number of documents ($n$) is small. Crucially, our method offers a significant computational advantage, with a fitting time that is 100 to almost 1,000 times smaller than SAR (even when not resorting to parallelization). 

We further visualize the first six canonical variates, colored by coarse category when fitting CCA on $r=10$ loadings, with $n=3000$ and $q = 1000$  (see Appendix C.5, Figure~27). 
We note that the variates recovered by our method and SAR \citep{wilms2015sparse} present similar structure (up to a sign flip). This is consistent with the fact that their MSEs indicate similar levels of prediction power. Interestingly, the variates exhibit clear streaks corresponding to different topics. By contrast, the approaches of both \cite{witten2009penalized} and \cite{parkhomenko2009sparse} exhibit considerably more mixing and do not seem to yield any clear and interpretable axis of variation in the data.

\begin{figure}[t]
    \centering
    \includegraphics[width=\linewidth]{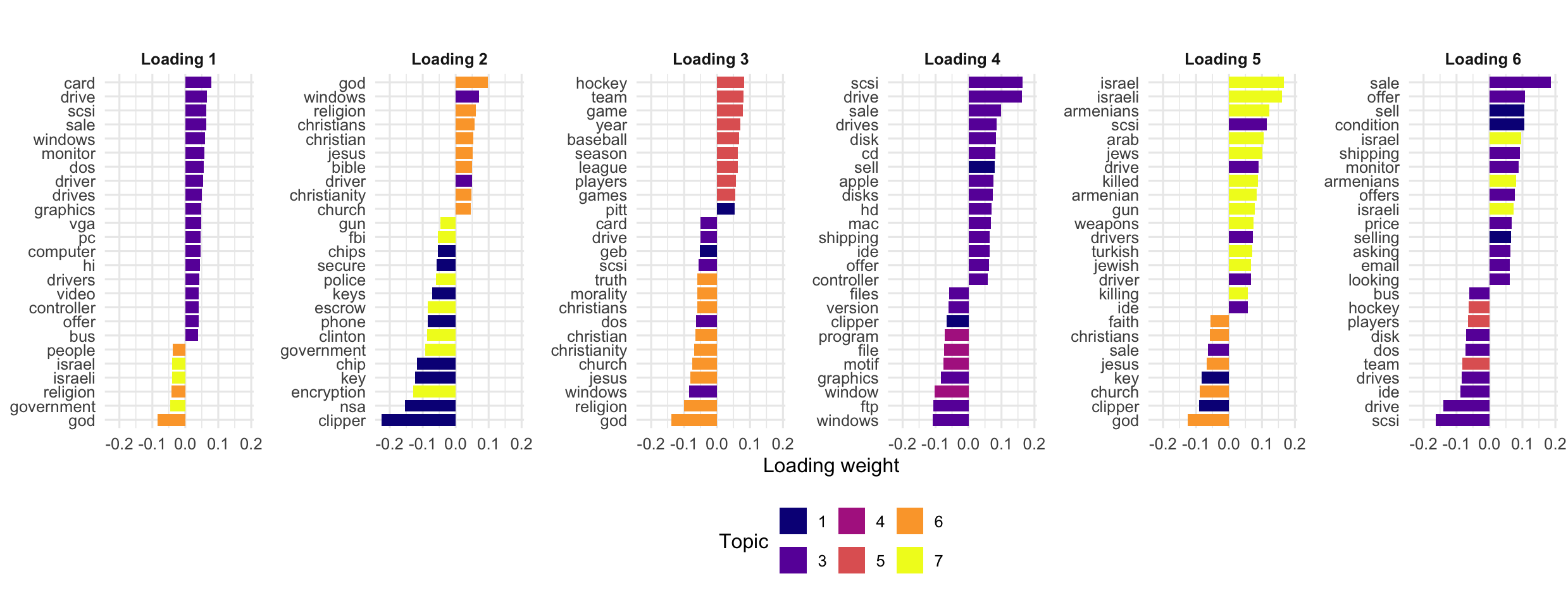}
    \caption{The top 25 loadings for the first 6 canonical directions obtained by ECCAR. Colors were added as a visual aid, and represent each word's estimated most frequent ``topic'', as estimated by the Latent Dirichlet Allocation algorithm.}
    \label{fig:loading_ecca_llm}
\end{figure}

We also present the loadings of the first 6 canonical directions obtained by ECCAR in Figure~\ref{fig:loading_ecca_llm}, where, to help visualization, each word is colored by its main topic (as fitted with a Latent Dirichlet Allocation Model on 7 topics). Overall, it is interesting to observe that the loadings recovered by our method seem to be dominated by one to two main categories (e.g. computers in Loading 4, computers and religion/politics in Loading 1). Interestingly, certain components seem to be able to differentiate between subtopics within each category (see Figure 29 in the Appendix). For instance, CCA direction 4 seems to be strongly associated with distinguishing between hardware-related articles and operating systems/graphics articles, which is reflected in the weights put on certain words (e.g. ``windows", ``graphics", ``version" vs ``disk", ``scsi", etc.). Likewise, CCA direction 5 divides political discussions about guns or general policy from those addressing Middle-East politics. This shows that the learned CCA axes are readily interpretable and align closely with the documents’ true labels.  For comparison, 
Figures~28(a)  and 28(b) in Appendix~C.5 
show the loadings recovered by the CCA variant of \cite{wilms2015sparse} and \cite{witten2009penalized}, respectively. We note again a strong resemblance between the loadings obtained for SAR and ECCAR.

\subsection{Beyond Gaussian data: A transcriptomics example}\label{subsec:stategra}
Finally, to demonstrate the utility of our method beyond Gaussian data, we consider deploying it for the analysis of the STATEGRA dataset \cite{gomez2019stategra}. This dataset profiles mouse pre-B-cell differentiation using multiple high-throughput technologies. We focus here on two transcriptomic layers: messenger RNA (mRNA) and microRNA (miRNA).  
Gene expression was measured in the B3 mouse pre-B cell line following tamoxifen-induced nuclear translocation of the transcription factor Ikaros, which triggers differentiation. Samples were collected at five time points (2, 6, 12, 18, and 24 hours) for both the Ikaros-induced condition and matched control samples (no Ikaros triggering), with three biological replicates per condition and time point, for a total of $n=30$ measurements. The resulting dataset provides paired measurements of mRNA and miRNA expression across time and experimental conditions, allowing us to study associations between the two modalities during differentiation. 

The data pre-processing (e.g., mapping, quantification, normalization) was carried out as described in \cite{gomez2019stategra}. To reduce the dimensionality of the data, the mRNA-seq data were filtered to keep at least 2,500 highly variable genes in each sample, resulting in a measurement matrix $X \in \mathbb{R}^{30 \times 3,869}$. The miRNA was processed using the STATEGRA pipeline \cite{gomez2019stategra}, resulting in a dataset $Y \in \mathbb{R}^{30 \times 476}$. Here, we examine the efficiency of our proposed ECCAR approach {to compute the first $r=2$ canonical directions (the canonical directions with strongest signal, as shown in Figure 33)} by comparing it to the sparse CCA approaches of Witten et al \cite{witten2009penalized} and of Wilms et al. \cite{wilms2015sparse}. Because the Fantope initialization of \cite{gao2017sparse} and \cite{gao2023sparse} was too time-consuming (over 12 hours), we restrict our analysis to methods with manageable computation times and compare their results, focusing on biological interpretation.

The results are shown in Figure~\ref{fig:stategra_res}, 
and loadings are shown in Figure 31 
of Appendix~C.6. The scatter plots of the canonical variates (Figure~\ref{fig:stategra_res}A, D, G) show a clear separation between control and Ikaros conditions across all methods, suggesting that the first canonical direction consistently captures factors that differentiate the two groups. 

Notably, this example demonstrates a failure mode in the application of Witten et al.'s sparse CCA  \cite{witten2009penalized}. In this scenario, the genes exhibit high correlation (Figure~32 of Appendix~C.6),
making the assumption of diagonal covariance matrices highly questionable. Consequently, the method fails to identify distinct canonical correlations. As seen in Figure~\ref{fig:stategra_res}, the canonical variates $Xu_1$ and $Xu_2$ are nearly collinear. This is corroborated by Figure 32 in the Appendix, which reveals correlations above 0.90 between the CCA variates. In contrast, the other approaches successfully extract orthogonal directions. ECCAR also appears to separate the Ikaros samples by time along the second canonical direction. By comparison, the approach of Wilms et al. \cite{wilms2015sparse} also seems to capture temporal variation, but primarily along the first direction, while the second direction does not appear to reflect any known structure.

To validate the different methods, we further inspected the canonical direction's loadings to verify whether they matched mRNA and miRNA genes known to be associated with B-cell differentiation. In particular, the lactate dehydrogenase A gene (\textit{ldha}) is a known target of Ikaros that is downregulated during Ikaros-induced differentiation of the B3 cell line \cite{gomez2019stategra}. Gomez et al \cite{gomez2019stategra} also highlight five miRNA genes that are known to have an impact on the \textit{ldha} expression levels: mmu-miR-342-3p, mmu-miR-532-5p, mmu-miR-532-3p, mmu-miR-363-5p, and  mmu-mir-188-3p. The first four genes are present in the top 20 loadings of the first CCA direction of Witten et al \cite{witten2009penalized}, with mmu-mir-188-3p appearing in the top 30 (Figure \ref{fig:stategra_res}C). However, comparing the remaining loadings in components 1 and 2 against known gene function databases (e.g., \textit{ensembl.org}) does not reveal strong evidence for the specific regulation of key steps in B-cell differentiation.

Both ECCAR and the sparse CCA approach of Wilms et al \cite{wilms2015sparse} feature much smaller sets of mRNA and miRNA genes with non-zero loadings (Figure~\ref{fig:stategra_res}E, F, H, I). However, Wilms et al.'s approach seems to detect miRNA genes that are overall associated with regulatory signals such as  mmu-miR-9-5p and mmu-miR-19a-3p present in the loading from component~2 (Figure~31B of the Appendix~C.6, rather than specifically associated with B-cell differentiation) .

Conversely, our approach highlights mmu-mir-188-3p as a key gene for differentiation, as seen in the miRNA analysis (Figure ~\ref{fig:stategra_res}I). From the RNA side, ECCAR's loadings along the first canonical direction emphasize the role of Ikzf1 (Ikaros, ENSMUSG00000018654, Figure~\ref{fig:stategra_res}H), which is a key transcription factor that regulates B-cell differentiation. The loadings of ECCAR’s second CCA direction are also biologically interpretable. In particular, gene ENSMUSG00000063229 (\textit{ldha}; Figure~31C), which also appears in component~1, and the miRNA mmu-miR-210-5p (Figure~31C), a regulator of metabolic adaptation and cell proliferation, receive high loadings, highlighting factors relevant to both transcriptional and post-transcriptional changes during B-cell differentiation. Thus, ECCAR successfully captures biologically meaningful signals relevant to B-cell differentiation.

\begin{figure}[h!]
    \centering
    \includegraphics[width=\linewidth]{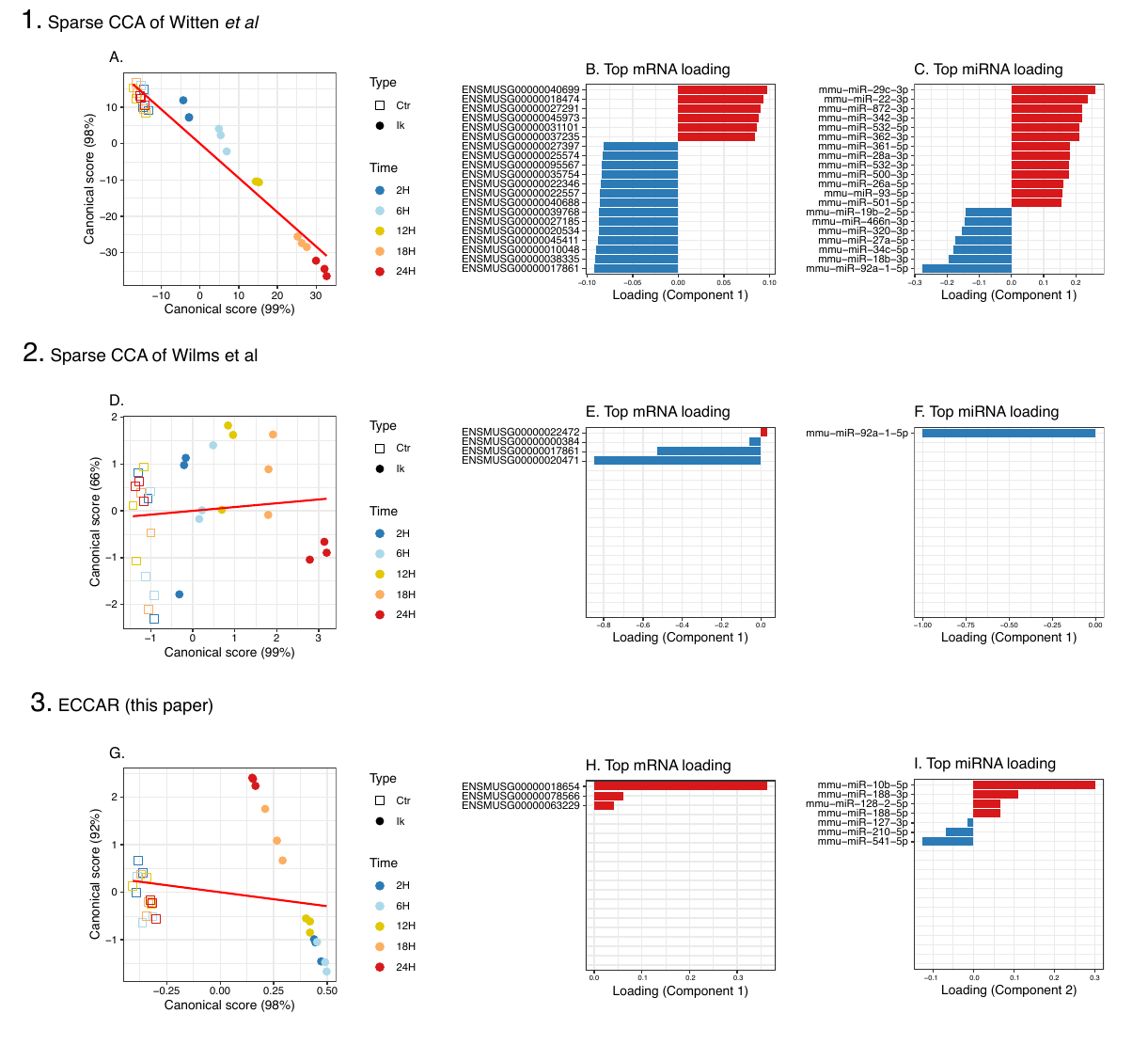}
    \caption{CCA integrating mRNA and miRNA expression profiles across time in control (Ctr) and Ik (Ikaros) conditions (STATEGRA dataset \cite{gomez2019stategra}). Panels 1 (A-C), 2 (D-F) and 3 (G-I) show results from three distinct CCA estimators applied to the same dataset: Sparse CCA of Witten et al \cite{witten2009penalized}, Sparse CCA of Wilms et al. \cite{wilms2015sparse} and ECCAR from this paper, respectively. Panels A, D, G show projections on the first pair of canonical components from each method. Points are colored by time and shape by conditions. The red lines show correlation trends between canonical scores. Panels B, E, H and C, F, I show top 20 mRNA and miRNA loadings to the first canonical component, respectively. 
    } 
    \label{fig:stategra_res}
\end{figure}

\section{Conclusion}
Understanding how CCA is used has important implications for method design. In particular, canonical correlation analysis should be {\it (a) scalable} to enable efficient processing of high-dimensional datasets; 
  {\it (b) interpretable} to facilitate meaningful scientific conclusions; and {\it (c) sparsistent} --- that is, the variables identified through CCA should come with provable guarantees. 

In this paper, we introduced ECCAR, the first estimator for sparse canonical correlation analysis that provably meets all three criteria. 
Beyond applications in genetics and neuroscience, we further showcased its practical utility by applying it to the challenging problem of interpreting Large Language Model embeddings.
This highlights the continued relevance of classical statistical methods in the era of modern data science. Rather than discarding foundational techniques when faced with new challenges, adapting them through tools such as regularization can lead to practical and powerful solutions. Our work provides not only a scalable and interpretable tool for practitioners, but also a blueprint for how classical multivariate analysis can evolve to meet contemporary demands. In particular, an important future direction is to extend the ECCAR methodology to settings involving more than two datasets, which is particularly relevant in multi-omics studies.

\section*{Data Availability \& Use of AI }

All the data used in this paper is publicly available. The Nutrimouse data can be found in the \texttt{R} package \texttt{mixomics}. The AUD dataset can be obtained from the  \texttt{R} package \texttt{CVR}. The ABIDE dataset can be loaded from the ABIDE website. The 20 Newsgroup dataset can be loaded from the Python package \texttt{scikit-learn}. The STATEGRA dataset can be downloaded directly using the links provided in the associated paper \cite{gomez2019stategra}.

The authors acknowledge the use of AI (ChatGPT) for improving \texttt{R} plot figures, as well as checking grammar, spelling, and typographical errors.

\section*{Acknowledgments}
C.D. was supported by the National Science Foundation under Award Number 2238616, as well as the resources provided by the University of Chicago’s Research Computing Center.
E.T. was supported by the Natural Sciences and Engineering Research Council of Canada under grant RGPIN-2023-04727; the University of Toronto Data Science Institute Catalyst grant; and the University of Toronto McLaughlin Center under grant MC-2023-05. 
This research was supported in part by grants from the NSF (DMS-2235451) and Simons Foundation (MPS-NITMB-00005320) to the NSF-Simons National Institute for Theory and Mathematics in Biology (NITMB).

\newpage
\setcounter{page}{1}

\begin{appendix}
    
\section{Related Works}\label{sec:lit}
Sparse CCA methods assume that only a subset of coefficients in the canonical directions are non-zero. In recent years, these methods have gained significant traction, particularly due to the greater interpretability of the results—a crucial feature given CCA's role in discovering associations between datasets. From a methodological perspective, sparse CCA methods can typically be classified into two categories: (a) \textit{heuristic-based} and (b) \textit{theory-based} approaches.

Heuristic-based approaches \citep{wilms2015sparse, wilms2016, parkhomenko2009sparse} build on the success of the lasso \citep{tibshirani1996regression} by adding an $\ell_1$-penalty to the objective function in Equation~\ref{eq:cca}. The canonical directions are then fitted in an alternating fashion, fixing one direction and solving for the other until convergence. When the number of canonical directions $r$ is greater than 1, they are typically computed sequentially by progressively deflating the matrices $X$ and $Y$. Although these methods are generally fast, questions regarding their accuracy remain unresolved. Notably, to our knowledge, the consistency of these methods has not been rigorously analyzed. Furthermore, due to the reliance on alternating fitting procedures, sensitivity to initialization has not been thoroughly investigated. To ensure computational efficiency, these methods (e.g., \cite{witten2009penalized}) can also make simplifying assumptions, such as neglecting correlations between variables and assuming that $\Sigma_X = I$ and $\Sigma_Y = I$. As we demonstrate in Section~\ref{sec:sim}, these assumptions can introduce substantial bias in the results.

In contrast, theory-based methods provide formal guarantees for estimation accuracy. To the best of our knowledge, the only theoretical treatment of sparse CCA consists in the series of works by Gao and coauthors \citep{chen2013sparse, gao2015minimax, gao2017sparse}, as well as more recent contributions by \cite{gao2023sparse} and \cite{donnat2024canonical}. 
These works attempt to formally characterize the estimation error with high-probability, and derive a corresponding algorithm. Framing the recovery of sparse components as a prediction problem, \citet{gao2017sparse} propose a sparsity-adaptive method that achieves optimal minimax rates for estimating both $U^\star$ and $V^\star$. They consider a regime where the true canonical directions are sparse, with support sizes satisfying $|\text{supp}(U^\star)| \leq s_u$ and $|\text{supp}(V^\star)| \leq s_v$, for some $s_u \leq p$ and $s_v \leq q$. In this setting, \citet{gao2015minimax} establish that the minimax rate for estimating $U^\star$ and $V^\star$ under the joint prediction loss
$\big\| \wh{U} \wh{V}^\top - U^\star {V^\star}^\top \big\|_F^2$
is given by
$$
\frac{1}{n{\lambda^\star}_r^2} \left( r(s_u + s_v) + s_u \log\left (\frac{ep}{s_u}\right ) + s_v \log\left (\frac{eq}{s_v}\right) \right),
$$
where $\lambda^\star_r$ denotes the $r$-th largest canonical correlation. 

Follow-up work further refines this minimax rate to the loss on each of the component which, in this case, is provided by:
$$\text{inf}_{W \in \mathcal{O}_r} \mathbb{E} \big\| W^\top\wh{U}^\top X - {U^\star}^\top X\big\|^2 \leq C \frac{s_u(r + \log(p))}{n\lambda_r^2}$$
To obtain this rate, the authors propose a {three-step procedure}. In step 1, they transform the sparse CCA problem into a convex optimization problem:
$$\operatorname{minimize} -\operatorname{Tr}(\widehat{\Sigma}_{XY} C) + \lambda \| C\|_{11}$$
The constraint set here is known as the Fantope, and had already been used in the context of sparse PCA estimation. In step 2, the estimators are refined using linear regression. The last step involves normalization of the result.
More recent work by \citet{gao2023sparse} expands upon the framework introduced in \citet{gao2015minimax, gao2017sparse}, replacing the second step of their procedure with a thresholded gradient descent algorithm. This modification has been shown to yield improved empirical performance over the original method. 

Finally, \citet{donnat2024canonical} propose a theoretically-grounded alternative to CCA that is computationally tractable and specifically tailored for settings in which only one of the datasets is high-dimensional. By leveraging the low-dimensional structure of the second dataset, they are able to reformulate the Fantope optimization problem as a more tractable regression problem. However, this approach is only applicable when one dataset has relatively low dimension.

\renewcommand{\arraystretch}{0.5}
\begin{table}[t]
\centering
\resizebox{\textwidth}{!}{
\begin{tabular}{llll}
\hline
\textbf{Feature}           & \textbf{Our method (ECCAR)} & \textbf{Gao et al. (2017)} & \textbf{Gao \& Ma (2023)} \\ \hline
\textbf{\begin{tabular}[c]{@{}l@{}}Statistical Rate \\ ($\min_{O \in \mathcal{O}_r}\|\wh{U}-UO \|_F$)\end{tabular}} &
  $O\left(\dfrac{1}{\lambda^{\star}_r }\sqrt{\dfrac{s_u s_v \log(p+q)}{n}}\right)$ &
  $O\left(\dfrac{1}{\lambda^{\star}_r }\sqrt{\dfrac{s_u (r+ \log(p))}{n}}\right)$ &
  $O\left(\dfrac{1}{\lambda^{\star}_r }\sqrt{\dfrac{(s_u + s_v) r \log(p + q)}{n}}\right)$ \\
  & & & \\
\textbf{Computational Cost} &
  $O(T(pn^2 + pqn))$ &
  \begin{tabular}[c]{@{}l@{}}$O(Tpq \min(p,q))$\\ Prohibitive due to Fantope \\ and multi-parameter tuning\end{tabular} &
  \begin{tabular}[c]{@{}l@{}}$O(T(p+q)^3)$\\ Prohibitive due to Fantope \\ and multi-parameter tuning\end{tabular} \\
  & & & \\
\textbf{Key Algorithmic Steps} &
  \begin{tabular}[c]{@{}l@{}}ADMM on a penalized \\ regression problem (step 1)\\ SVD (step 2)\end{tabular} &
  \begin{tabular}[c]{@{}l@{}}Fantope Projection (step 1) \\ Regression-based Refinement (step 2) \\ Normalization (step 3)\end{tabular} &
  \begin{tabular}[c]{@{}l@{}}Fantope Projection (step 1)\\ Thresholded Gradient \\ Descent (step 2)\end{tabular} \\
  & & & \\
\textbf{Sample Splitting?} & No & Yes (3-fold) & No \\ \hline
\end{tabular}
}
\caption{Comparison of theoretical guarantees and computational costs of sparse CCA. Here, we assume the high-dimensional setting $p,q \gg n$.}
\label{tab:comp}
\end{table}

\section{Algorithms}\label{app:alg}

\subsection{Algorithm for the sparse CCA setting}
We detail the steps of the ADMM Algorithm. To this end, we first transform our original loss function into the following ADMM Lagrangian:
\begin{equation*}
      \mathcal{L}(B, Z, H) =  \frac{1}{2} \Big\| \frac1n X B Y^\top  - I_n\Big\|_F^2 + \rho \mathcal{P}(Z) + \mu \langle H, B - Z \rangle + \frac{\mu}{2} \| B - Z \|_F^2.
    \end{equation*} 
    where $\mathcal{P}(Z)$ represents a penalty function (for instance, an $\ell_{11}$ penalty or a group penalty). Below we demonstrate how to solve the ADMM steps efficiently.
     \begin{description}
         \item[Pre-computations:]  Before running the algorithm, we first compute the eigen-decompositions   $\wh{\Sigma}_{X} = U_X \Lambda_1 U_X^\top$, and  $\wh{\Sigma}_{Y} = U_Y \Lambda_2 U_Y^\top$, where $U_X \in \mathbb{R}^{p \times \mathrm{rank}(\widehat{\Sigma}_X)}, U_Y \in \mathbb{R}^{q \times \mathrm{rank}(\widehat{\Sigma}_Y)}$. Since $\mathrm{rank}(\widehat{\Sigma}_X) \le p \wedge n$ and $\mathrm{rank}(\widehat{\Sigma}_Y) \le  q \wedge n$,
         these operations have cost $O(p^2 \times (p \wedge n))$ and $O(q^2 \times (q \wedge n))$, respectively.
\item[ADMM procedure:] We then repeat the following steps until convergence.

1.  {\it Updating $B$.}  Setting $\nabla_B \mathcal{L}(B, Z, H) = 0$, we have
\begin{align*}
                     \wh{\Sigma}_{X} B \wh{\Sigma}_{Y} - \wh{\Sigma}_{XY} + \mu H +  \mu (B-Z) &= 0  \\
          U_X \Lambda_1 U_X^\top B U_Y \Lambda_2 U_Y^\top + \mu B & = \wh{\Sigma}_{XY} + \mu (Z - H)\\
           \Lambda_1 U_X^\top B U_Y \Lambda_2  + \mu U_X^\top B U_Y &= U_X^\top \big(\wh{\Sigma}_{XY} +\mu (Z - H)\big) U_Y.
\end{align*}
Let $\tilde{B} = U_X^\top B U_Y$. Rewriting the equation, we obtain
\begin{align*}
           \Lambda_1 \tilde{B} \Lambda_2  + \mu \tilde{B} =U_X^\top\big(\wh{\Sigma}_{XY} +\mu (Z - H)\big) U_Y.
\end{align*}
Equivalently, for each entry $i \in [r_X], j \in [r_Y]$,
\[ \big( (\Lambda_1)_{ii}(\Lambda_2)_{jj} + \mu \big) \tilde{B}_{ij} =  \big( U_X^\top(\wh{\Sigma}_{XY} +\mu (Z - H)) U_Y\big)_{ij}.\]
Thus, $\tilde{B}$ can be easily computed through element-wise division of the entries of the matrix $U_X^\top( \wh{\Sigma}_{XY} +\mu (Z - H)) U_Y$ by the corresponding values $(\Lambda_1)_{ii}(\Lambda_2)_{jj} + \mu.$ The operations for computing this matrix is $O(pq)$ for the matrix additions,  $O( (p \wedge n)pq)$ and $O( (p \wedge n) q (q \wedge n))$ for the multiplications by $U_X$ and $U_Y$, respectively. 
{It follows that $$B = U_X\tilde{B}U_Y^\top + B^\perp$$
for some $B^\perp \in \mathbb{R}^{p \times q}$ satisfying $U_X^\top B^\perp U_Y = 0$. Substituting this decomposition into the optimality condition $$\nabla_B \mathcal{L}(B, Z, H) = \wh{\Sigma}_{X} B \wh{\Sigma}_{Y} - \wh{\Sigma}_{XY} + \mu H +  \mu (B-Z) = 0$$ yields $$ \mu B^\perp = \mu \left( (Z - H) - U_X U_X^\top  (Z - H) U_Y U_Y^\top \right),$$ which implies \begin{align*}
    B &= U_X \tilde{B} U_Y^\top + B^\perp \\&= U_X \tilde{B} U_Y^\top +  (Z - H) - U_X U_X^\top  (Z - H) U_Y U_Y^\top \\
    &= U_X (\tilde{B} - U_X^\top  (Z - H) U_Y)U_Y^\top + Z - H
\end{align*} The matrix $B$ can then be computed in $O( p (p \wedge n)(q \wedge n)  )$ operations. 

Intuitively, note the computation of $\tilde{B}$ is based on element-wise scaling of
$$
U_X^\top \widehat{\Sigma}_{XY} U_Y
\approx
U_X^\top \widehat{\Sigma}_X \, U^\star \Lambda^\star V^{\star \top} \widehat{\Sigma}_Y U_Y
\approx
\Lambda_1 \, U_X^\top \!\left( U^\star \Lambda^\star V^{\star \top} \right) U_Y \, \Lambda_2,
$$
which can be viewed as a scaled rotation of the target matrix $U^\star \Lambda^\star V^{\star \top}$ expressed in the principal-component coordinates of $X$ and $Y$. The update for $\tilde{B}$ then applies element-wise shrinkage to these components by factors $(\Lambda_1)_{ii}(\Lambda_2)_{jj} + \mu$, acting as a ridge-type regularization that downweights directions associated with small variances while preserving dominant ones. The additional term involving $Z - H$ pulls the solution toward the current ADMM anchor, thereby balancing fidelity to the empirical cross-covariance with proximity to the previous sparse iterate.

In contrast, the complement $B^\perp$ lies in directions that are invisible to one or both covariance operators, meaning that the data provides no information about these components. In those directions, the optimality condition reduces to simply matching $Z - H$. Overall, the update performs a data-driven, ridge-type estimation on the identifiable subspace while copying the anchor on the unidentifiable orthogonal complement, which helps ensure numerical stability in high-dimensional settings where $p$ or $q$ may exceed the sample size.}

2. {\it  Updating $Z$.} We consider the objective function as a function of $Z$: \begin{equation*}
        \mathcal{L}(Z) = \begin{cases}
            \rho \|Z\|_{11} + \frac{\mu}{2} \|H + B - Z \|_F^2,& \text{with $\ell_1$ sparsity;} \\
            \rho \sum_g \sqrt{T_g} \|Z\|_F + \frac{\mu}{2} \|H + B - Z \|_F^2,& \text{with group sparsity.} 
        \end{cases} 
    \end{equation*}
 The solution, thus, has a closed form:
\begin{align*}
    Z_{ij}: &= S_{\rho / \mu}(H_{ij} + B_{ij}) =  \big( 1 - \rho / (\mu |H_{ij} + B_{ij}|) \big)_+ (H_{ij} + B_{ij})\ \ \text{with $\ell_1$ sparsity;}  \\
    Z_{g}: &= S_{\rho / \mu}(H_g + B_g)  =  \big( 1 - \sqrt{T_g} \rho / (\mu \|H_{g} + B_{g} \|_F) \big)_+  (H_{g} + B_{g})\ \  \text{with group sparsity.}
\end{align*}

3.  {\it Updating $H.$} $H: = H + B - Z$

\end{description}

The same algorithm can be adapted to the group-sparse setting by simply replacing the thresholding step of Step 2 by a group-thresholding step.

\newpage

\section{Simulation and real data experiments}\label{app:real_data}

\subsection{Synthetic Experiments}\label{app:subsec:simu}

Below, we present a figure comparing all existing sparse CCA methods based on simulation results.

 \begin{figure}[!htbp]
     \centering
     \includegraphics[width= \textwidth]{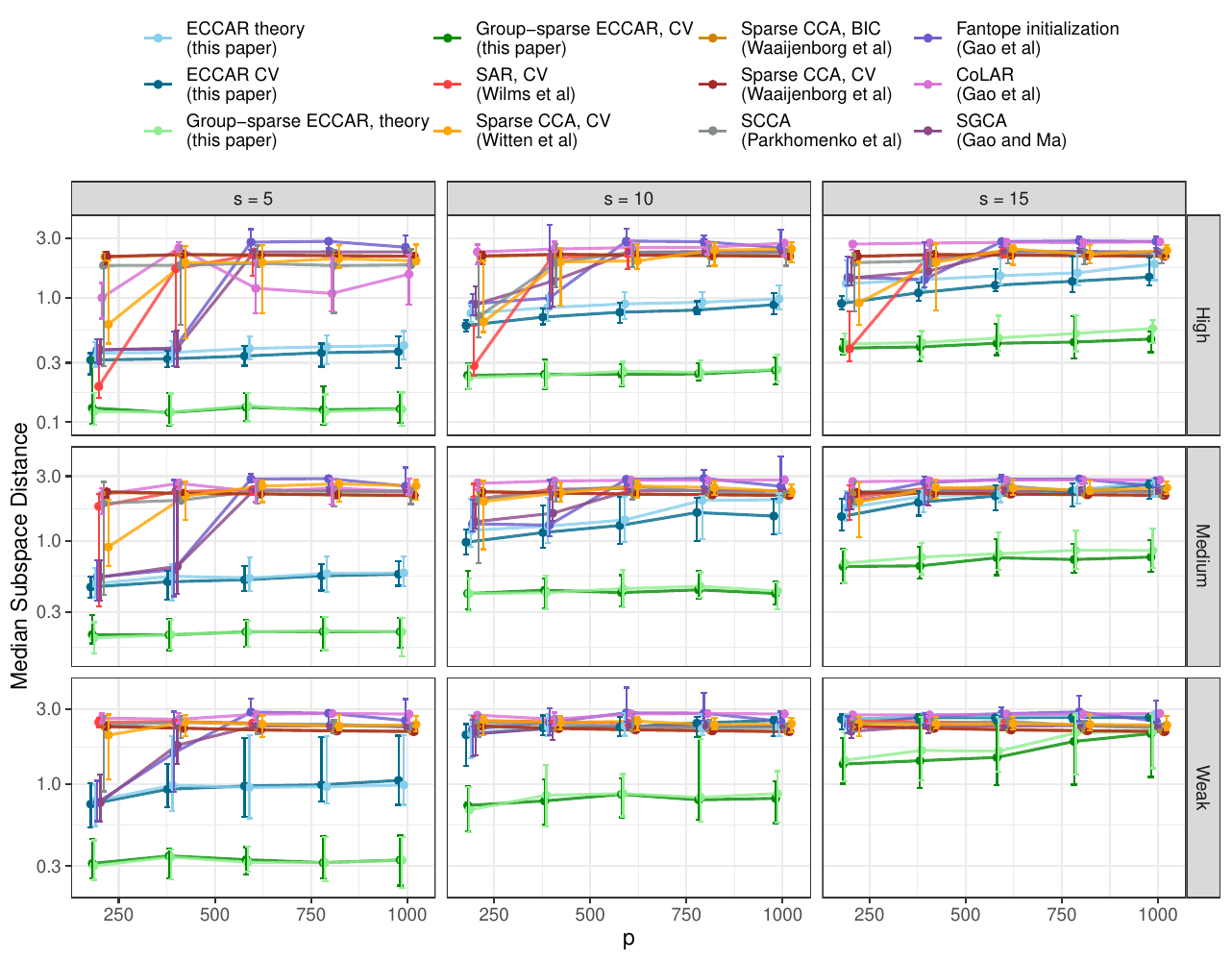}
\caption{Estimation error as a function of $p$ and $q$, and the support sizes $s_u=s_v = s.$ Points indicate the median subspace distance  over 25 independent experiments with error bars representing the 5th and 95th percentiles.}
 \end{figure}
 
 \begin{figure}[!htbp]
     \centering
     \includegraphics[width= \textwidth]{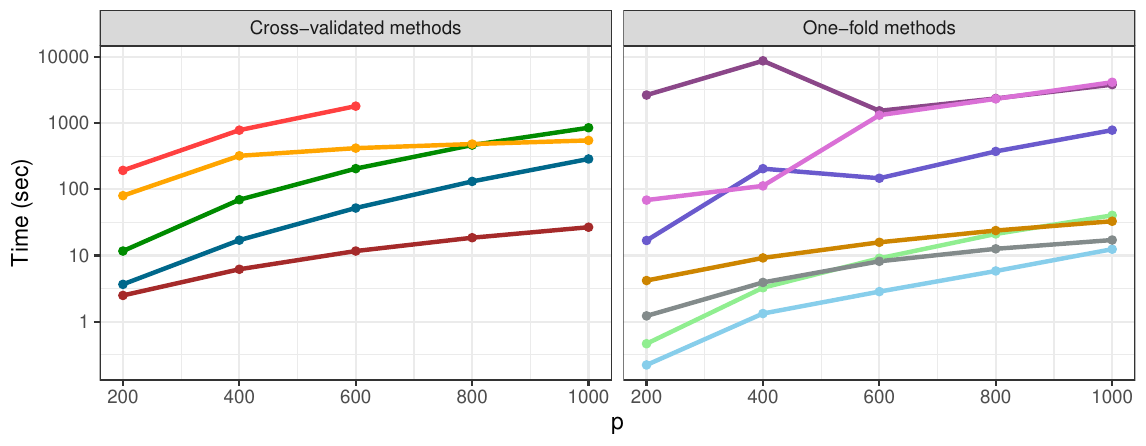}
\caption{Estimation time as a function of $p$ and $q$, and the support sizes $s_u=s_v = 5.$ Points indicate the median running time over 25 independent experiments.}
 \end{figure}

  \begin{figure}[!htbp]
     \centering
     \includegraphics[width= \textwidth]{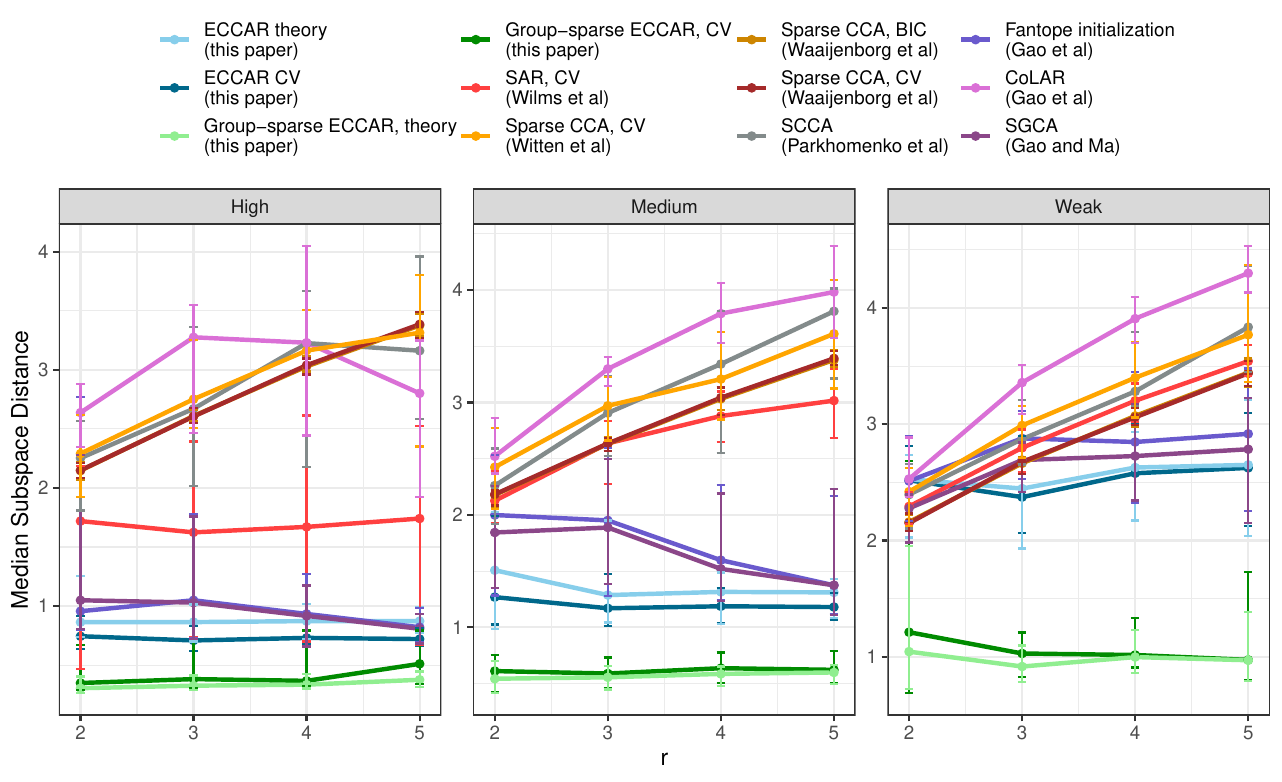}
\caption{ Estimation error as a function of $r$ with $s=10$, $p=200$, and $n=400$. Median distance with error bars representing the 95th and 5th percentiles. }
     \label{fig:sim(r)}
 \end{figure}

  \begin{figure}[!htbp]
  \label{fig:sim_t(p)}
     \centering
     \includegraphics[width= \textwidth]{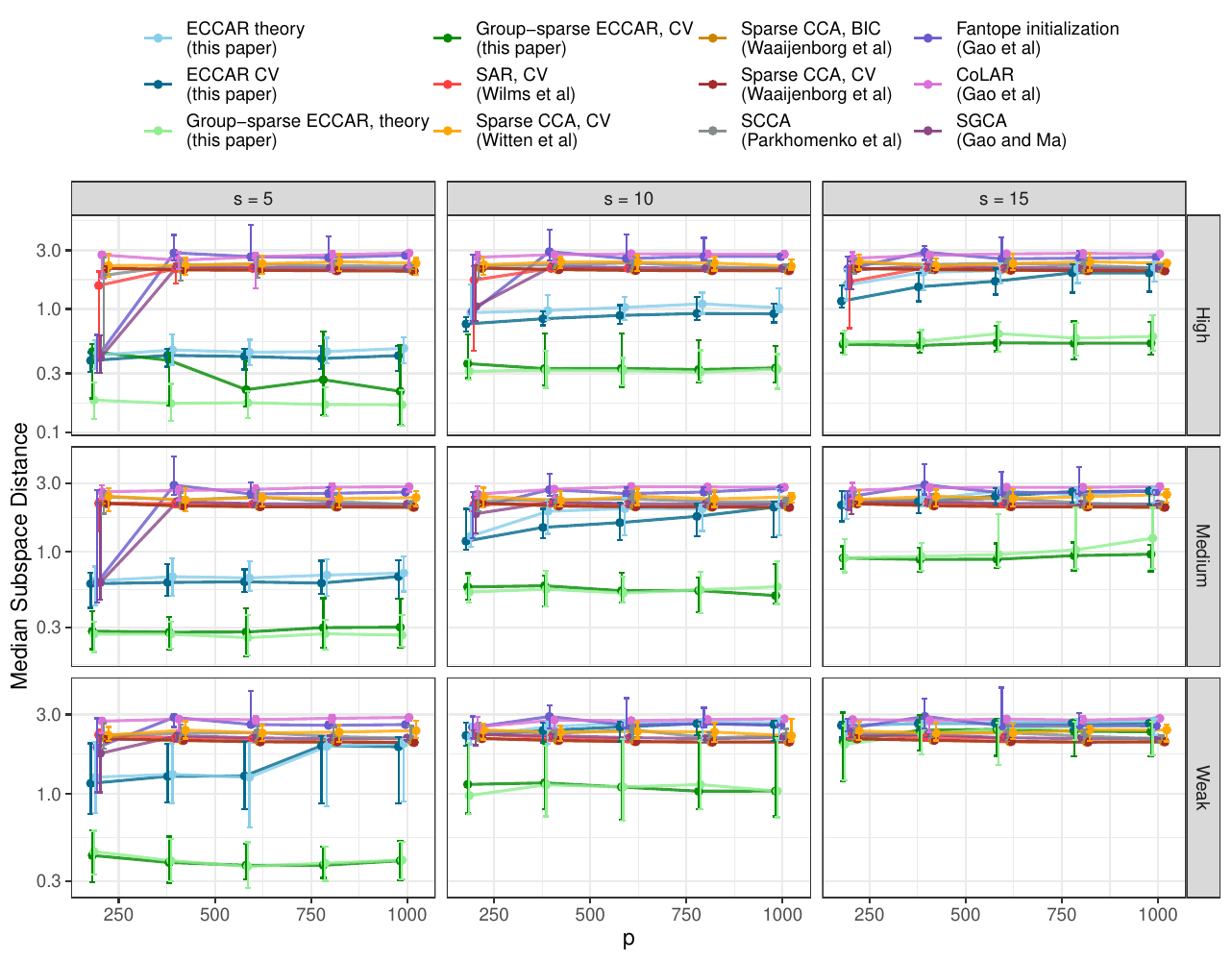}
\caption{{Estimation error as a function of $p$ and $q$ when $X$ and $Y$ follow $t$ distributions with degree of freedom 6 and the support sizes $s_u=s_v = s$. Points indicate the median subspace distance over 25 independent experiments with error bars representing the 5th and 95th percentiles.}}
 \end{figure}
 

 \clearpage 
 
\subsection{The Nutrimouse data}\label{app:subsec:nutrimouse}

The Nutrimouse dataset includes the expression measurements of $p=120$ genes potentially involved in nutritional issues and the concentrations of $q=21$ hepatic fatty acids for $n=40$ mice. Additionally, it contains two categorical variables: genotype, indicating whether the mouse is wildtype or genetically
modified (PPAR), and diet, indicating the oils used in the experimental diets.  We test our proposed method on the Nutrimouse dataset and compare
its performance to the benchmarks introduced in the previous section except SGCA by \cite{gao2023sparse} due to its long running time. We set the number
of canonical directions to $r = 5$ based on the scree plot showing the singular values of $\Sigma_{XY}$.

When applying our sparse ECCAR method to the genes \(X\) and fatty acids \(Y\), we determine the appropriate regularization parameter for our method using 8-fold
cross-validation: six folds  were used for training, one for validation, and one for testing. The optimal penalty $\rho$ is chosen using the average validation mean squared error (MSE) between canonical variates $X\widehat{U}$ and $Y\widehat{V}$. For other methods, the hyper-parameters are chosen by the default implementation. The performance is assessed using four metrics computed through cross-validation:  the average MSE  between the canonical variates \(X\widehat{U}\) and \(Y\widehat{V}\);
the average correlation between \(X\widehat{U}\) and \(Y\widehat{V}\); the classification loss for genotype using transformed genes; the classification loss for diet using transformed fatty acids. Classification is performed using support vector machines (SVM) or random forests (RF).

\begin{figure*}[h]
    \centering
    \begin{subfigure}[b]{0.43\textwidth}
        \centering
        \includegraphics[width = 0.9\textwidth, trim = {0 0 5cm 0}, clip]{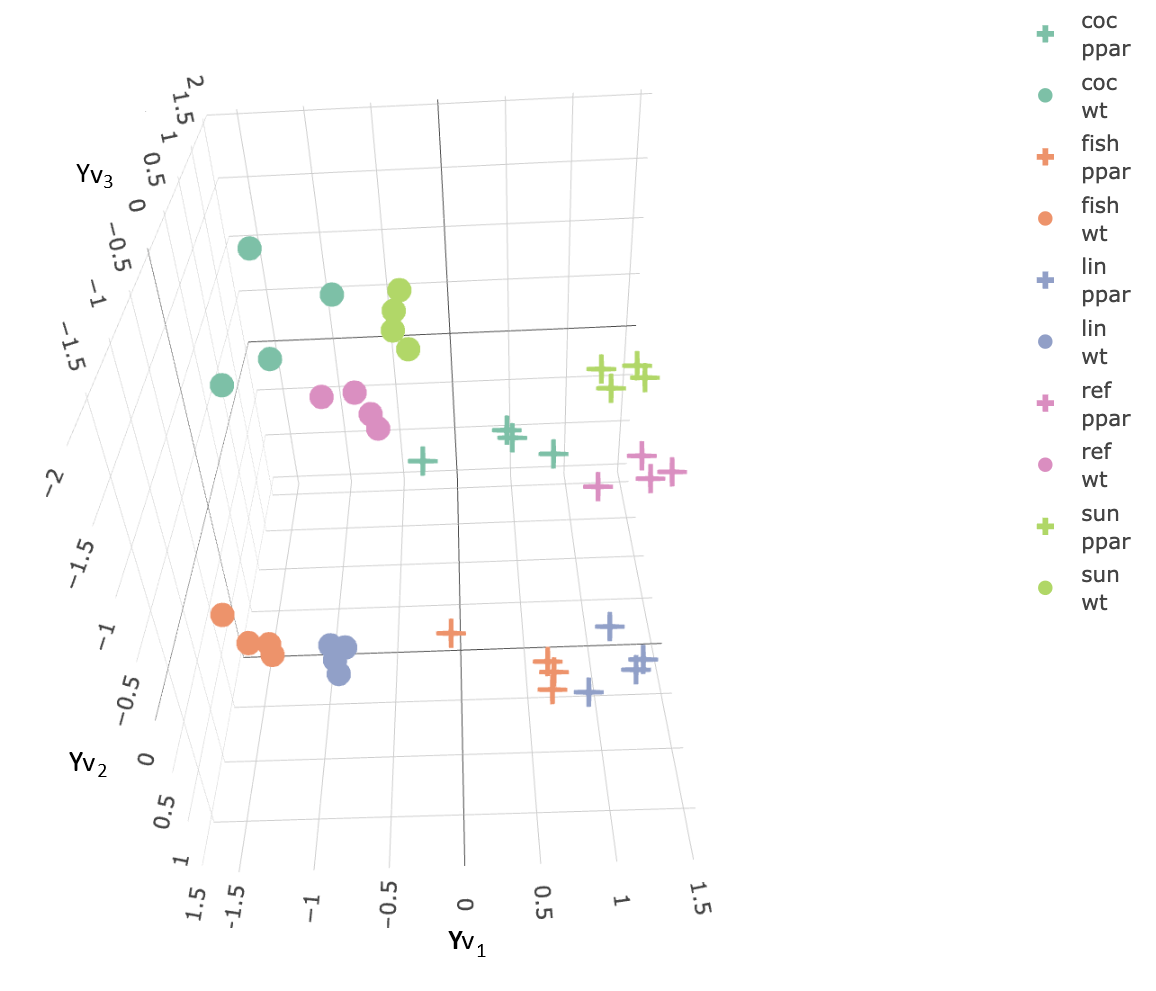}
        \caption{Sparse ECCAR}
    \end{subfigure}
    ~ 
    \begin{subfigure}[b]{0.52\textwidth}
        \centering
        \includegraphics[width = 0.9\textwidth]{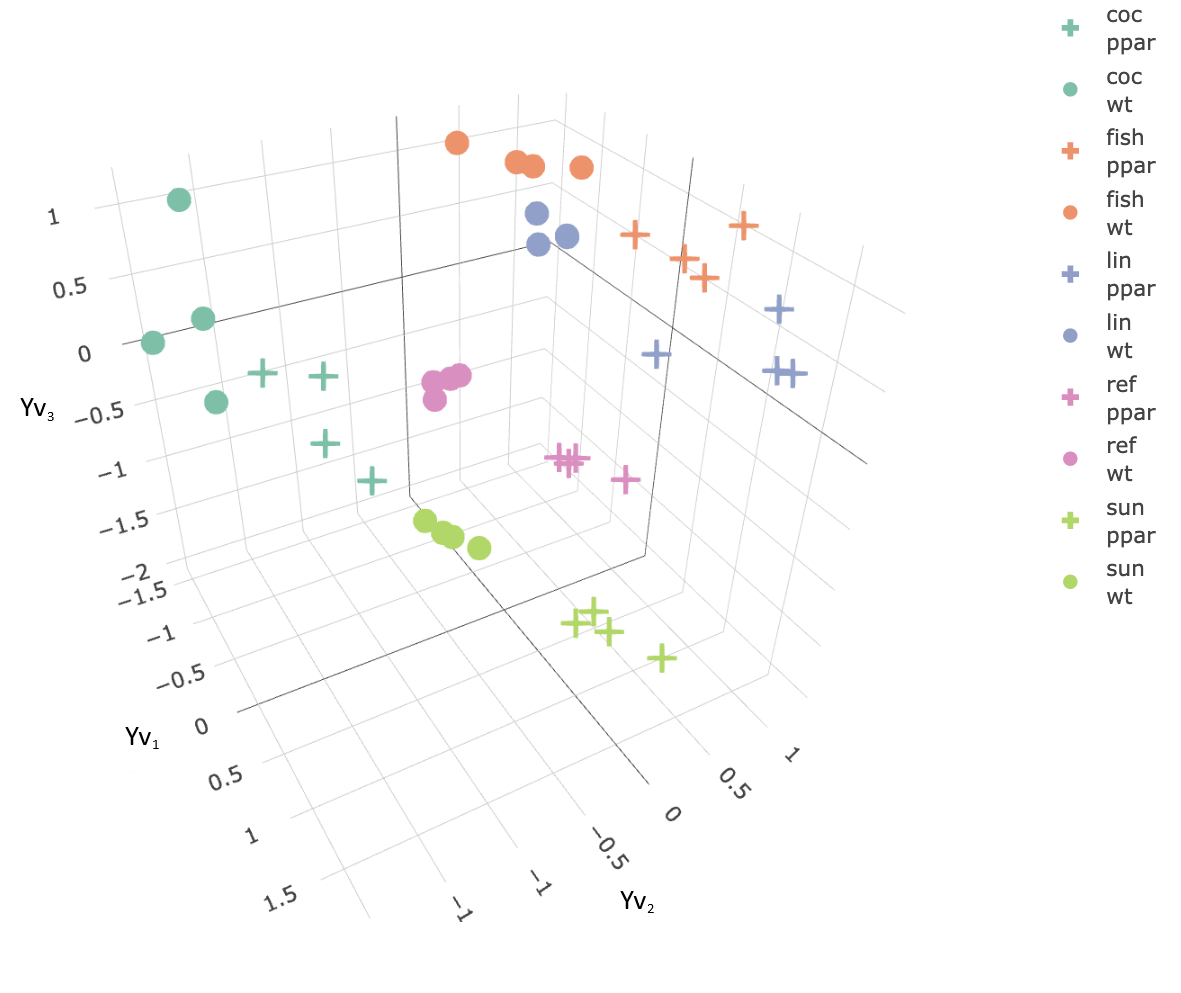}
        \caption{SAR by \citet{wilms2015sparse}}
    \end{subfigure}
    \caption{Scatter plots for first three canonical variates produced for the Nutrimouse dataset. Colors represent diet and shapes represent genotype.}
      \label{fig:nutri:xu}
\end{figure*}

As shown in Table~\ref{tab:nutrimouse}, our method outperforms the competing approaches in terms of test MSE, test correlation, and prediction accuracy for diet. For genotype prediction, our method performs nearly as well as the best-performing alternative, SAR. We visualize the clustering patterns produced by our method and SAR (see Figure~\ref{fig:nutri:xu}). Notably, our method provides a clear separation of both genotype and diet groups.

\begin{table}[h]
\centering
\begin{tabular}{lcccc}
\toprule
\textbf{Method} & \textbf{Test} & \textbf{Test} & \textbf{Genotype Acc}  &  \textbf{Diet Acc}  \\
& \textbf{MSE} & \textbf{Correlation} & \textbf{SVM / RF}  &  \textbf{SVM / RF}  \\
\midrule
Sparse ECCAR, CV & $\boldsymbol{0.827}$ & $\boldsymbol{0.529}$ & ${0.969}$ / 0.927  & $\boldsymbol{0.727}$ / $0.563$ \\
(this paper) &  &  &   &  \\ \hline
SAR, CV & 0.844 & 0.524 & $\boldsymbol{0.979}$ / 0.969 & 0.656 / 0.696 \\ 
(\citet{wilms2015sparse}) &  &  &   &  \\ \hline
SAR, BIC & 0.935 & 0.443 & $\boldsymbol{0.979}$ / 0.902 & 0.521 / 0.540\\
(\citet{wilms2015sparse}) &  &  &   &  \\ \hline
SCCA, CV & 0.945 & 0.418 & 0.958 / 0.875 & 0.317 / 0.417 \\ 
(\citet{parkhomenko2009sparse}) &  &  &   &  \\ \hline
Sparse CCA, CV & 1.470 & 0.301 & 0.969 / 0.948 & 0.338 / 0.288 \\
(\citet{witten2009penalized}) &  &  &   &  \\ \hline

Sparse CCA, permuted & 1.350 & 0.362 & 0.969 / 0.913 & 0.423 / 0.542 \\ 
(\citet{witten2009penalized}) &  &  &   &  \\ \hline
Sparse CCA, CV & 2.140 & -0.275 & 0.969 / 0.892 & 0.458 / 0.417 \\
(\citet{waaijenborg2009}) &  &  &   &  \\ \hline
Sparse CCA, BIC & 2.140 & -0.261 & 0.969 / 0.917 & 0.490 / 0.473 \\ 
(\citet{waaijenborg2009}) &  &  &   &  \\ \hline
Fantope initialization & 1.320 & 0.384 & 0.954 / 0.885 & 0.310 / 0.433 \\ 
(\citet{gao2017sparse}) &  &  &   &  \\
\bottomrule
\end{tabular}
\caption{Performance comparison of different methods on the Nutrimouse dataset using 8-fold cross-validation.}
\label{tab:nutrimouse}
\end{table}

 We evaluate the concordance between the genes and fatty acids selected by sparse ECCAR and the findings reported by \citet{martin2007novel}. Based on the first component, our method identifies the following fatty acids as most influential: C20:2n-6, C20:1n-9, C18:3n-3, C18:3n-6, C16:1n-9, C14:0, C22:5n-3, C18:1n-7, C16:0, and C20:3n-9. Among these, Martin et al.\ reported that C18:3n-3 exhibited a ``robust increase'' in expression in PPAR$\alpha$ genotypes compared to the wild type---consistent with our observation that this loading effectively separates genotypes. They also noted that changes in C16:1n-9 levels mirrored diet-induced changes within each genotype, and that C22:5n-3 was a major constituent of the fish oil used in their diet intervention.
Regarding gene selection, our method highlights \textit{CAR1}, \textit{HPNCL}, \textit{CPT2}, \textit{CYP3A11}, \textit{CYP4A10}, and \textit{GSTmu} as the most influential. Martin et al.\ observed a consistent downregulation of \textit{CYP3A11} in PPAR$\alpha$ livers, alongside an overexpression of \textit{CAR1}, that is in agreement with our findings.

\begin{figure*}[h]
    \centering
    \begin{subfigure}[b]{0.47\textwidth}
        \centering
        \includegraphics[width = \textwidth]{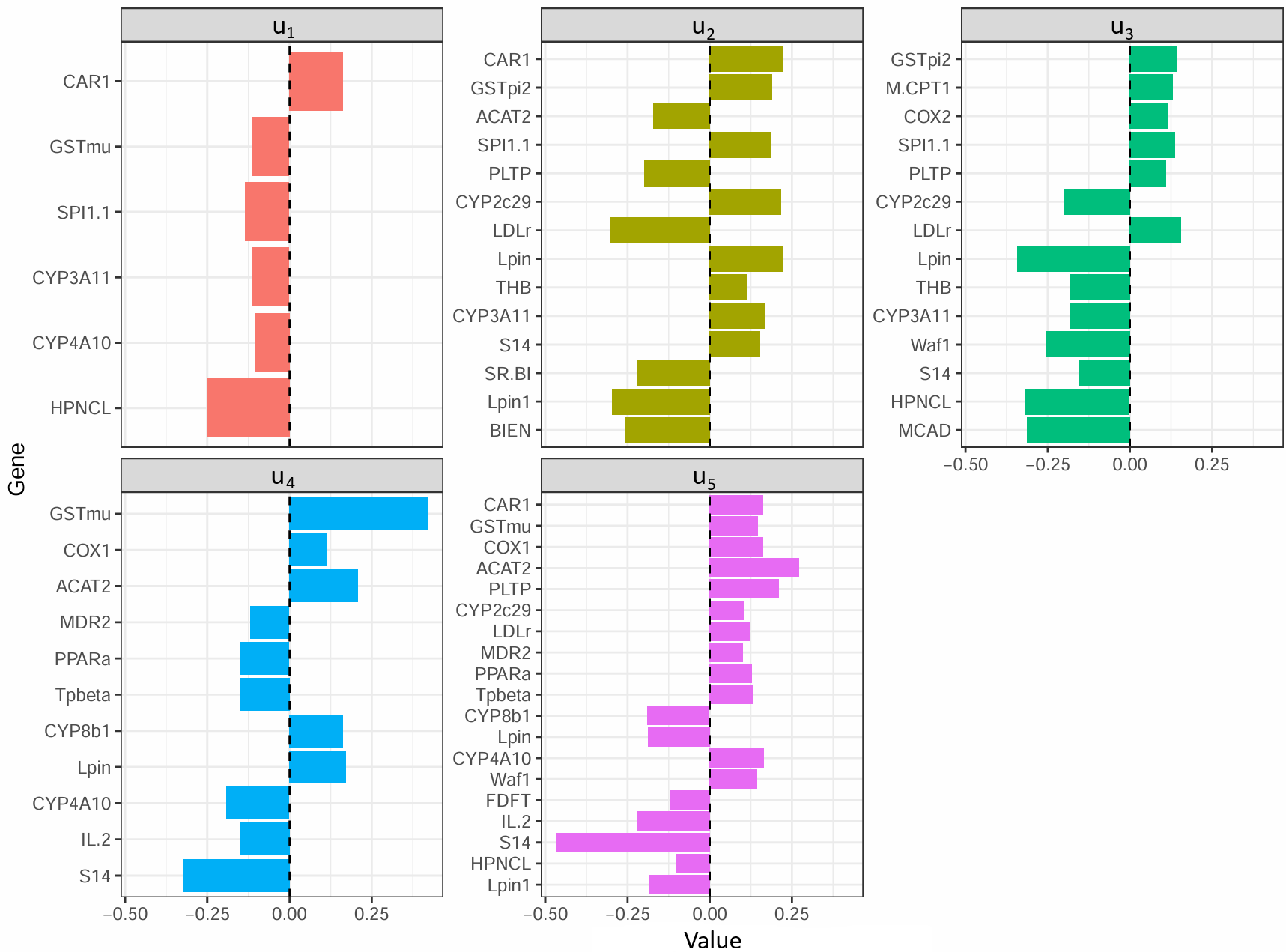}
        \caption{Genes}
    \end{subfigure}
    ~ 
    \begin{subfigure}[b]{0.47\textwidth}
        \centering
        \includegraphics[width = \textwidth]{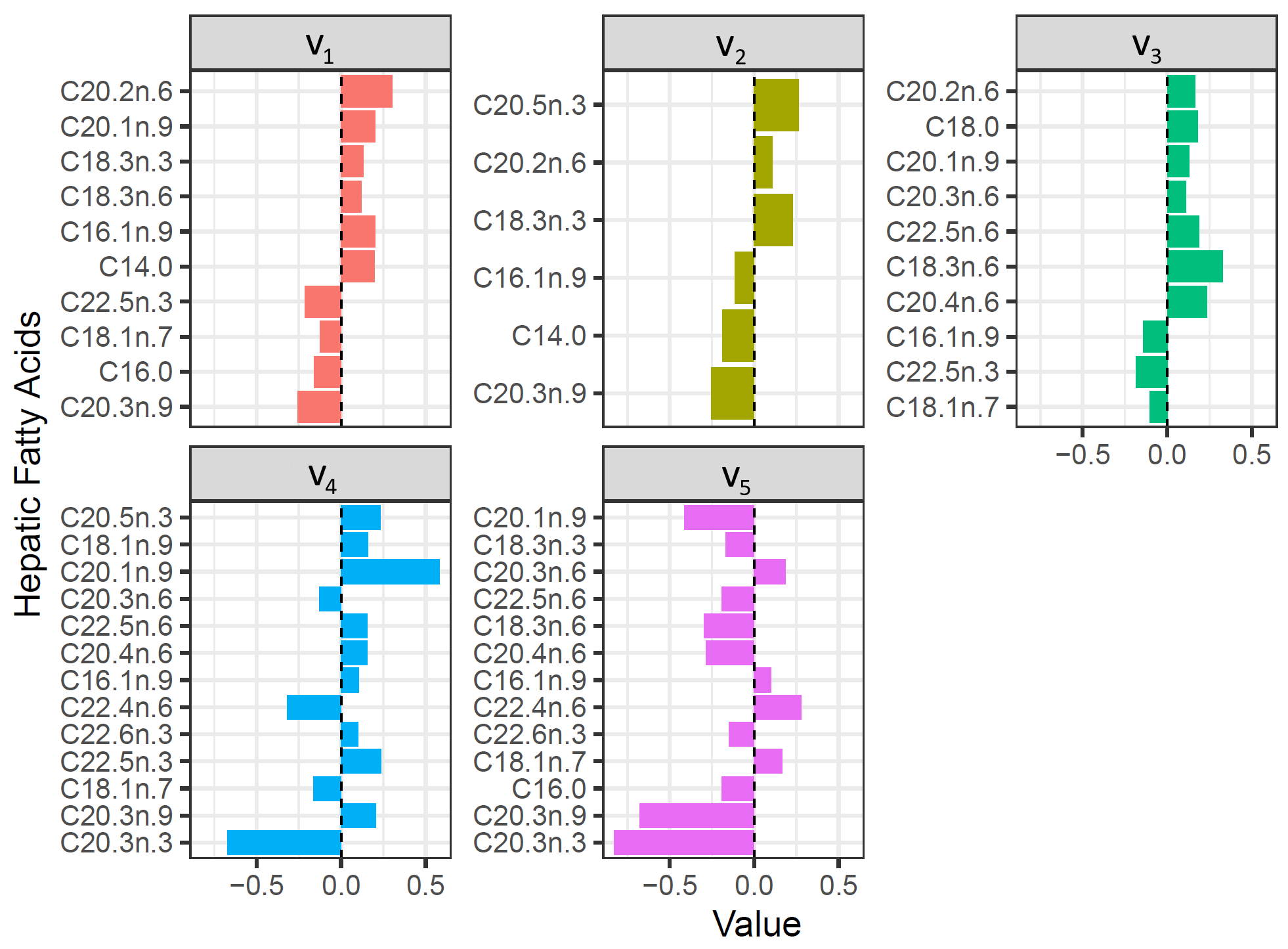}
        \caption{Acids}
    \end{subfigure}
    \caption{Five canonical directions produced for the Nutrimouse dataset obtained by sparse ECCAR.}
      \label{fig:nutri:uv}
\end{figure*}

\clearpage
{
\subsection{Alcohol Use Disorder}
\paragraph{Figures} We include a plot of singular values of the cross-covariance matrix $X^\top Y$ to support the choice of $r = 2$ in the analysis.
\begin{figure*}[h]
    \centering
        \includegraphics[width = 0.8\textwidth]{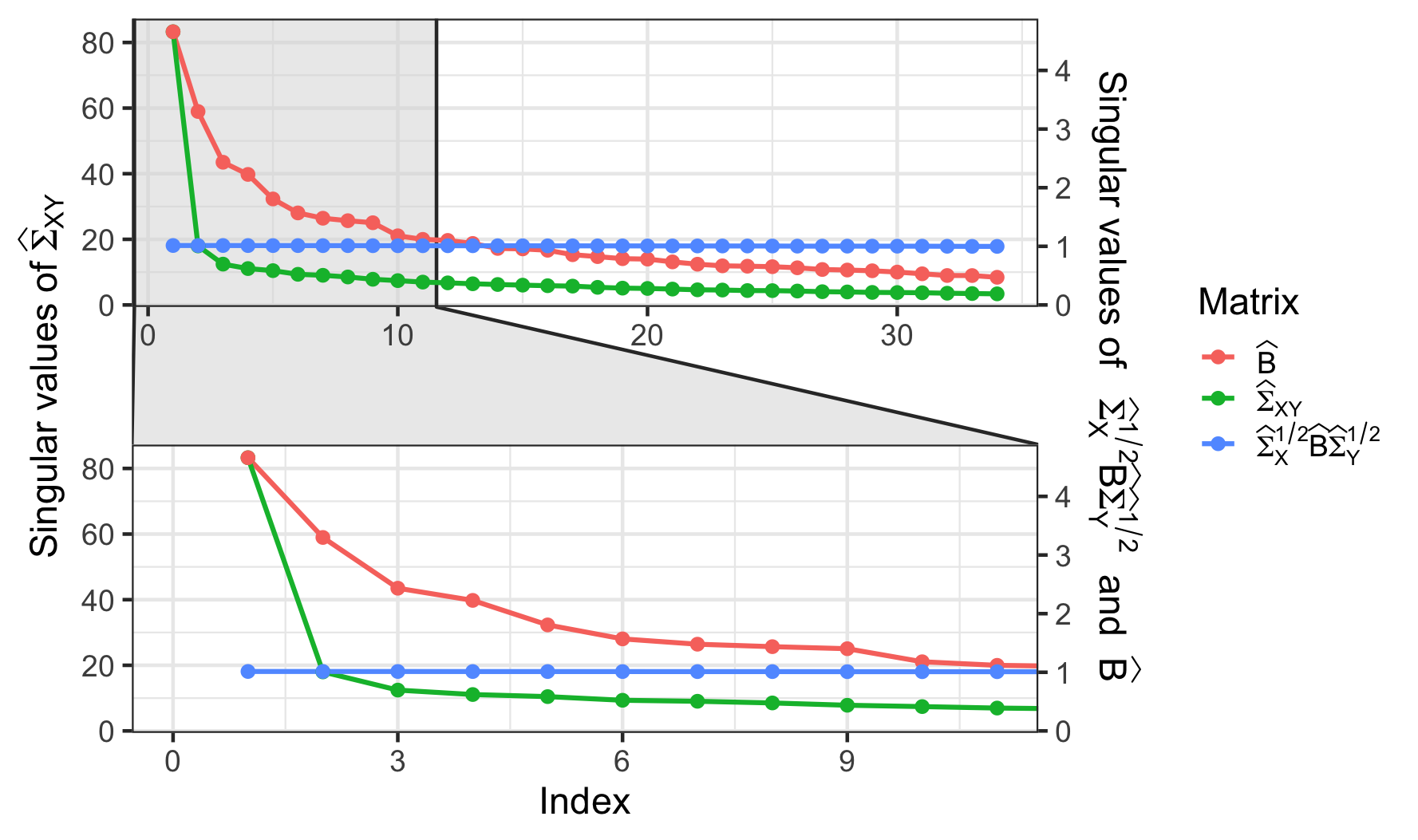}
\caption{Singular values of the empirical cross-covariance matrix between CpG methylation features and gene expression features in the AUD data, its denoised version and $\hat{B}$. The sharp decrease after the first two singular values for $\hat{\Sigma}_{XY}$ and  $\hat{\Sigma}_{X}^\frac12 \hat{B} \hat{\Sigma}_{Y}^\frac12 $ supports the choice of rank $r=2$ used for ECCAR (and other CCA benchmarks) in this paper.}
        \label{fig:aud_rank}
\end{figure*}}
\subsection{The Autism Brain Imaging Data Exchange}\label{app:subsec:abide}

\paragraph{Preprocessing.} We preprocess the ABIDE dataset using the Configurable Pipeline for the Analysis of Connectomes (C-PAC; \cite{craddock2013towards}), a standardized workflow for fMRI data. The initial steps include slice-timing correction, realignment for motion, and normalization of signal intensities. To reduce noise from non-neuronal sources, we apply nuisance regression \citep{lund2006non}, targeting several confounding factors such as physiological artifacts (e.g., cardiac and respiratory cycles), head motion, and scanner drift.

Head motion artifacts are modeled with 24 motion parameters, while scanner drift is addressed using both linear and quadratic terms. Physiological noise is controlled using the CompCor approach \citep{behzadi2007component}, which extracts the top five principal components from signals within white matter and cerebrospinal fluid (CSF). After nuisance correction, the data undergo band-pass filtering in the 0.01–0.1 Hz range and global signal regression to further isolate neural fluctuations.

For spatial normalization, functional images are aligned to anatomical scans using FSL’s Boundary-Based Registration (BBR), followed by non-linear warping to MNI space with the Advanced Normalization Tools (ANTs; \cite{avants2009advanced}).

\begin{figure*}[h]
    \centering
    \begin{subfigure}[b]{0.43\textwidth}
        \centering
        \includegraphics[width = \textwidth]{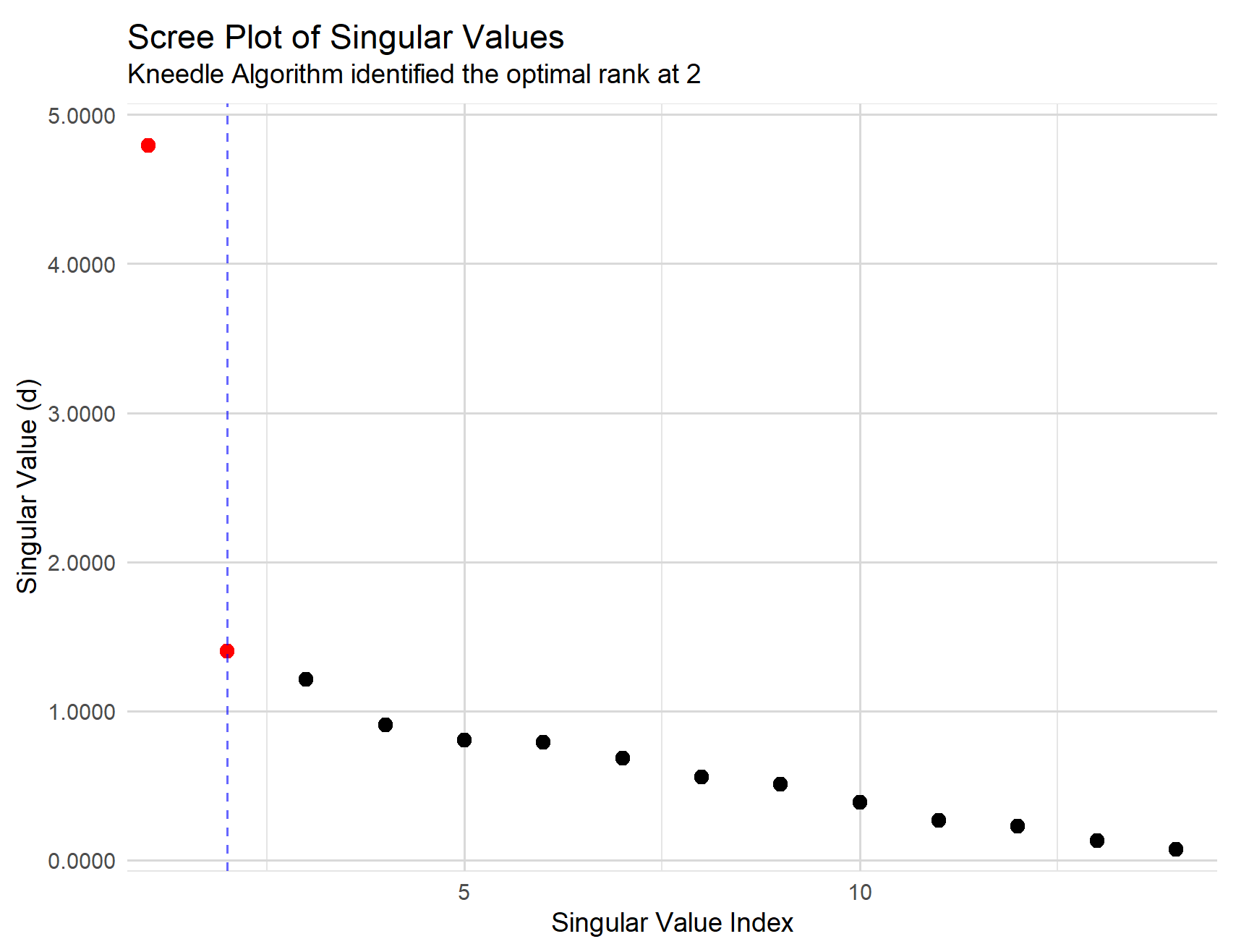}
        \caption{Screeplot for the matrix $\hat{\Sigma}_{XY}$ in the ABIDE dataset. Dashed line indicates rank selected by the Kneedle algorithm.}
    \end{subfigure}
    ~ 
    \begin{subfigure}[b]{0.43\textwidth}
        \centering
        \includegraphics[width = \textwidth]{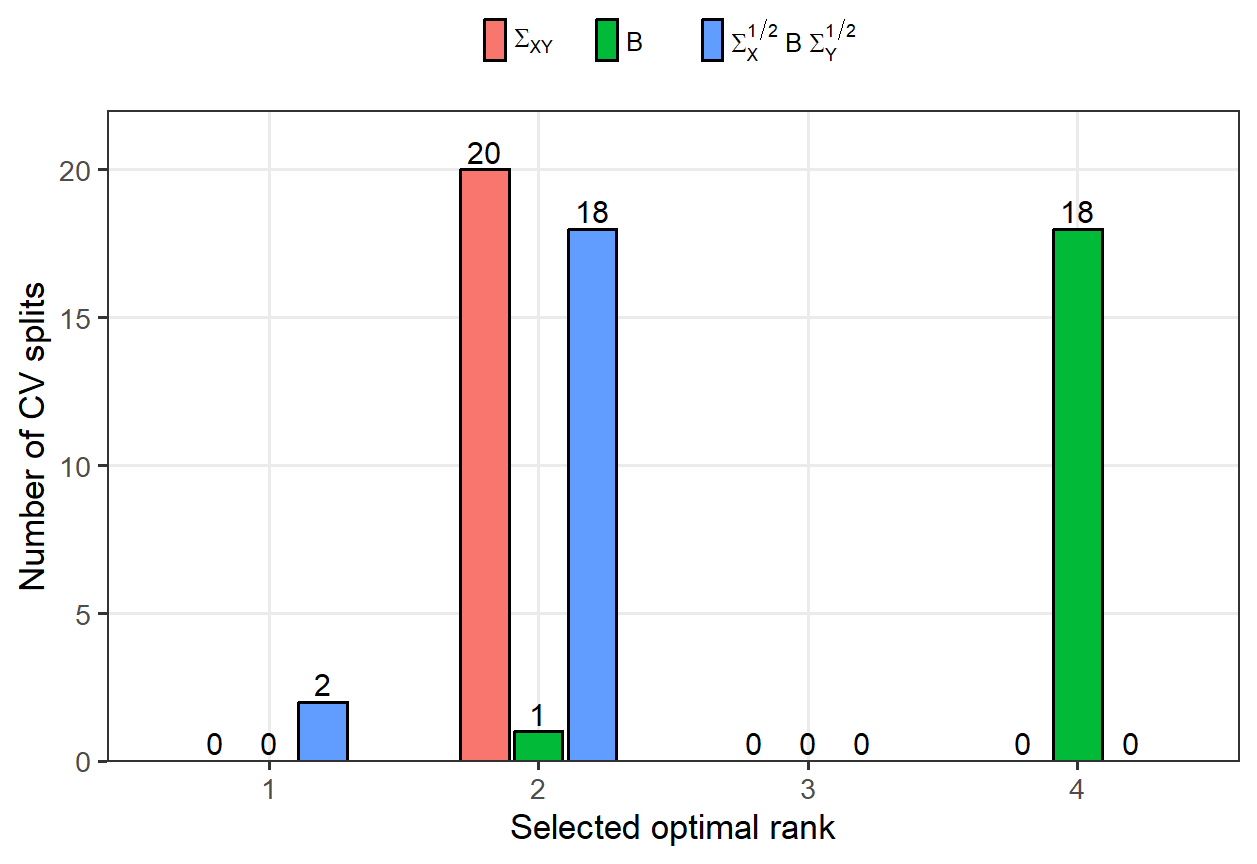}
        \caption{Optimal rank chosen by the Kneedle algorithm over 20 random CV splits of the data (to choose the appropriate $\rho$).}
    \end{subfigure}
    \caption{Diagnostic Plots for the ABIDE data for selecting the appropriate rank. In the rest of this analysis, we focus on $r=2$, since both the screeplots for  $\hat{\Sigma}_{XY}$  and  $\hat{\Sigma}_{X}^\frac12 \hat{B}  \hat{\Sigma}_{Y}^\frac12$ concur.}
    \label{fig:abide:rank}
\end{figure*}

\paragraph{Figures.} We include additional plots for the analysis of the ABIDE data, such as the heatmaps for the first canonical direction $u_1$ (Figures~\ref{fig:abide:u:rowsparse}-\ref{fig:abide:u:witten})
and the supplementary graphs for the sparse CCA benchmarks of \cite{parkhomenko2009sparse} and \cite{witten2009penalized}(Figure~\ref{fig:abide:competitors}).

\begin{figure*}[h!]
    \centering
    \begin{subfigure}[b]{0.43\textwidth}
        \centering
        \includegraphics[width = \textwidth]{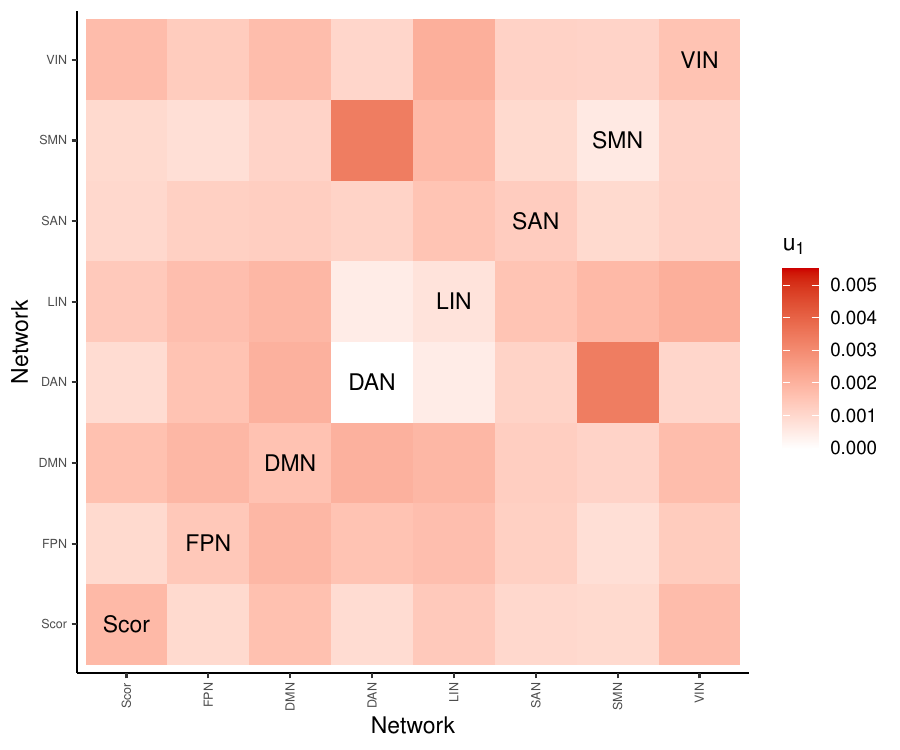}
        \caption{row-sparse ECCAR}
    \end{subfigure}
    ~ 
    \begin{subfigure}[b]{0.43\textwidth}
        \centering
        \includegraphics[width = \textwidth]{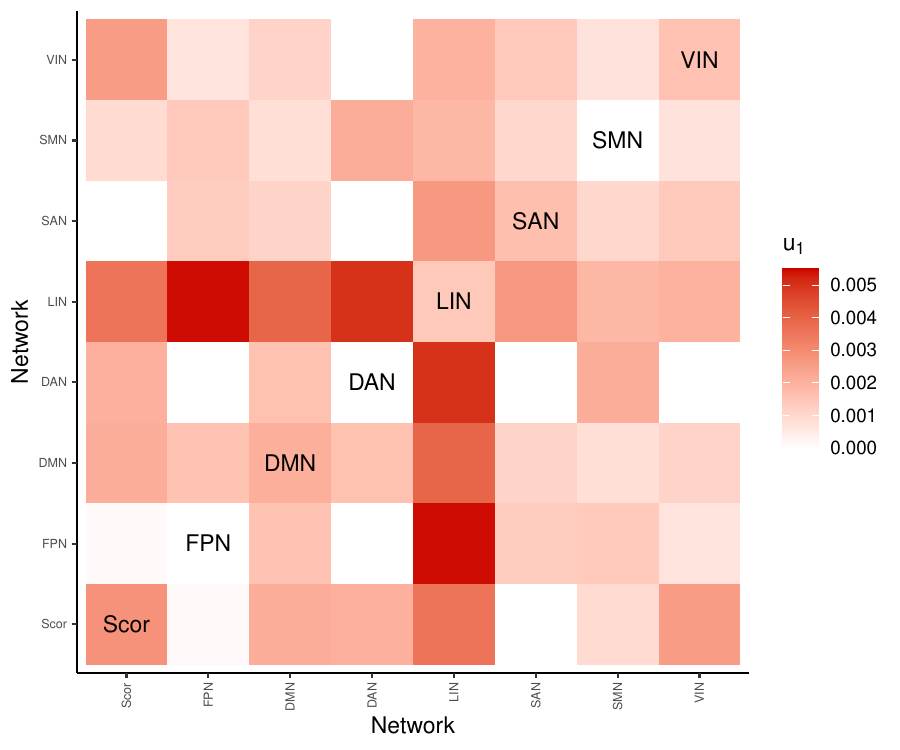}
        \caption{group ECCAR}
    \end{subfigure}
    \caption{The heatmap for the first canonical direction vector $u_1$ corresponding to the brain signal. The plot represents the average of absolute values of loadings within each network.}
    \label{fig:abide:unet}
\end{figure*}

\begin{figure*}[h!]
    \centering
        \includegraphics[width = 0.6\textwidth]{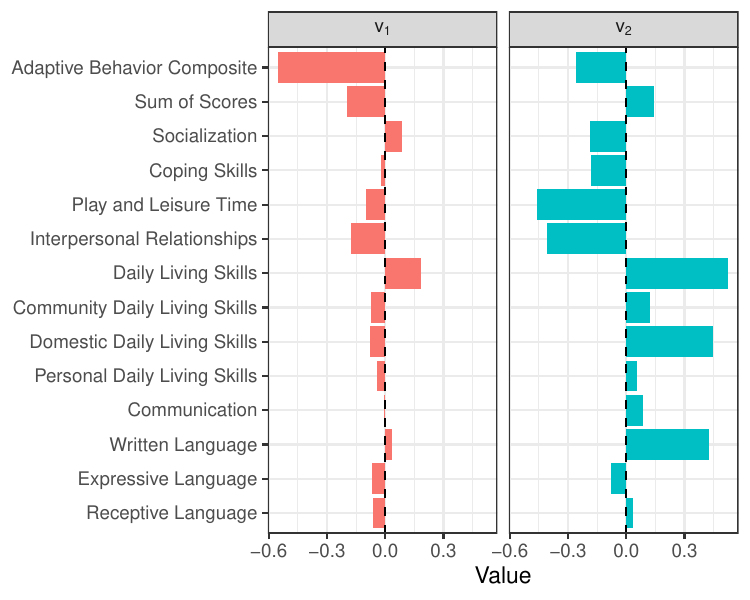}
        \caption{The barplot for the first and second canonical direction vectors $v_1$ and $v_2$ corresponding to the Vineland behavioral scores obtained by row-sparse ECCAR.}
        \label{fig:abide:v}
\end{figure*}

\begin{figure}[h!]
    \centering
    \includegraphics[width=0.8\textwidth]{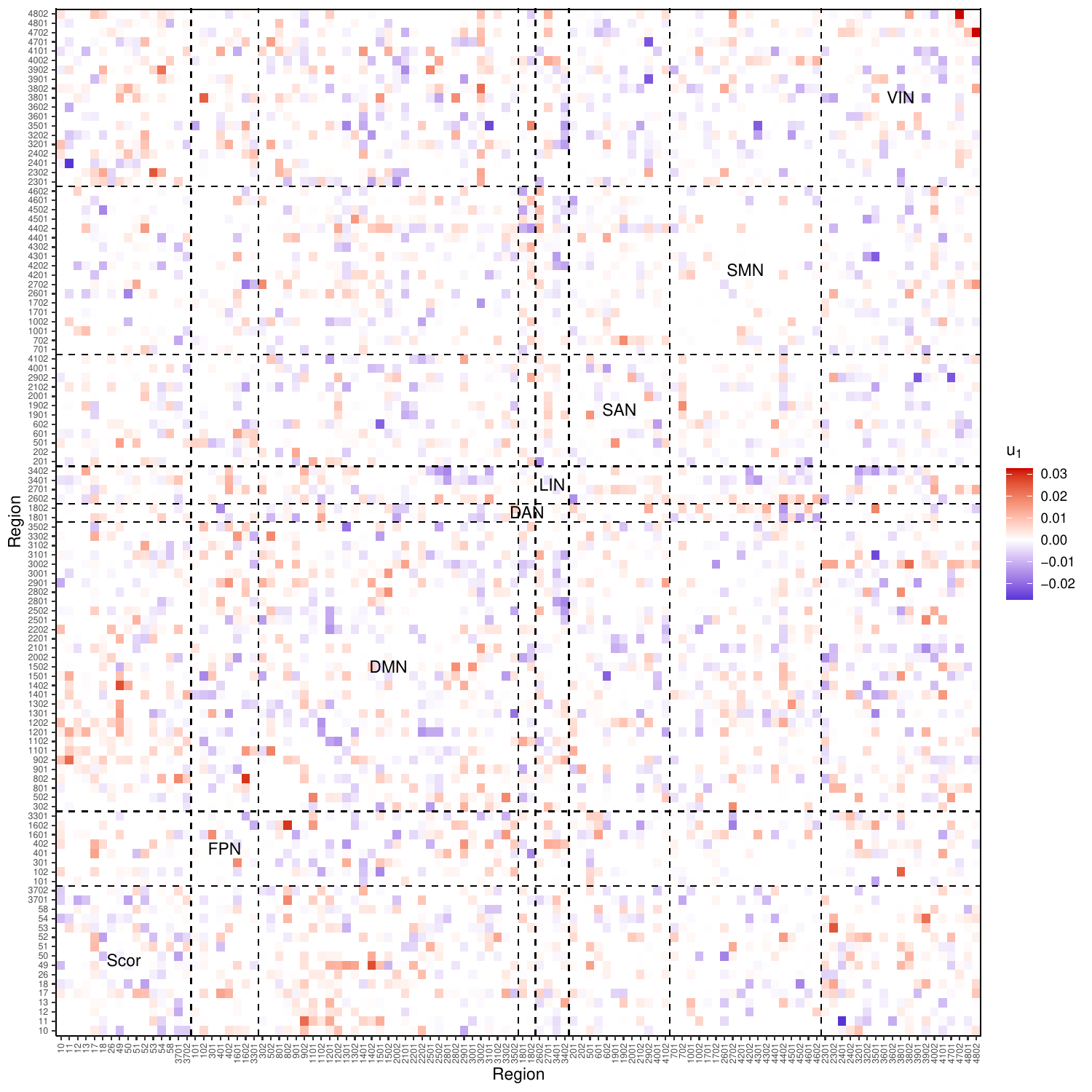}
    \caption{Heatmap of the first canonical direction vector $u_1$ from row-sparse ECCAR on ABIDE data.
The x- and y-axes indicate brain region indices. Horizontal and vertical lines separate the eight brain networks, the name of each network is represented in the diagonal blocks.}
    \label{fig:abide:u:rowsparse}
\end{figure}

\begin{figure}[h!]
    \centering
    \includegraphics[width=0.8\textwidth]{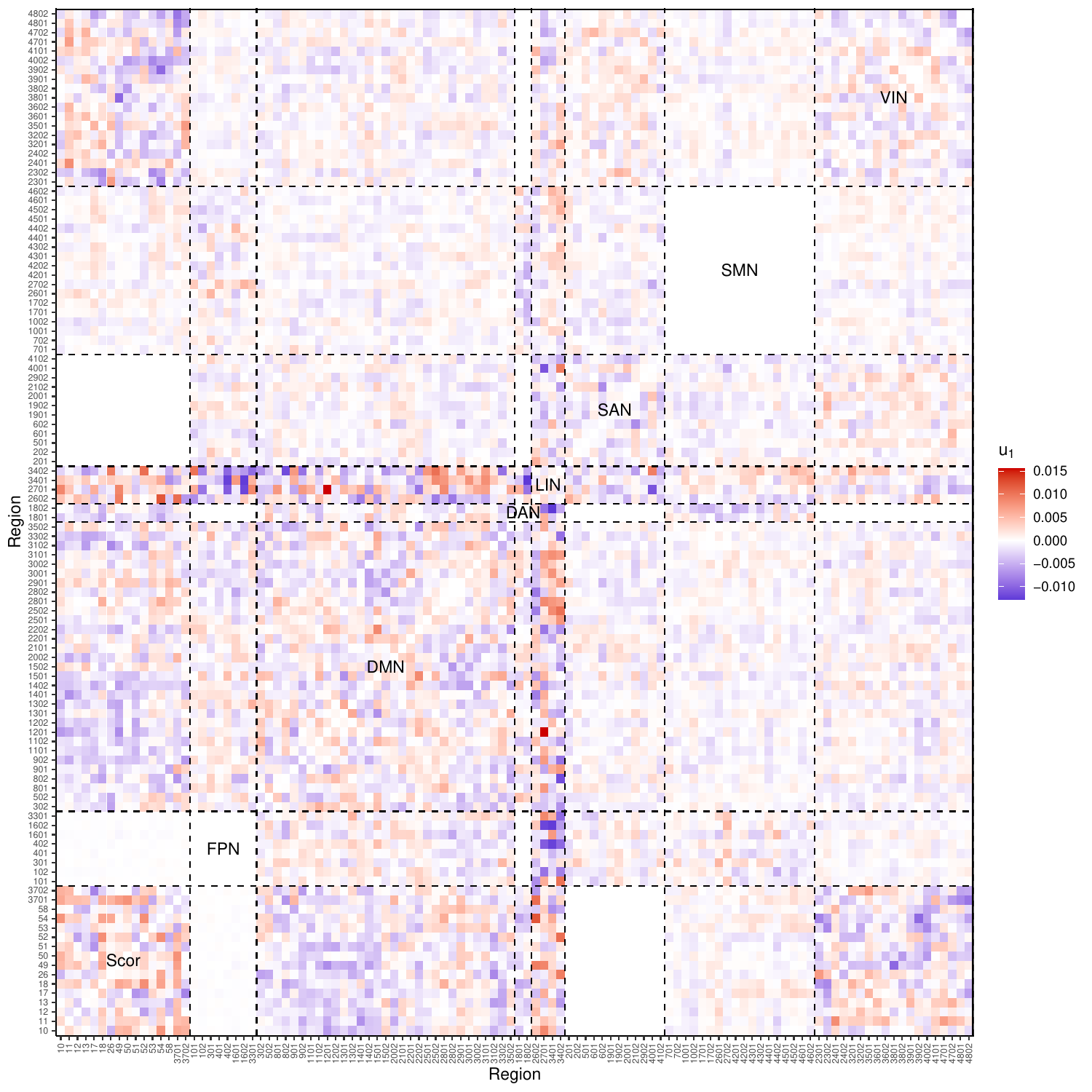}
    \caption{Heatmap of the first canonical direction vector $u_1$ from group-sparse ECCAR on ABIDE data.
The x- and y-axes indicate brain region indices. Horizontal and vertical lines separate the eight brain networks, the name of each network is represented in the diagonal blocks.}
    \label{fig:abide:u:group-sparse}
\end{figure}

\begin{figure}[h!]
    \centering
    \includegraphics[width=0.8\textwidth]{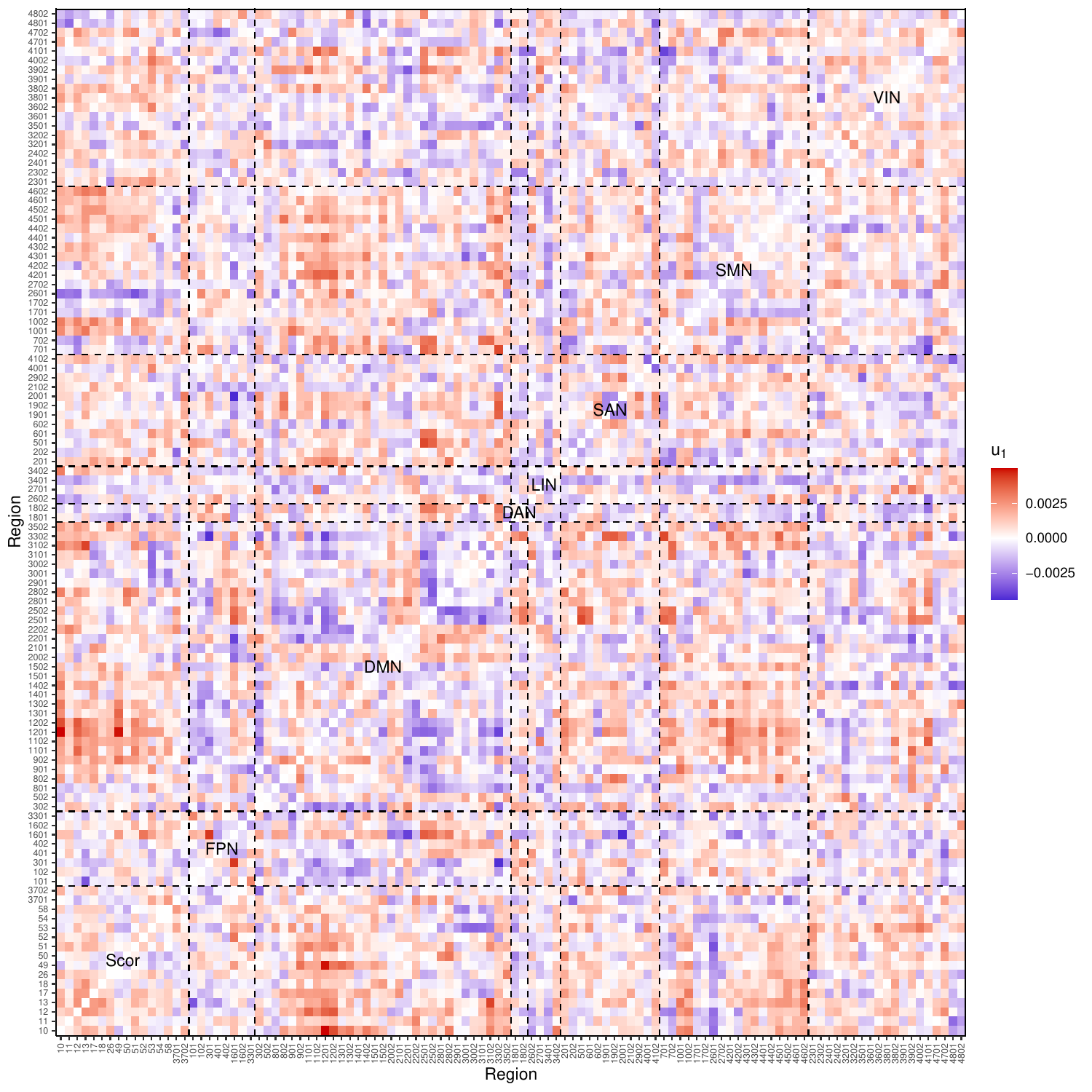}
    \caption{Heatmap of the first canonical direction vector $u_1$ from sparse CCA by \citet{parkhomenko2009sparse} on ABIDE data.
The x- and y-axes indicate brain region indices. Horizontal and vertical lines separate the eight brain networks, the name of each network is represented in the diagonal blocks.}
    \label{fig:abide:u:parkhomenko}
\end{figure}

\begin{figure}[h!]
    \centering
    \includegraphics[width=0.8\textwidth]{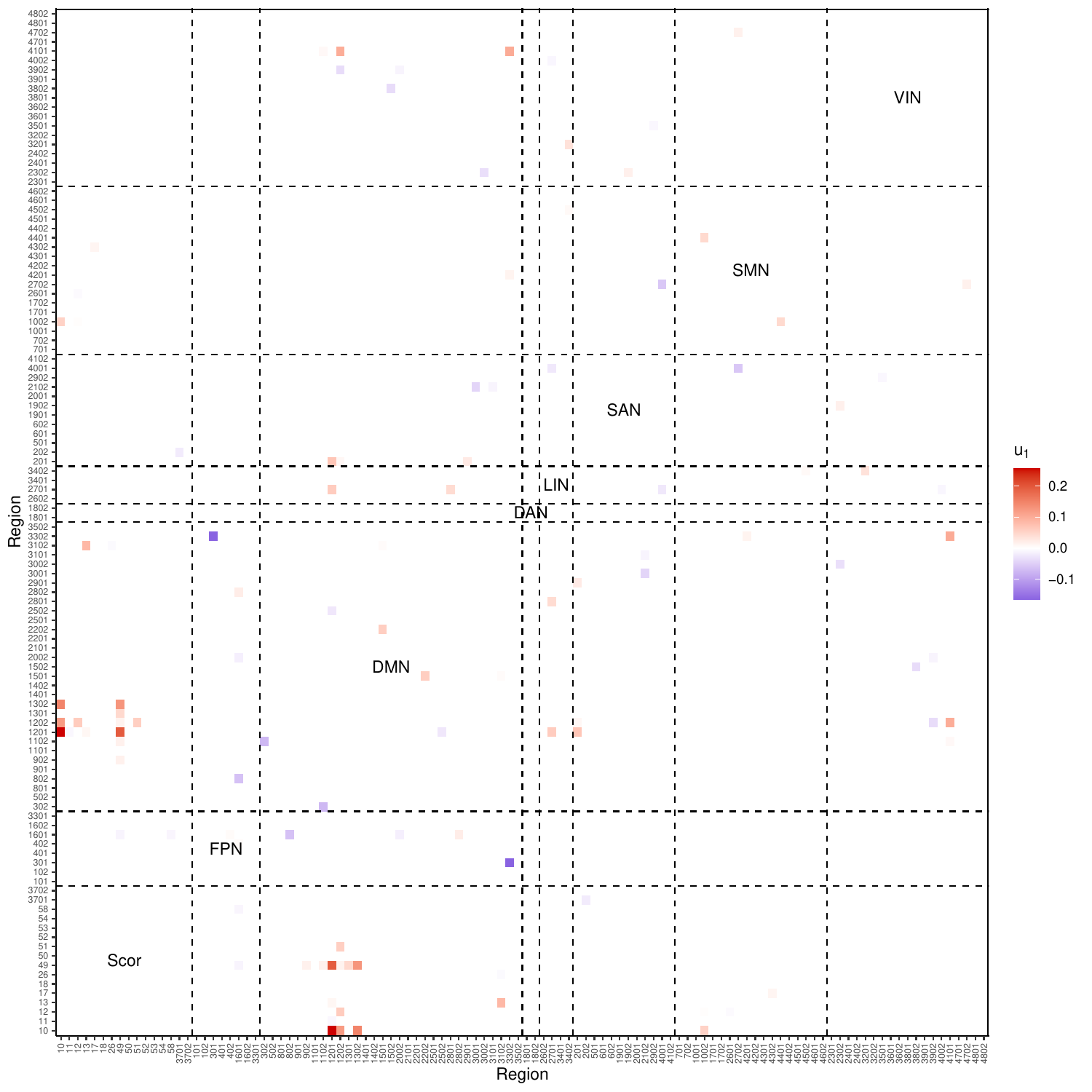}
    \caption{Heatmap of the first canonical direction vector $u_1$  from sparse CCA by \citet{witten2009penalized} on ABIDE data.
The x- and y-axes indicate brain region indices. Horizontal and vertical lines separate the eight brain networks, the name of each network is represented in the diagonal blocks.}
    \label{fig:abide:u:witten}
\end{figure}

\begin{figure*}[h!]
    \begin{subfigure}[b]{0.43\textwidth}
        \includegraphics[width = \textwidth]{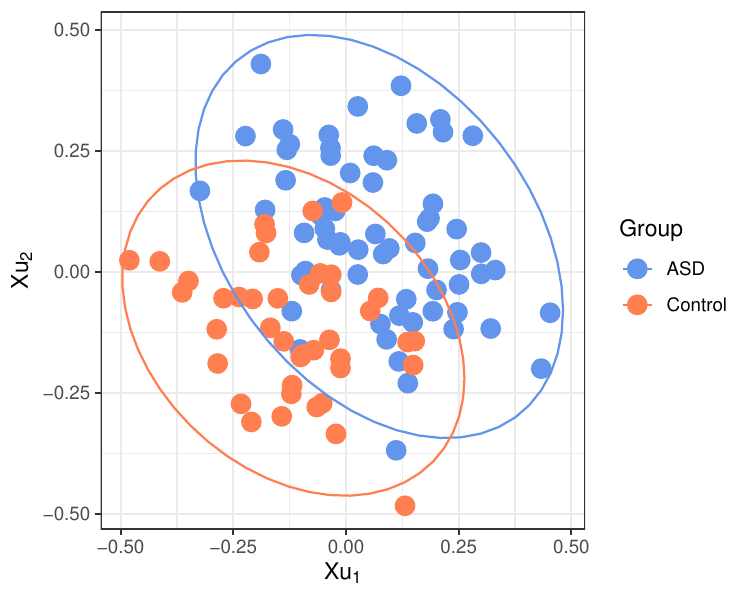}
    \end{subfigure}
    ~ 
    \begin{subfigure}[b]{0.34\textwidth}
        \includegraphics[width = \textwidth, trim = {0, 0, 2.5cm, 0}, clip]{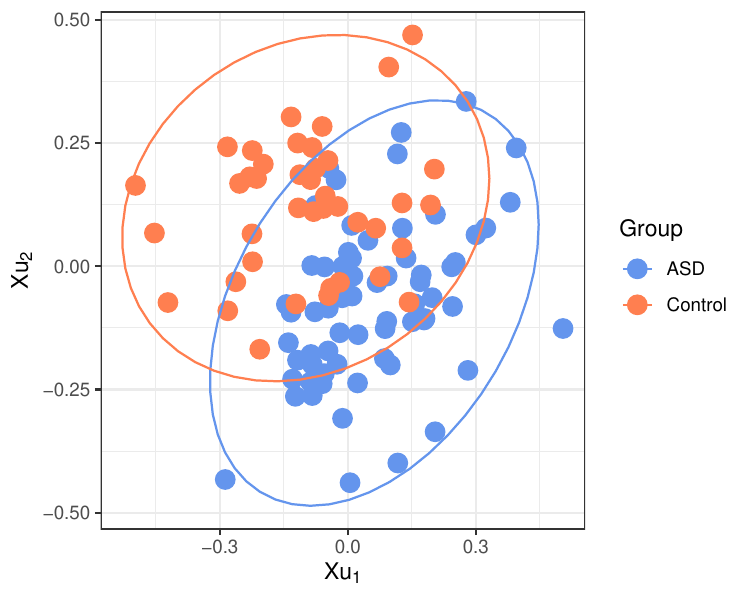}
    \end{subfigure}
    \centering
    \begin{subfigure}[b]{0.43\textwidth}
        \includegraphics[width = \textwidth]{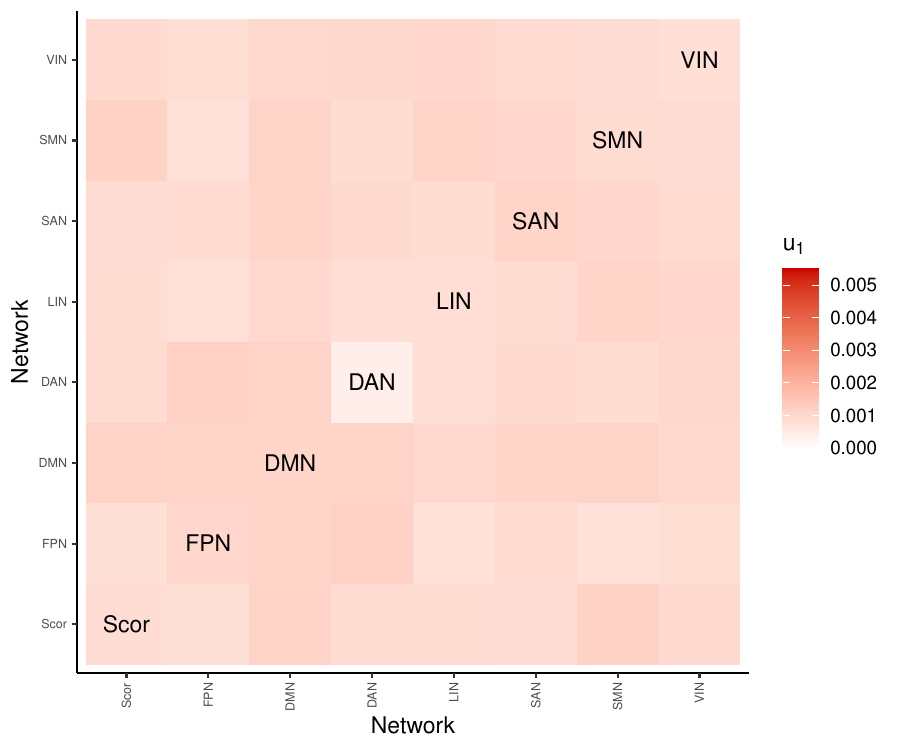}
    \end{subfigure}
    ~ 
    \begin{subfigure}[b]{0.37\textwidth}
        \includegraphics[width = \textwidth, trim = {0, 0, 2cm, 0}, clip]{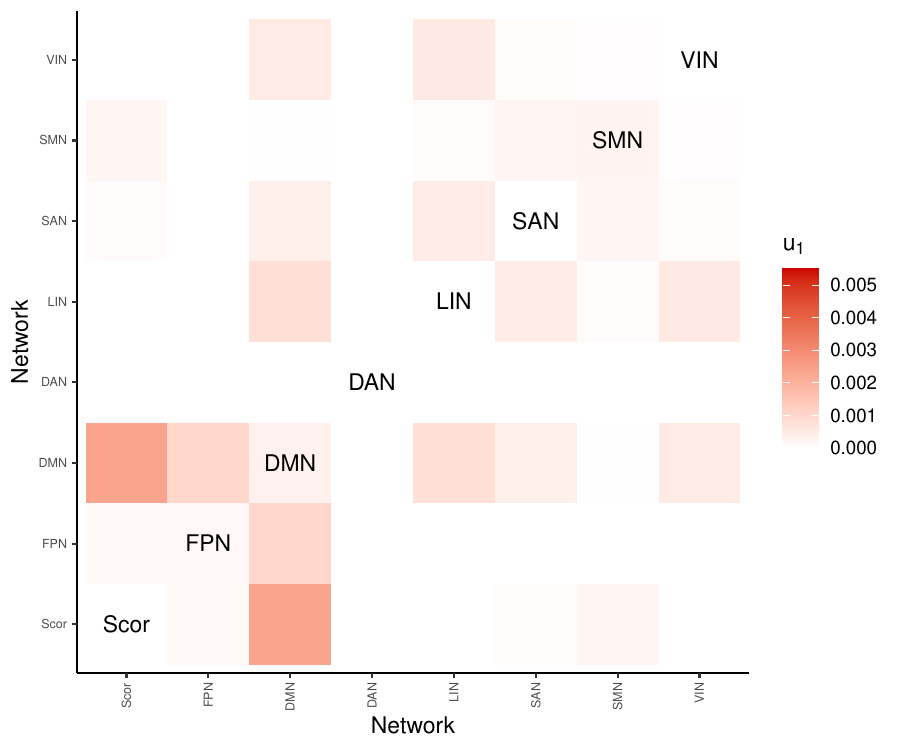}
    \end{subfigure}
    \begin{subfigure}[b]{0.37\textwidth}
        \centering
        \includegraphics[width = \textwidth, trim = {4.5cm, 0, 0.2cm, 0}, clip]{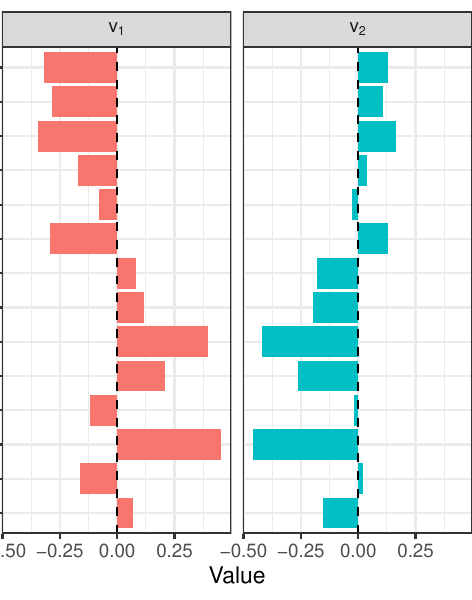}
        \caption{SCCA by \citet{parkhomenko2009sparse}}
    \end{subfigure}
        ~ 
    \begin{subfigure}[b]{0.58\textwidth}
        \centering
        \includegraphics[width = \textwidth, trim = {0.2cm, 0, 0, 0}, clip]{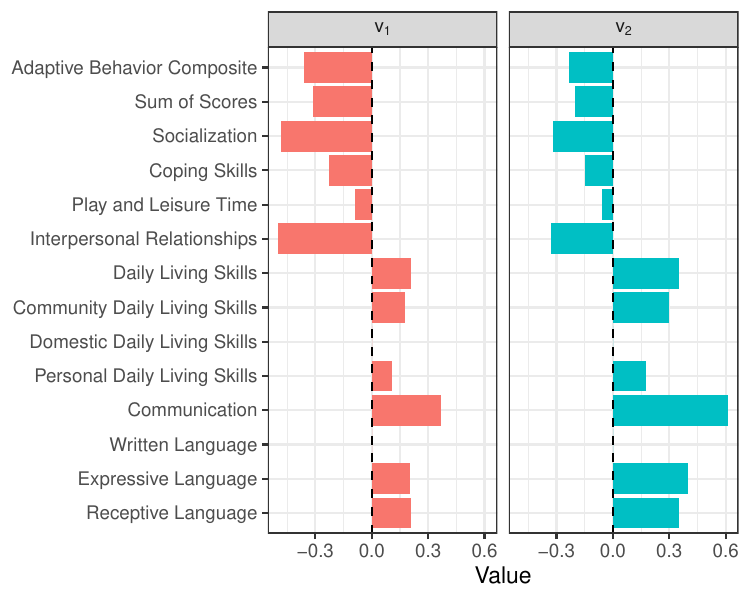}
        \caption{Sparse CCA by \citet{witten2009penalized}}
    \end{subfigure}

    \caption{The results for the ABIDE data. Top: the projection of the data onto the space of the first two canonical variates colored by the disease status.
    Middle: the heatmap for $u_1$ (average of absolute values within each network). Bottom: the bar plot for $v_1$ and $v_2$.}
\label{fig:abide:competitors}
\end{figure*}

\clearpage

\subsection{LLM Interpretability}
\label{app:subsec:llm}

In this subsection, we present additional tables and plots for our LLM interpretability example.

\begin{table}[h!]
\centering
\begin{tabular}{ll}
\toprule
\textbf{Category} & \textbf{Fine-grain topics}              \\ \midrule
Religion & alt.atheism; soc.religion.christian; talk.religion.misc   \\ \\
Computer & comp.graphics; comp.os.ms-windows.misc; comp.sys.ibm.pc.hardware;\\
  &  comp.sys.mac.hardware; comp.windows.x                    \\ \\
Misc     & misc.forsale     \\  \\                                     
Cars     & rec.autos; rec.motorcycles                               \\ \\
Sport    & rec.sport.baseball; rec.sport.hockey                     \\ \\
Science  & sci.crypt; sci.electronics; sci.med; sci.space           \\ \\
Politics & talk.politics.guns; talk.politics.mideast; talk.politics.misc    \\
\bottomrule
\end{tabular}%
\caption{20–Newsgroups coarse categories and constituent topics.}
\label{tab:20ng_compact}
\end{table}

\begin{figure*}[h!]
    \begin{subfigure}[b]{0.48\textwidth}
        \includegraphics[width = \textwidth, trim = {0, 0, 2.5cm, 0}, clip]{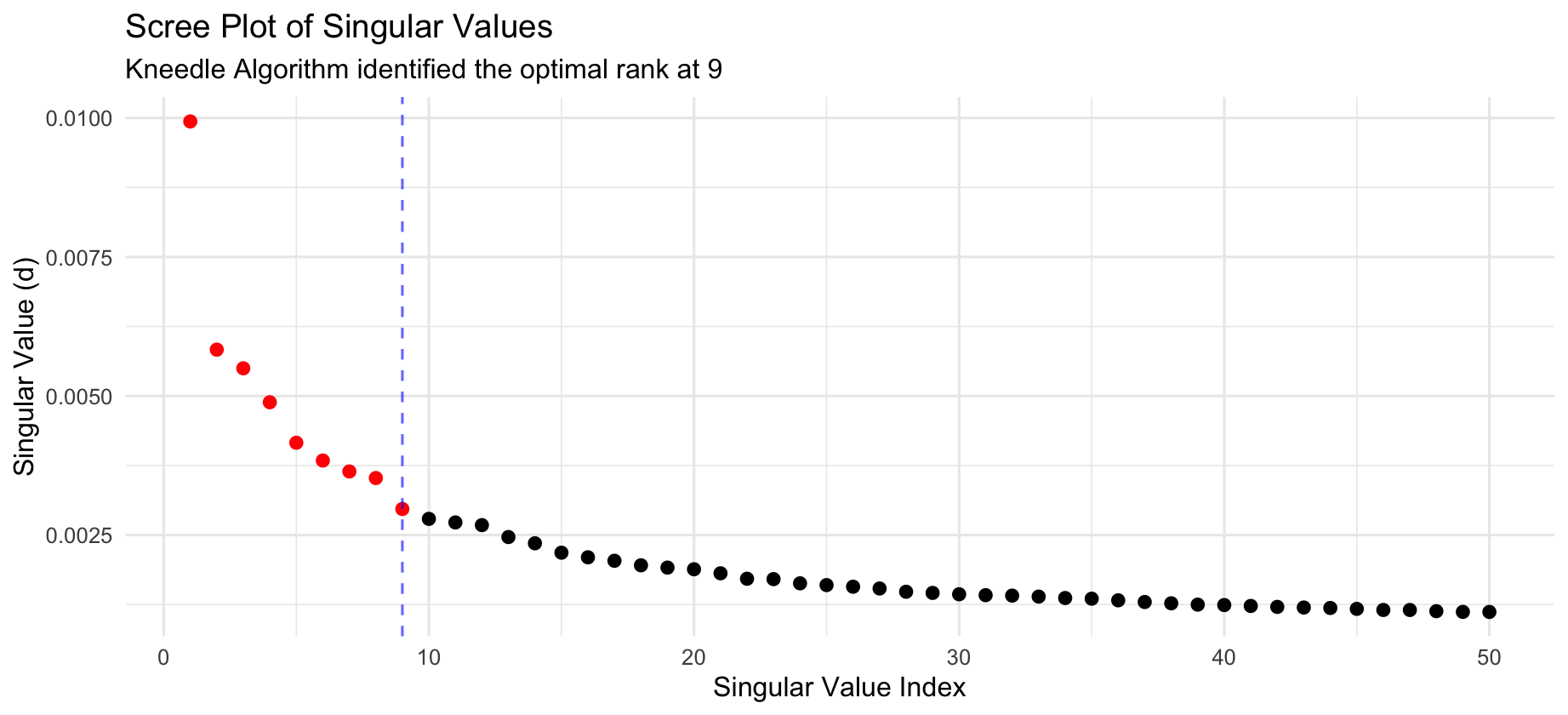}
    \end{subfigure}
    \centering
    \begin{subfigure}[b]{0.48\textwidth}
        \includegraphics[width = \textwidth]{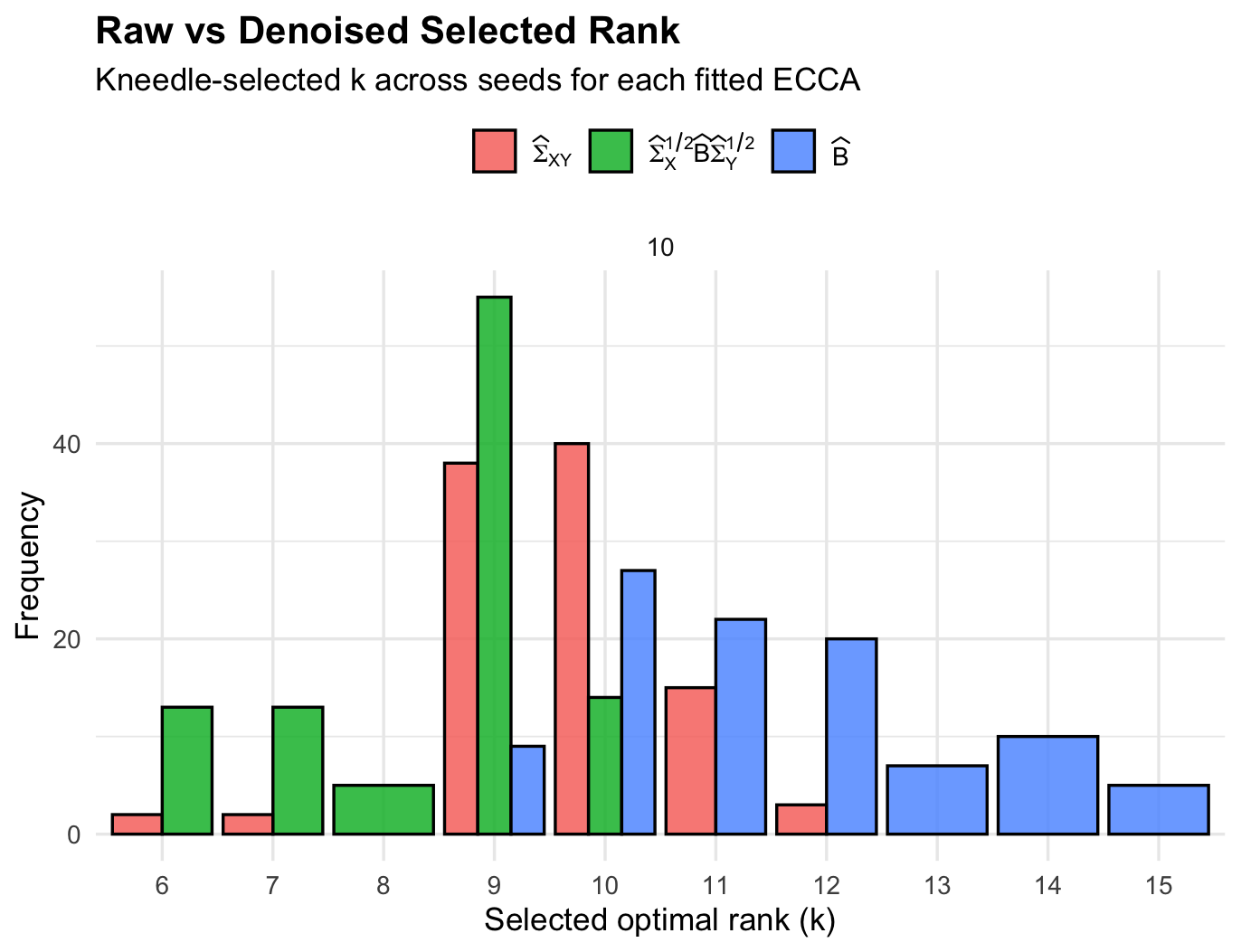}
    \end{subfigure}
    \caption{Screeplot of the original LLM data used to choose the number of canonical directions. Left plot shows an example of a screeplot of the raw covariance $\widehat{\Sigma}_{XY}$ on one draw of 3,000 samples. The first four singular values (left plot) seem more separated from the bulk. Right plot shows the number of directions chosen through the Kneedle Algorithm on the raw covariance matrix $\widehat{\Sigma}_{XY}$ (red), denoised covariance $ \widehat \Sigma_{X}^{\frac12} \widehat B \widehat \Sigma_{Y}^{\frac12}$ (green), and estimated $\widehat B$ (blue). The medians (over 100 random samples of 3,000 datapoints) show that the elbow of the raw spectrum tends to concentrate around the $10^{th}$ singular value, whilst that of the denoised one has a stronger peak at 9. Throughout the paper $r=4$ is used as a coarse resolution and $r=10$ as the finer resolution supported by the rank diagnostic.}
\label{fig:llm:screeplot}
\end{figure*}

\begin{figure*}[h!]
\centering
    \begin{subfigure}[b]{0.7\textwidth}
        \includegraphics[width = \textwidth]{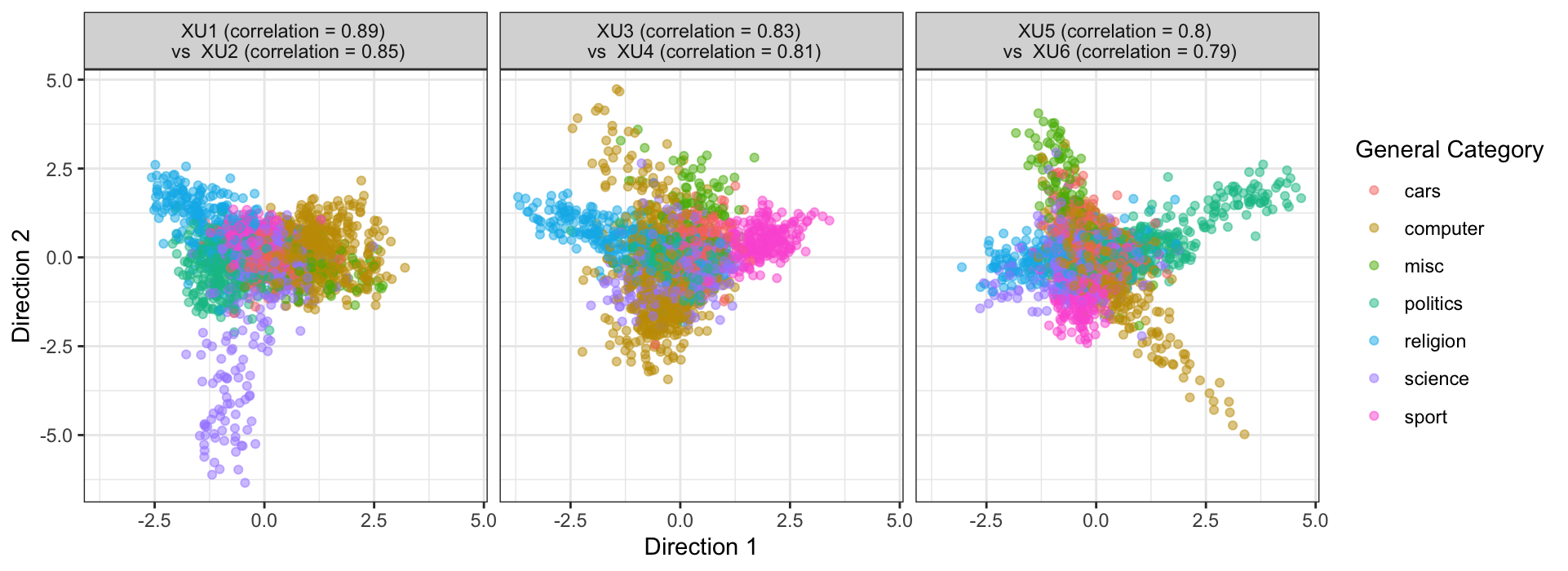}
        \caption{Sparse ECCAR.}
        \label{fig:llm_ecca}
    \end{subfigure}
  \begin{subfigure}[b]{0.7\textwidth}
        \includegraphics[width = \textwidth]{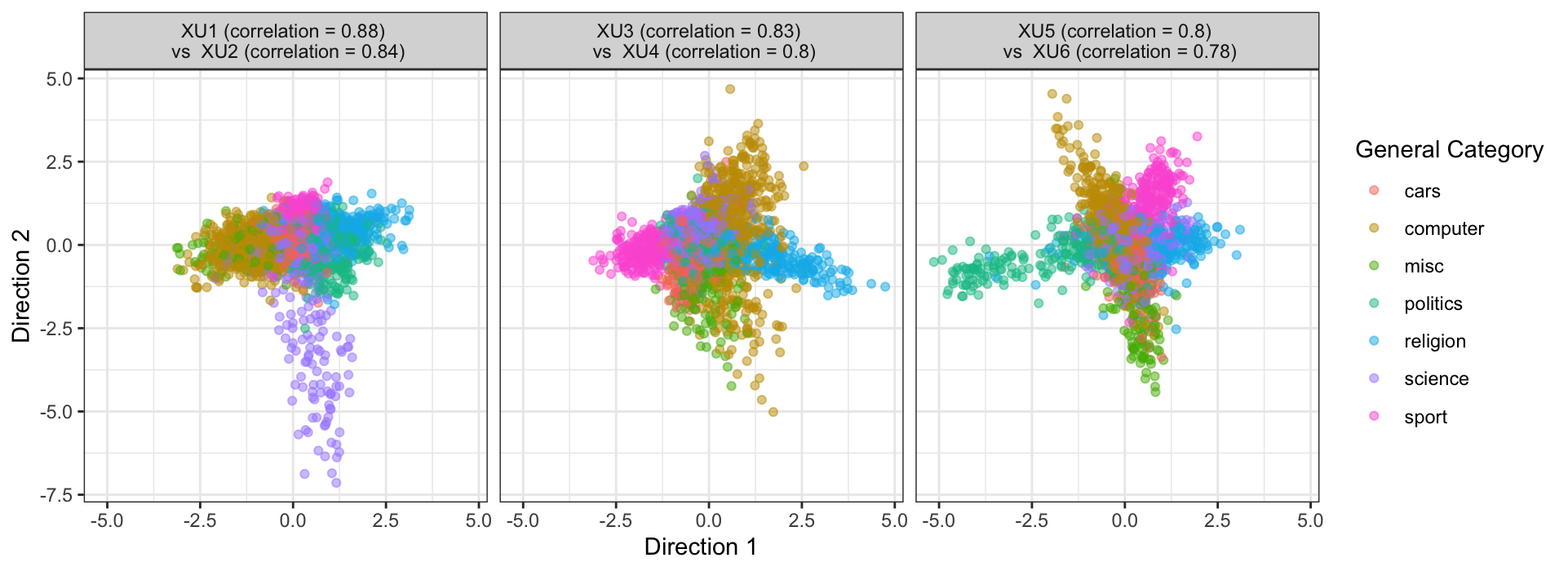}
        \caption{SAR \citet{wilms2015sparse}}
        \label{fig:llm_sar}
    \end{subfigure}
    \begin{subfigure}[b]{0.7\textwidth}
        \includegraphics[width = \textwidth]{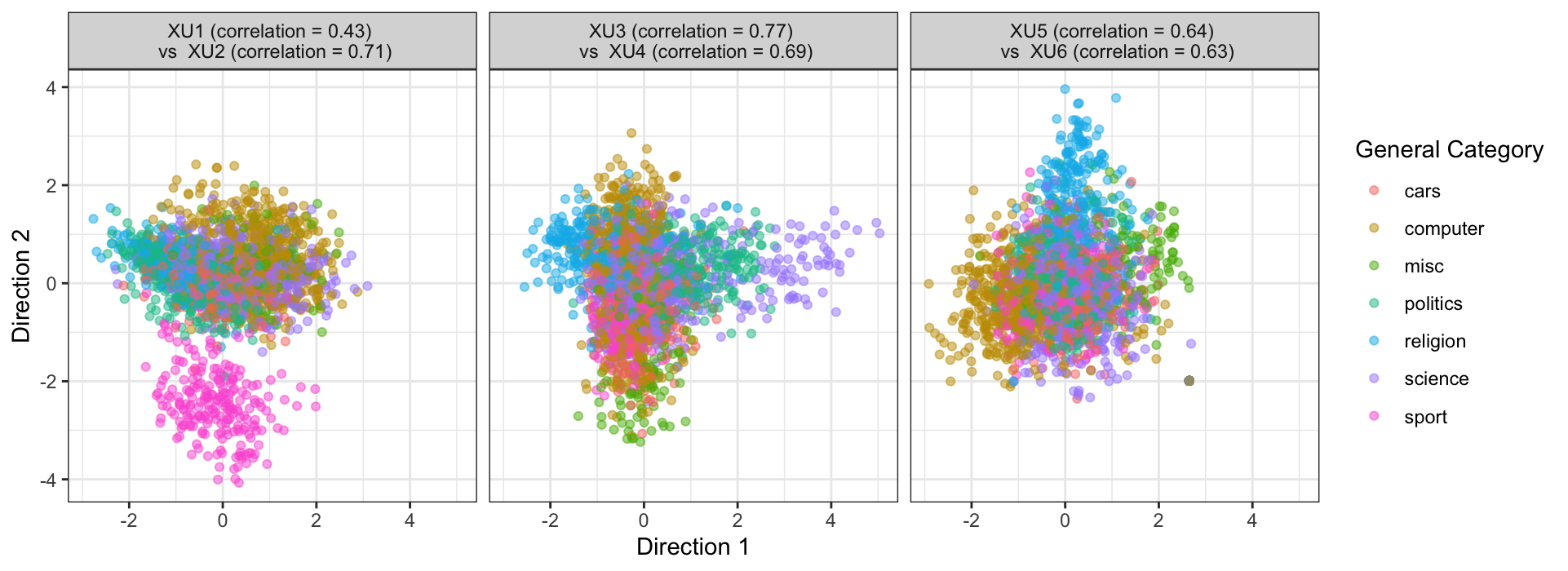}
        \caption{Sparse CCA method of \citet{witten2009penalized}.}
        \label{fig:llm_witten1}
    \end{subfigure}
        \begin{subfigure}[b]{0.7\textwidth}
        \includegraphics[width = \textwidth]{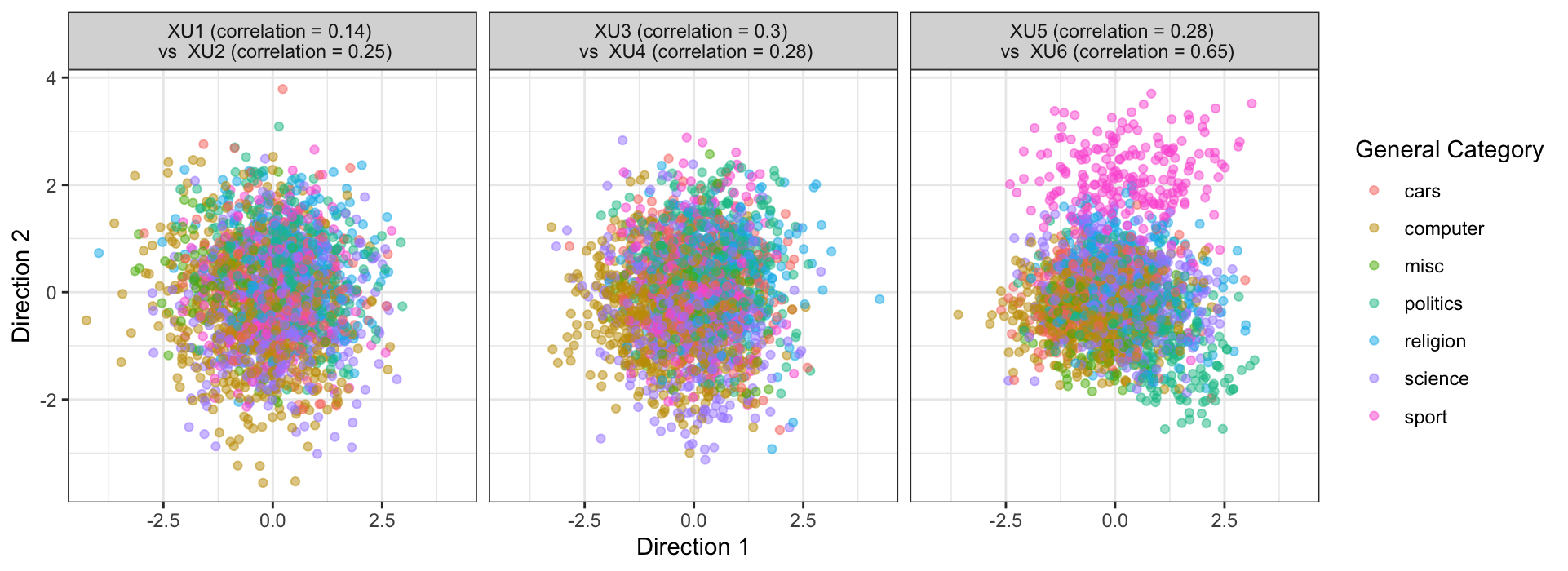}
        \caption{Sparse CCA of \citet{parkhomenko2009sparse}.}
        \label{fig:llm_witten2}
    \end{subfigure}
    \caption{CCA variates produced by sparse CCA for our LLM interpretability example: Alignment $XU$ of the LLM document embeddings with the TF-IDF frequencies.} 
\label{fig:llm:others}
\end{figure*}

\begin{figure*}[h!]
\centering
     \begin{subfigure}[b]{\textwidth}
        \includegraphics[width = \textwidth]{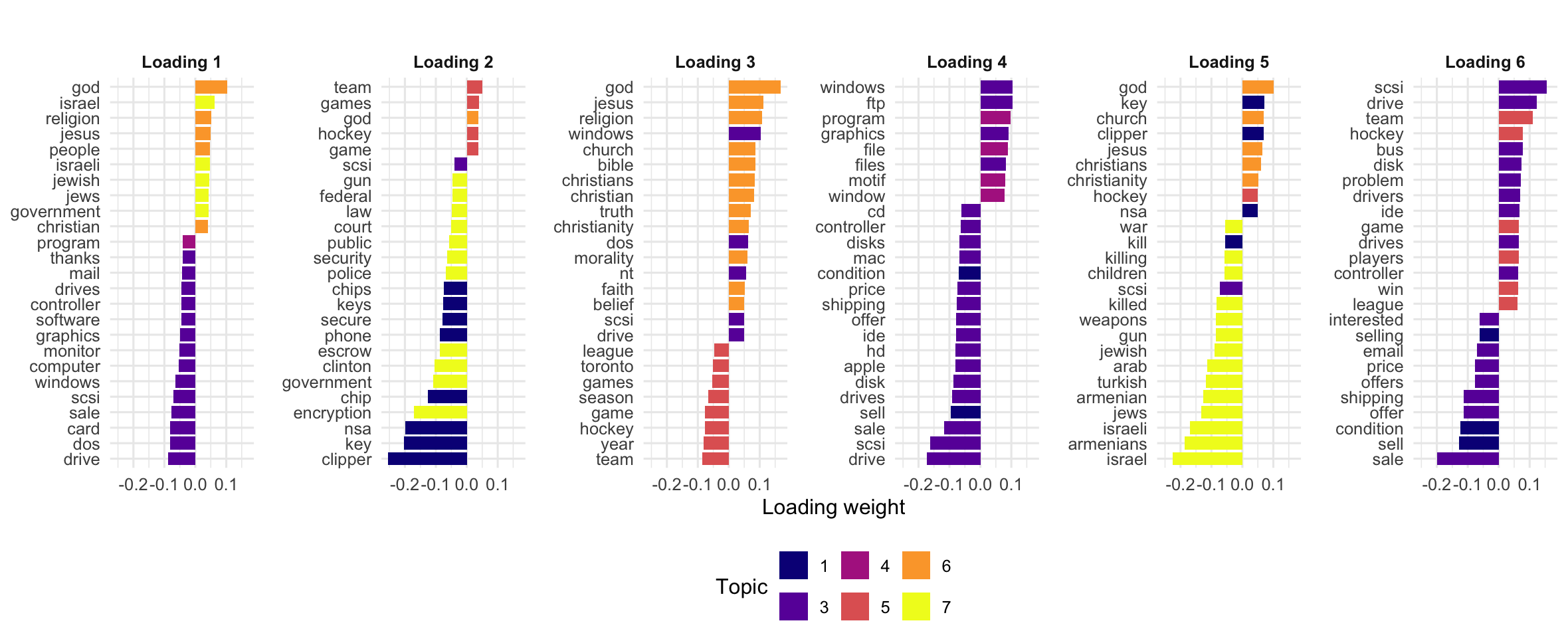}
        \caption{Loadings obtained using SAR by \citet{wilms2015sparse}}
        \label{fig:loading_sar_llm}
    \end{subfigure}
    \begin{subfigure}[b]{\textwidth}
        \includegraphics[width = \textwidth]{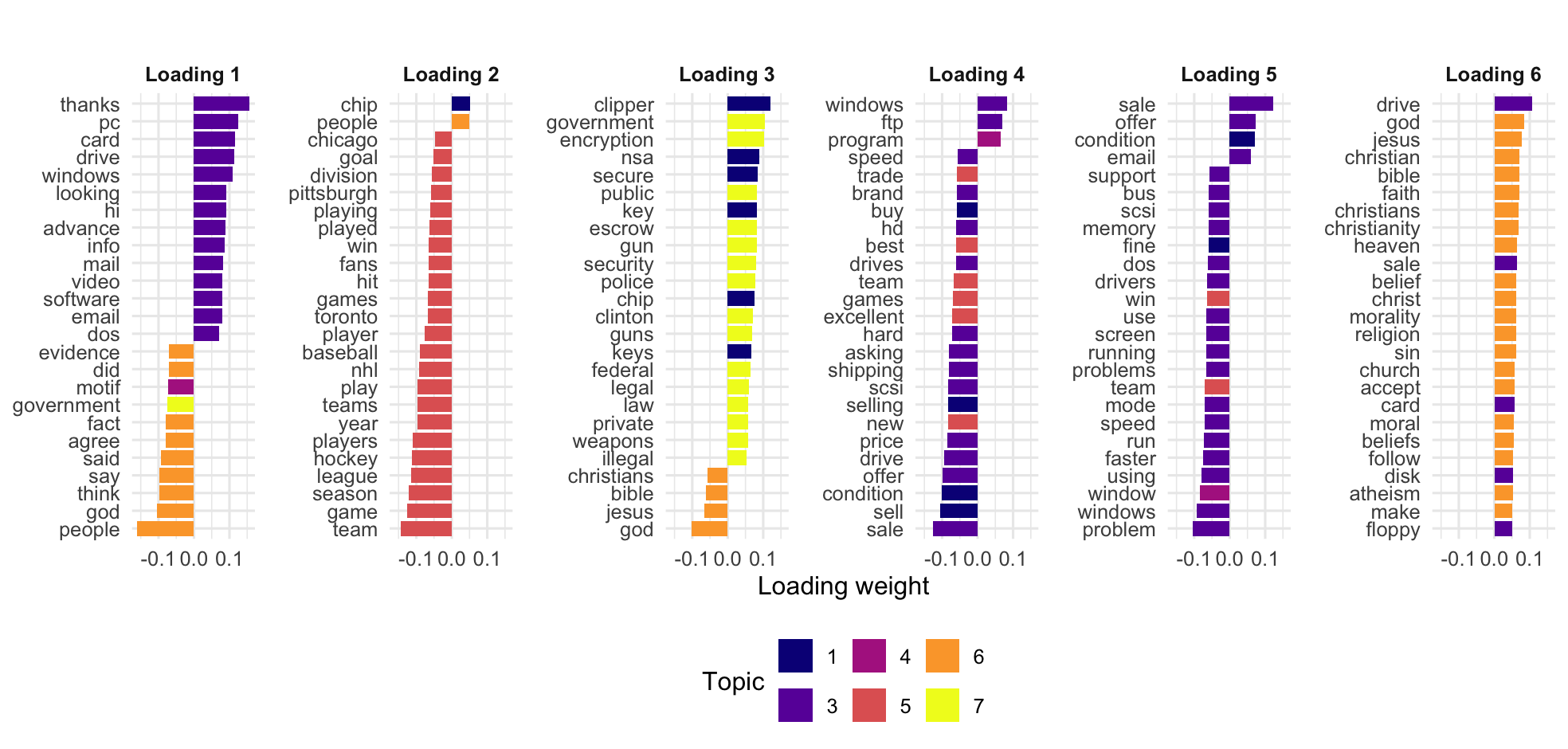}
        \caption{Loadings obtained using the Sparse CCA by \citet{witten2009penalized}}
         \label{fig:loading_witten_llm}
    \end{subfigure}
    \caption{CCA loadings as recovered by the sparse CCA method for the first 6 canonical directions. Colors indicate the word's main topics, as estimated by the Latent Dirichlet Allocation algorithm. }
\label{fig:llm:loading:others}
\end{figure*}

\begin{figure*}[phtb]
\centering
    \begin{subfigure}[b]{0.8\textwidth}
        \includegraphics[width = \textwidth]{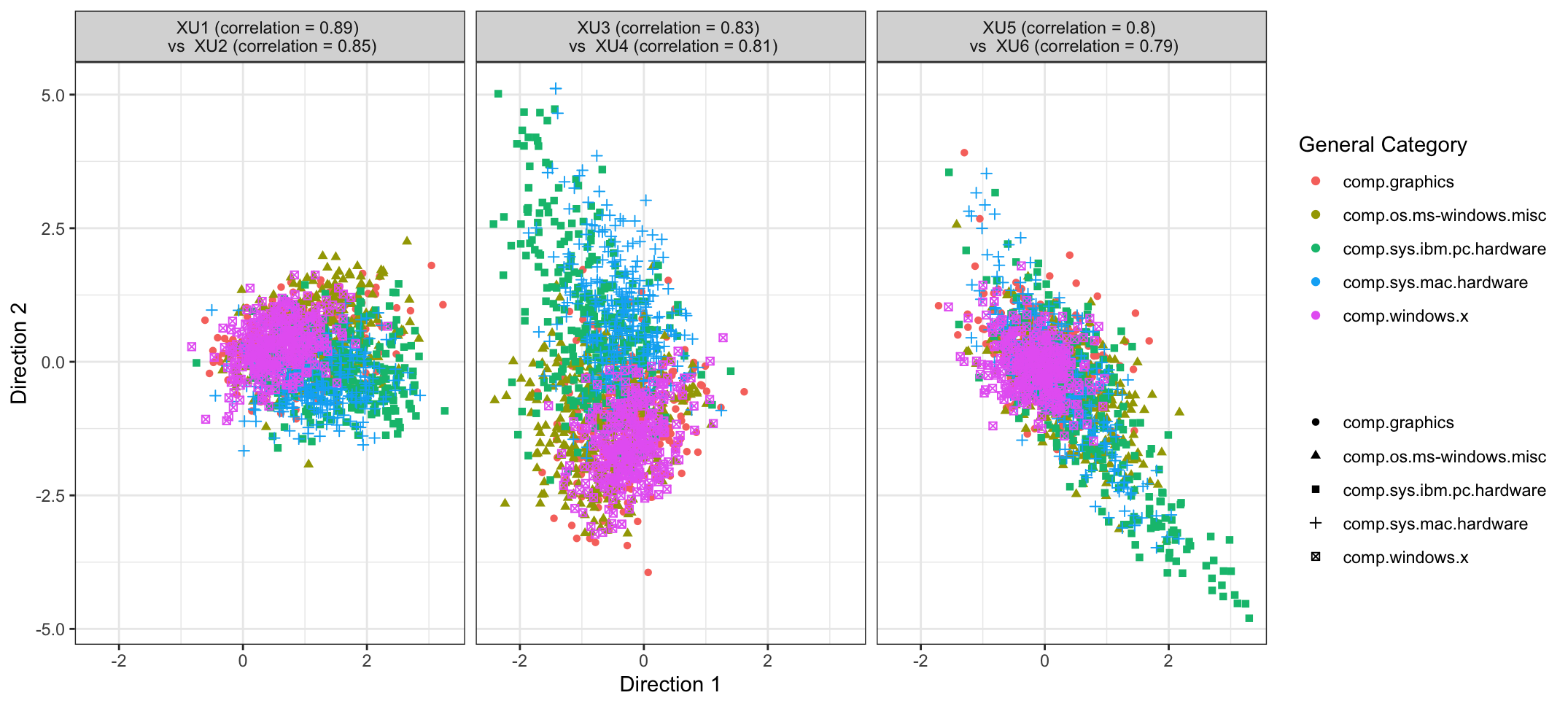}
        \caption{Computers}
         \label{fig:ecca_comp1}
    \end{subfigure}
  \begin{subfigure}[b]{0.8\textwidth}
        \includegraphics[width = \textwidth]{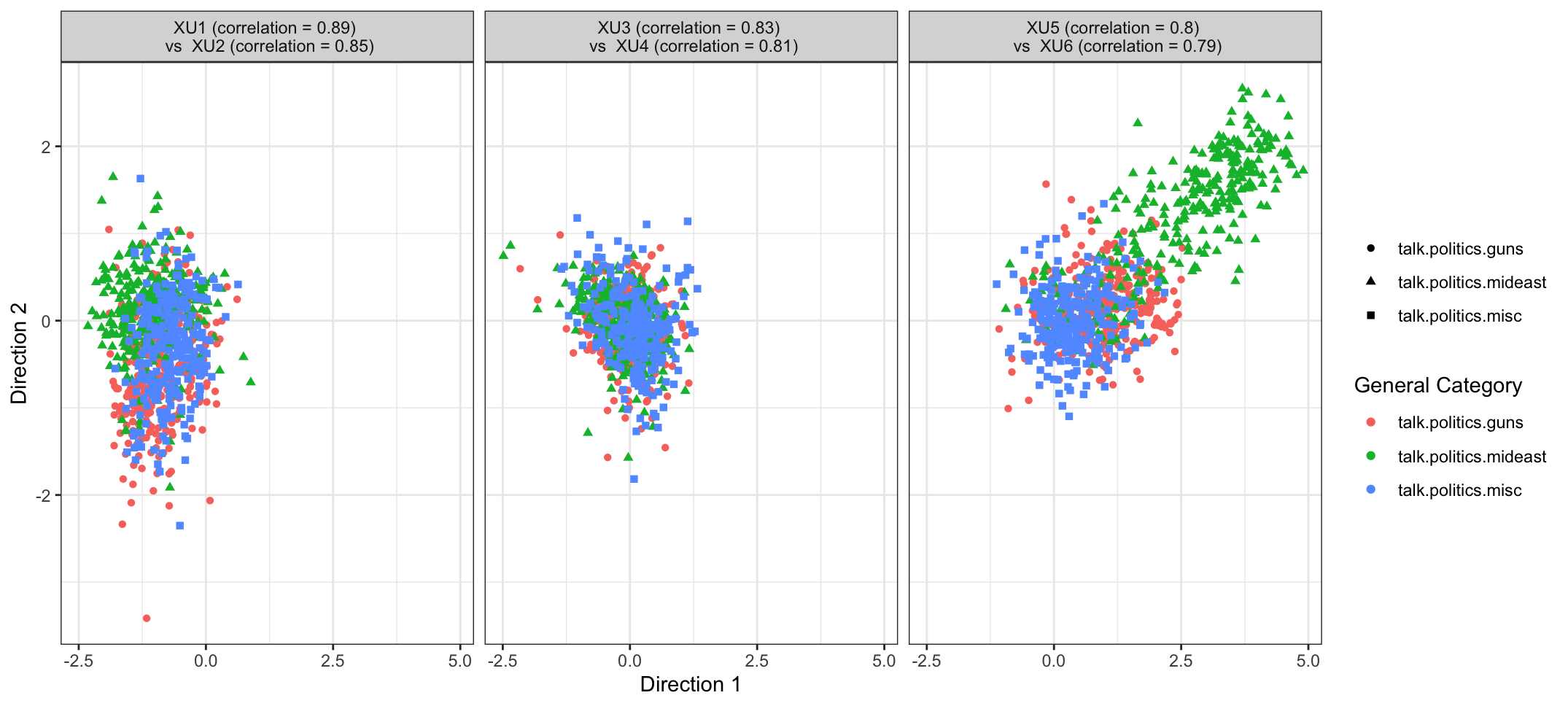}
         \label{fig:ecca_comp2}
        \caption{Politics}
    \end{subfigure}
    \begin{subfigure}[b]{0.8\textwidth}
        \includegraphics[width = \textwidth]{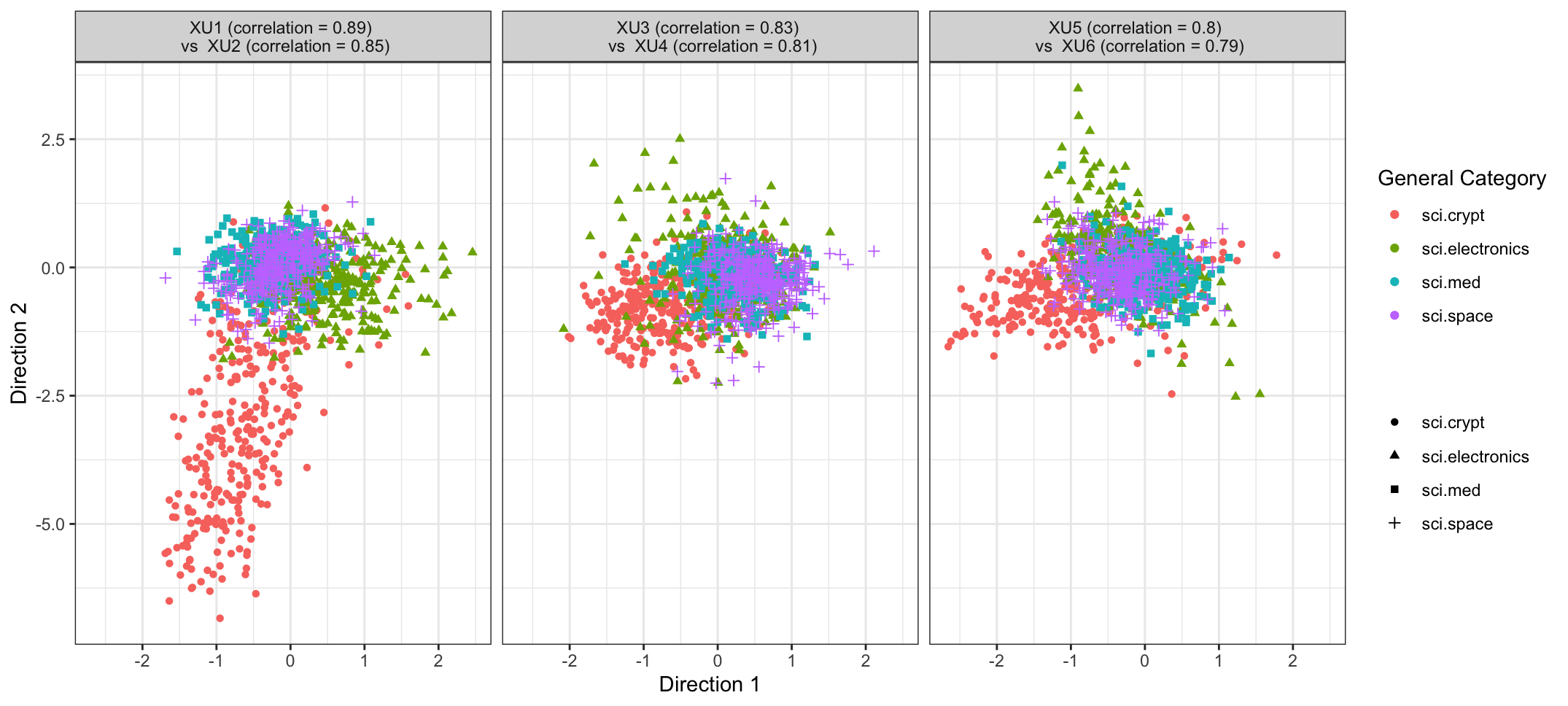}
        \caption{Science}
        \label{fig:ecca_science}
    \end{subfigure}
    \caption{CCA projections of points belonging to a category, using \texttt{ECCAR}.}
\label{fig:llm:ecca}
\end{figure*}

\clearpage

\subsection{Beyond Gaussian data: A transcriptomics example}
\label{app:subsec:gene}

\begin{figure}[H]
    \centering
    \includegraphics[width=\linewidth]{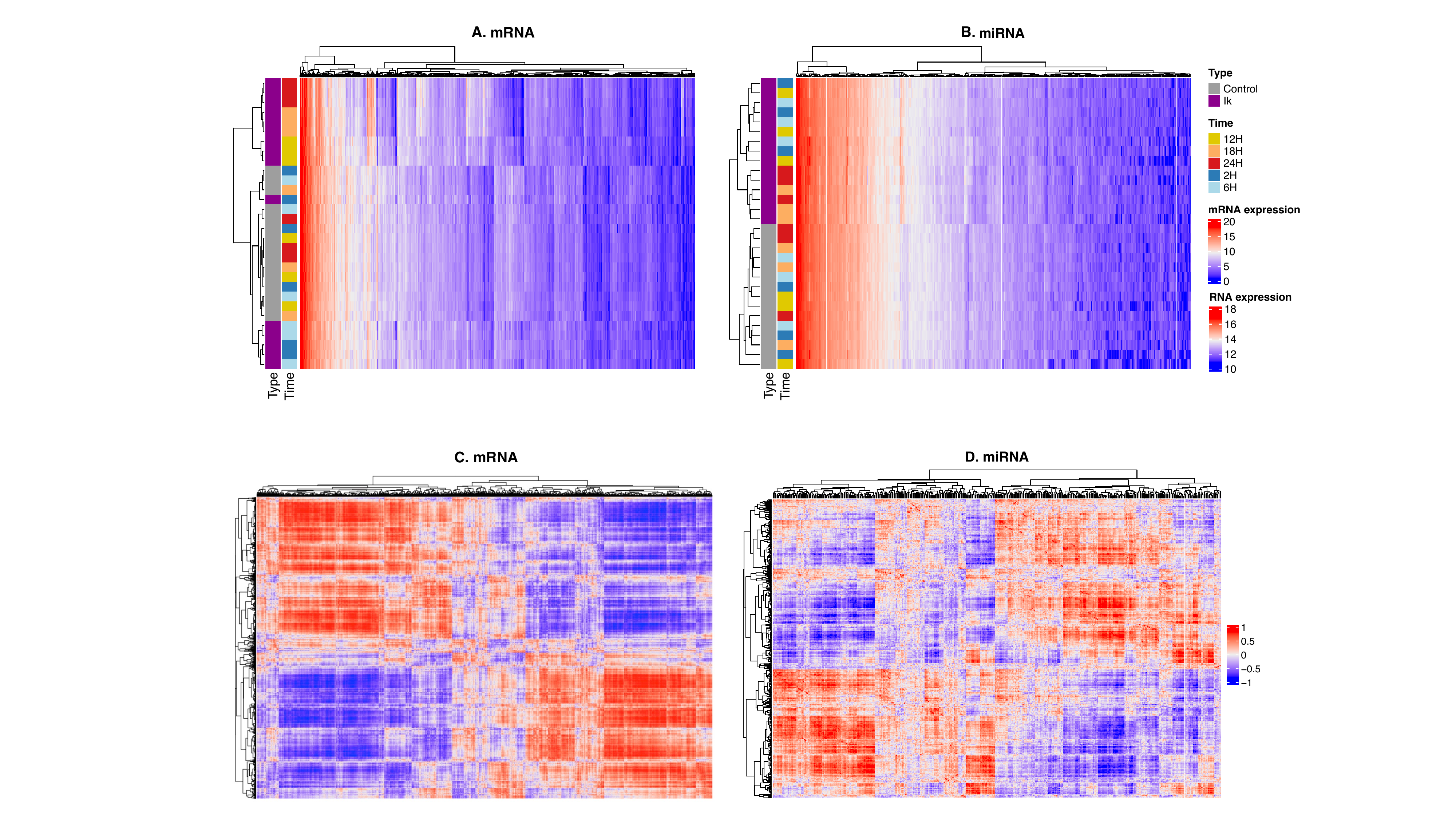}
    \caption{Visualization of the STATEGRA dataset.}
    \label{fig:heatmap_stategra}
\end{figure}

Figure \ref{fig:heatmap_stategra} shows a heatmap of the raw STATEGRA data highlighting strong correlations between groups of features.

\begin{figure}
    \centering
    \includegraphics[width=\linewidth]{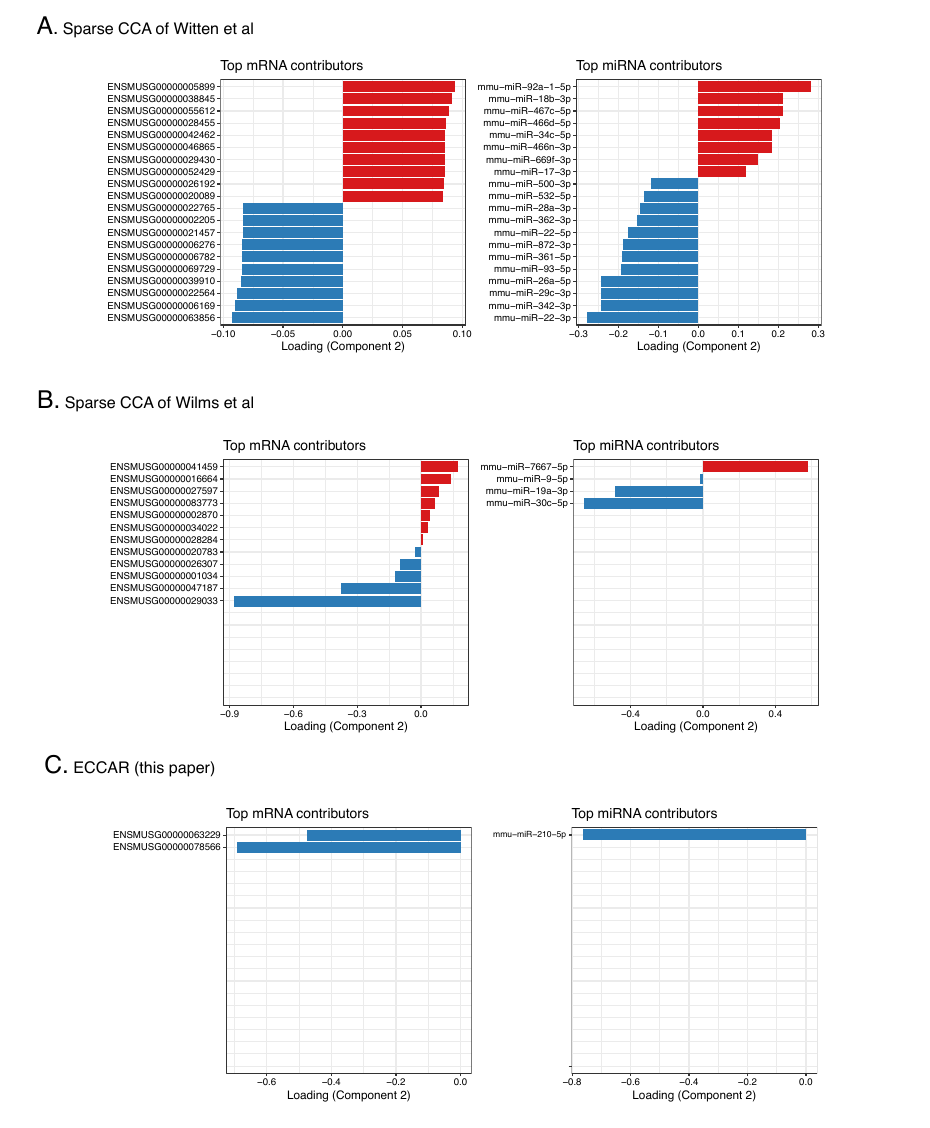}
    \caption{Visualization of the 2nd CCA directions for the different methods.}
    \label{fig:stategra_res2}
\end{figure}

\begin{figure}
    \centering
    \includegraphics[width=\linewidth]{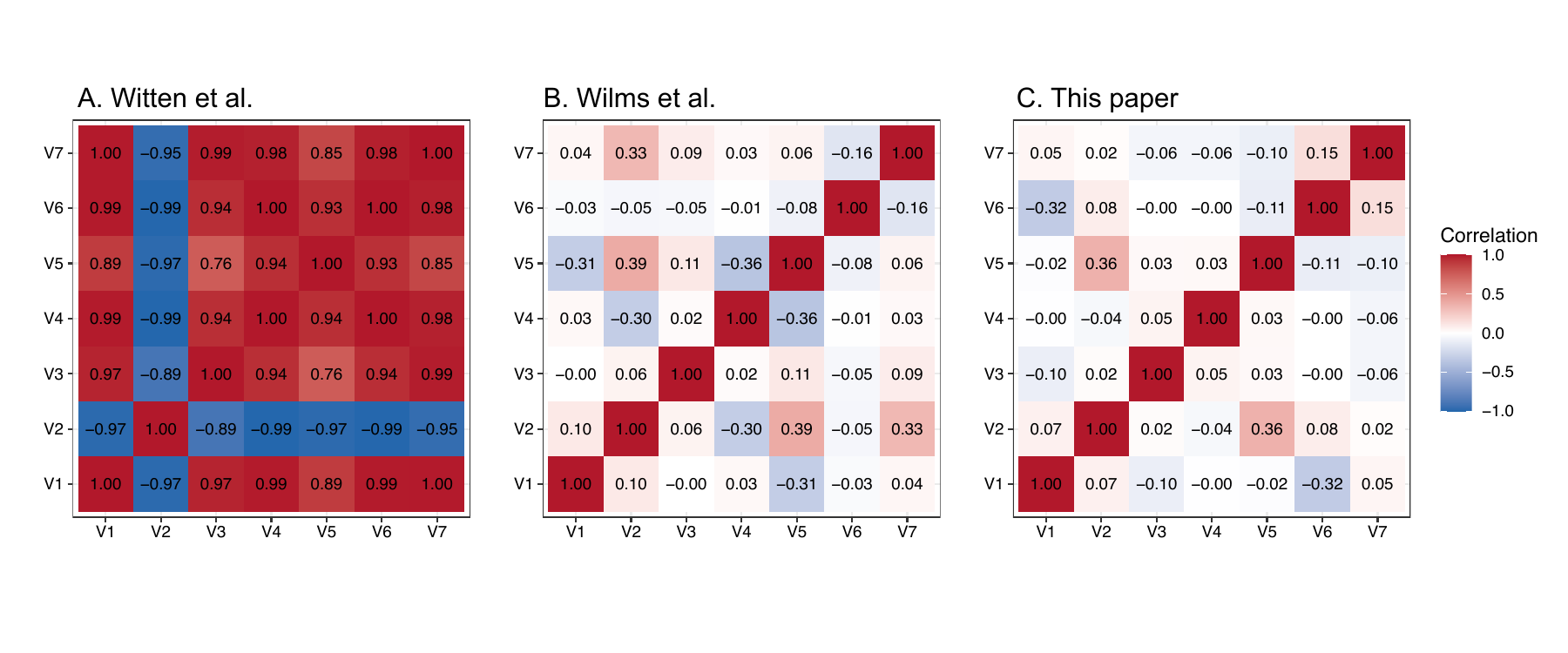}
    \caption{Visualization of the orthogonality of the estimated CCA variates $XU$ on the STATEGRA dataset.}
    \label{fig:ortho_stategra1}
\end{figure}

\begin{figure}
    \centering
    \includegraphics[width=\linewidth]{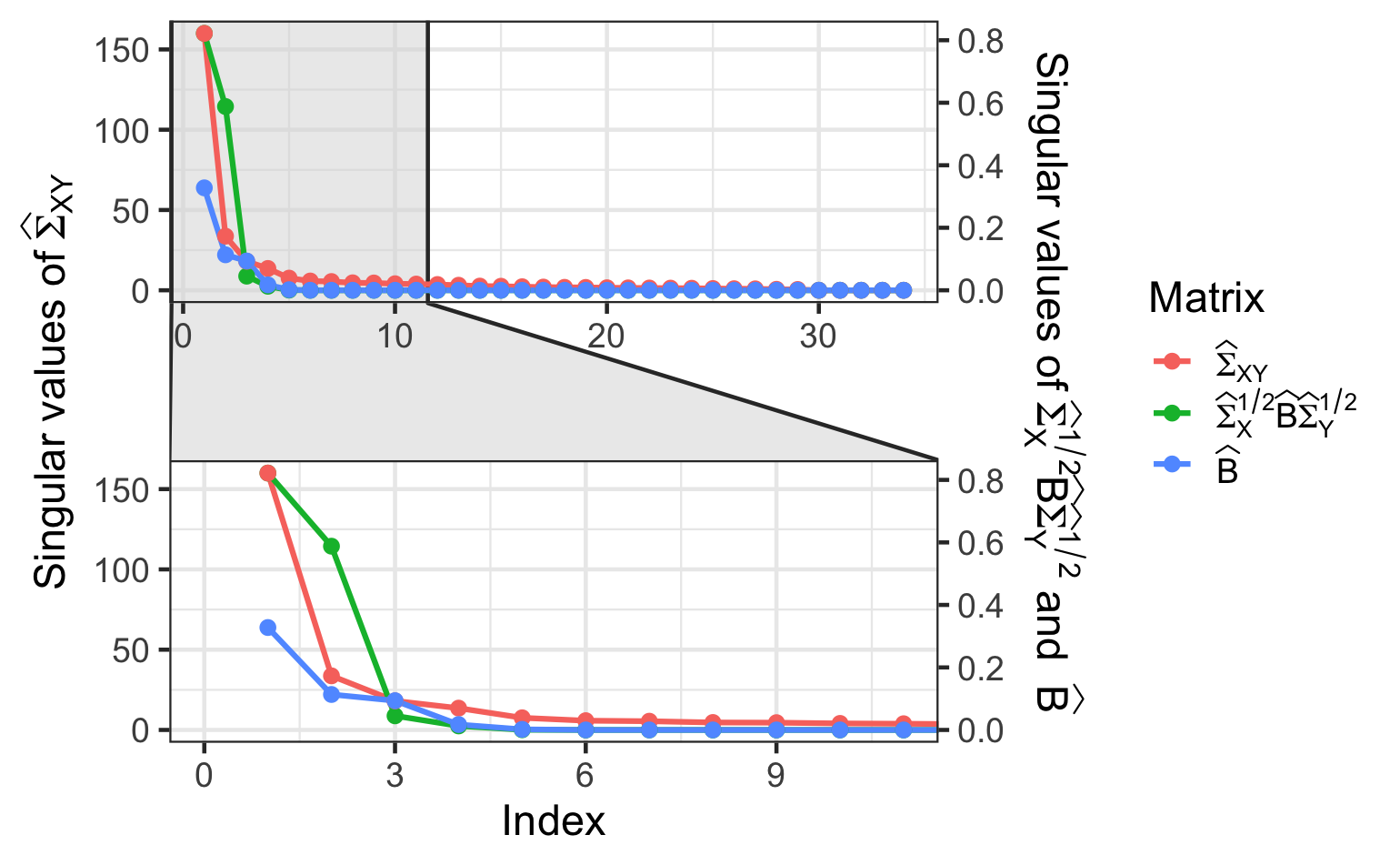}
    \caption{Visualization (in red) of the singular values of the covariance matrix $\widehat \Sigma_{XY}$, as well as those of the matrix $ \widehat \Sigma_{X}^{\frac12} \widehat B \widehat \Sigma_{Y}^{\frac12}$ (green) and $\widehat B$ (blue). Using a simple elbow rule, we choose to analyze the first two principal canonical directions, which are well separated from the remainder of the spectrum. In this case, both curves agree on the value of the elbow, but the separation is more striking for the singular values of the denoised matrix   $ \widehat \Sigma_{X}^{\frac12} \widehat B \widehat \Sigma_{Y}^{\frac12}$.}
    \label{fig:ortho_stategra2}
\end{figure}

\clearpage

\section{Technical lemmas}\label{app:technical_lemmas}
\subsection{Matrix Perturbation Bounds}
\label{subsec:perturbation}
  In this section, we provide facts about matrix perturbation that we have used in the proofs. 
  We start with the following theorem, which is a variant of Davis-Kahan theorem.

  \begin{theorem}[Theorem 2 in~\cite{yu2015useful}]
  \label{prop:Davis-Kahan}
   Let \({M} \in \R^{n \times n}\) be a rank \(K\) symmetric matrix with smallest nonzero eigenvalue \(\sigma_{K}({M})\), and let $\wh M \in \R^{n \times n}$ be any symmetric matrix. Let $\wh U(\wh M)\in \R^{n \times K}$ and $U(M)\in \R^{n \times K}$ be the matrices of \(K\) leading  eigenvectors of \(\widehat{M}\) and \({M}\), respectively. Then there exists a \(K \times K\) orthogonal matrix $O$ such that
   \begin{equation}
     \|\wh U(\wh M) - U({M}) O\|_{F} \le \frac{2 \sqrt{2} \|\widehat{M} - {M}\|_F}{\sigma_{K}({M})}.
   \end{equation}
  \end{theorem}

The following corollary and lemma are taken from \cite{klopp2021assigning} and very slightly adapted in terms of the Frobenius norm. The proof is provided to make this manuscript self contained, but it is essentially the same as the original proof of  \cite{klopp2021assigning}.
  \begin{corollary}[Adapted from Corollary 4 of \cite{klopp2021assigning}]\label{lemma:singvectorConsistency}
    Let $M$ and $X$ be matrices with singular value decompositions given by $M = U\Lambda V^\top $ and $X = \widehat{U} \widehat{\Lambda} (\widehat{V})^\top $ where $\wh U$ and $\wh V$ are the left and right singular vectors of $X$ corresponding to its $K$ leading singular values.
    Then there exists  \(K \times K\) orthogonal matrices \(O\) and \(\tilde{O}\) such that
    \begin{equation}\label{DKSingularVectors}
    \begin{split}
      \|\wh U - UO\|_{F} 
     & \leq \frac{2\sqrt{2}(\|X\|_{op} + \|M\|_{op})\|X - M\|_F}{\sigma_{K}^2(M)}\\
      \|\wh V - V\tilde{O}\|_{F} 
      &\leq \frac{2\sqrt{2}(\|X\|_{op} + \|M\|_{op})\|X - M\|_F}{\sigma_{K}^2(M)}
          \end{split}
    \end{equation}
    Furthermore, if \(\|X-M\|_{op} \le \frac{1}{2} \sigma_{K}(M)\) then 
    \begin{equation}\label{DKSingularVectors-bis}
    \max \Big\{\|\wh U - UO\|_{F}, \|\wh V-V\tilde{O}\|_{F}\Big\}
      \leq 
      \frac{5\sqrt{2} \kappa(M)\|X - M\|_F}{\sigma_{K}(M)}.
    \end{equation}
  \end{corollary}
  \begin{proof}
    Applying Theorem~\ref{prop:Davis-Kahan} to matrices $M M^\top $ and $X X^\top $ we get 
    \begin{equation}
      \|\wh U - UO\|_{F} \leq \frac{2\sqrt{2}\|M M^\top -X X^\top  \|_F)}{\sigma_{K}(M M^\top )}
      \leq 
      \frac{2\sqrt{2}(\|X\|_{op} + \|M\|_{op})\|X - M\|_F}{\sigma_{K}^2(M)}.
    \end{equation}
    The last line follows by observing that:
    \begin{equation}
        \begin{split}
     \|M M^\top -X X^\top  \|_F 
     &= \|M M^\top  - M X^\top  + M X^\top -X X^\top  \|_F    = \|M(M^\top  -X^\top ) + (M -X)X^\top  \|_F \\
     &\leq  \|M(M^\top  -X^\top )\|_F + \|(M -X)X^\top  \|_F \\
          &\leq  \|M\|_{op}\|M^\top  -X^\top \|_F + \|X^\top  \|_{op}\|M -X\|_F \\
        \end{split}
    \end{equation}
    Similarly, the inequality on $\widehat{V}$ is obtained by applying Theorem~\ref{prop:Davis-Kahan}
    to matrices $M^\top  M$ and $X^\top  X$.
    Next, if  \(\|X-M\|_{op} \le \frac{1}{2} \sigma_{K}(M)\) then due to the triangle inequality
   we have $\|X\|_{op} \le \|M\|_{op} + \frac{1}{2} \sigma_{K}(M) \le \frac{3}{2}\|M\|_{op}$. Combining this fact with~\eqref{DKSingularVectors}, we obtain~\eqref{DKSingularVectors-bis}. 
  \end{proof}
  
{
\begin{lemma}[{Remark 4.7.3 of \cite{vershynin2018high}}]
Let $X$ be a sub-Gaussian random vector in $\mathbb{R}^n$. 
More precisely, assume that there exists $K \ge 1$ such that
$$
\|\langle X, x\rangle\|_{\psi_2}
\le
K\,\|\langle X, x\rangle\|_{L^2}
\quad \text{for all } x \in \mathbb{R}^n .
$$
Let $\Sigma = \mathbb{E}[XX^\top]$ be the covariance matrix, and let
$
\Sigma_m = \frac{1}{m}\sum_{i=1}^m X_i X_i^\top
$
be the sample covariance matrix based on i.i.d.\ copies $X_1,\dots,X_m$ of $X$.
Then for every $u \ge 0$, with probability at least
$1 - 2e^{-u}$, we have
$$\|\Sigma_m - \Sigma\|
\le
C K^2
\left(
\sqrt{\frac{n+u}{m}} + \frac{n+u}{m}
\right)
\|\Sigma\|,$$
where $C>0$ is an absolute constant.
\label{lemma:covariance}
\end{lemma}

\begin{lemma}[{Lemma 16 of \cite{gao2015minimax}}]
Let $A,B$ be positive semidefinite matrices. Then for any unitarily
invariant norm $\|\cdot\|$,
$$
\|A^{1/2} - B^{1/2}\|
\le
\frac{1}{
\sigma_{\min}\!\left(A^{1/2}\right)
+
\sigma_{\min}\!\left(B^{1/2}\right)
}
\,\|A - B\|.
$$
\label{lemma:root_difference}
\end{lemma}
}
The following Lemmas \ref{lemma1}, \ref{lemma2}, and \ref{lemma3} correspond to Lemmas 6.4 and 6.5 in \citet{gao2017sparse}, and Lemma 9.7 in its supplementary material.

\begin{lemma}
    Assume $r\sqrt{\log(p + q)/n} \le c$ for some sufficiently small constant $c \in (0, 1)$
Then there exist some constants $a, a' > 0$ only depending on M and c such that $\|\wh{\Sigma}_{XY} - \wh{\Sigma}_{X} B^\star \wh{\Sigma}_{Y}\|_\infty \le a\sqrt{\log(p + q)/n}$ with probability at least $1 - (p + q)^{-a'}$. \label{lemma1}
\end{lemma}

For any positive semi-definite matrix $B$, define 
$$\phi_{max}^B(k) = \underset{\|u\|_0 \le k, \|u\|_2 = 1}{\max} u^\top  B u, \ \ \phi_{min}^B(k) = \underset{\|u\|_0 \le k, \|u\|_2 = 1}{\min} u^\top  B u$$
\begin{lemma}
Assume $\frac{1}{n} \left( (k_{u} \wedge p) \log\left(\frac{ep}{k_{u} \wedge p}\right)  + (k_{v} \wedge q) \log\left(\frac{eq}{k_{v} \wedge q}\right)  \right) \le c
$ for some sufficiently small constant \( c > 0 \). Then there exist some constants  $b, b' > 0  $ only depending on  $M$  and  $c$  such that for
\[
\delta_{u}(k_{u}) = \sqrt{\frac{1}{n} \left( (k_{u} \wedge p) \log\left(\frac{ep}{k_{u} \wedge p}\right)
\right) }\] 
and
\[
\delta_{v}(k_{v}) = \sqrt{\frac{1}{n} \left( (k_{v} \wedge q) \log\left(\frac{eq}{k_{v} \wedge q}\right) \right) }
\]
we have
\begin{align*}
M^{-1} - b\delta_{u}(k_{u}) &\le \phi_{min}^{\wh{\Sigma}_{X}}(k_u) \le \phi_{max}^{\wh{\Sigma}_{X}}(k_u) \leq M + b\delta_{u}(k_{u}), \\
M^{-1} - b\delta_{v}(k_{v}) &\le \phi_{min}^{\wh{\Sigma}_{Y}}(k_v) \le \phi_{max}^{\wh{\Sigma}_{Y}}(k_v) \leq M + b\delta_{v}(k_{v}), 
\end{align*}
with probability at least $1 - \exp \left( -b' (k_{u} \wedge p) \log\left(\frac{ep}{k_{u} \wedge p}\right) \right) - \exp \left( -b' (k_{v} \wedge q) \log\left(\frac{eq}{k_{v} \wedge q}\right) \right)  $ \label{lemma2}
\end{lemma}

\begin{lemma}
Assume $n \geq c (s_u + s_v + \log(ep/s_u) + \log(eq/s_v))) $ for some sufficiently large constant $c$. Then there exists constant $C, C'$ only depending on $c$ such that
\begin{align*}
    \|U^{\star \top}  \wh{\Sigma}_{X} U^\star - I\|_{op} \vee \| (U^{\star \top}  \wh{\Sigma}_{X} U^\star)^{1/2} - I\|_{op}  &\le C \sqrt{\frac{1}{n} (s_u + \log(ep/s_u)) } \\
        \|V^{\star \top}  \wh{\Sigma}_{Y} V^\star - I\|_{op} \vee \| (V^{\star \top}  \wh{\Sigma}_{Y} V^\star)^{1/2} - I\|_{op}  &\le C \sqrt{\frac{1}{n} (s_v + \log(eq/s_v)) }
\end{align*}
with probability at least $1 - \exp(-C'(s_u + \log(ep/s_u))) - \exp(-C'(s_v + \log(eq/s_v)))$
\label{lemma3}
\end{lemma}

\begin{lemma}
     Consider the parameter spaces $F(s_u, s_v, p, q, r; M )$ of  all covariance matrices $\Sigma$ that satisfy the condition. Assume $
         n \geq \max\{s_u , s_v\}$. Then the solution of  $$     \min_{B \in \mathbb{R}^{s_u \times s_v }} \frac{1}{2}\| \frac{1}{n}X_{S_u}BY_{S_v}^\top - I_n\|_F^2 + \rho \|B\|_1, $$ satisfies 
$$\|\tilde{\Delta}\|_F \le \frac{2(\| \wh{\Sigma}_{XY} -  \wh{\Sigma}_{X} B^\star \wh{\Sigma}_{Y}\|_\infty + \rho )}{\sigma_{\min}( (\wh{\Sigma}_{X})_{S_uS_u}) \sigma_{\min}( (\wh{\Sigma}_{Y})_{S_vS_v}) } \sqrt{s_u s_v}$$
    \label{lem:constrained}
\end{lemma}

\begin{proof}[Proof of Lemma \ref{lem:constrained}]
Following the step 1 of the proof of Theorem \ref{theorem1}, we have

$$\frac{1}{2} \|\frac{1}{n}X_{S_u} \tilde{\Delta} Y_{s_v}\|_F^2 \le \| \wh{\Sigma}_{XY} -  \wh{\Sigma}_{X}  B^\star \wh{\Sigma}_{Y}\|_\infty \|\tilde{\Delta}\|_1 + \rho   \|\tilde{\Delta}\|_1 $$
Therefore 
$$  \|\frac{1}{n} X_{S_u} \tilde{\Delta} Y_{S_v}\|_F^2 \le 2 (\| \wh{\Sigma}_{XY} -  \wh{\Sigma}_{X}  B^\star \wh{\Sigma}_{Y}\|_\infty + \rho )\|\tilde{\Delta}\|_1  \le 2 (\| \wh{\Sigma}_{XY} -  \wh{\Sigma}_{X}  B^\star\wh{\Sigma}_{Y}\|_\infty + \rho ) \sqrt{s_u s_v}\|\tilde{\Delta}\|_F $$
On the other hand,
\begin{align*}
    \|\frac{1}{n} X_{S_u} \tilde{\Delta} Y_{S_v}\|_F^2 &\geq \sigma_{\min}( (\wh{\Sigma}_{X})_{S_uS_u}) \sigma_{\min}( (\wh{\Sigma}_{Y})_{S_vS_v}) \|\tilde{\Delta}\|_F^2 ,
\end{align*}
Therefore
\begin{align*}
      \sigma_{\min}( (\wh{\Sigma}_{X})_{S_uS_u}) \sigma_{\min}( (\wh{\Sigma}_{Y})_{S_vS_v}) \|\tilde{\Delta}\|_F^2 \le  \|\frac{X_{S_u} \tilde{\Delta} Y_{s_v}}{n}\|_F^2 \le  2 (\| \wh{\Sigma}_{XY} -  \wh{\Sigma}_{X}  B^\star \wh{\Sigma}_{Y}\|_\infty + \rho ) \sqrt{s_u s_v}\|\tilde{\Delta}\|_F 
\end{align*}
This implies
$$\|\tilde{\Delta}\|_F \le \frac{2(\| \wh{\Sigma}_{XY} -  \wh{\Sigma}_{X}  B^\star \wh{\Sigma}_{Y}\|_\infty + \rho )}{\sigma_{\min}( (\wh{\Sigma}_{X})_{S_uS_u}) \sigma_{\min}( (\wh{\Sigma}_{Y})_{S_vS_v}) } \sqrt{s_u s_v}$$
\end{proof}

\section{Proofs of the consistency of ECCAR in low dimensions}\label{app:proof_consistency_low_d}

\subsection{Proof of Theorem~\ref{theorem_consistency_product}}

\begin{proof}
A derivation of the Karush–Kuhn–Tucker conditions (KKT) for minimizing $\mathcal{L}(B)$ reveals that the solution $\wh{B}$ can be expressed as a function of the sample covariance matrices $\wh \Sigma_X$, $\wh \Sigma_Y$ and $\wh \Sigma_{XY}$:
    \begin{equation}
\nabla_B         \mathcal{L}(\wh B)= 0  \iff \frac{X^\top X}{n}  \wh B \frac{Y^\top Y}{n} - \frac{X^\top Y}{n} = 0 \iff    \wh B  = \wh{\Sigma}_X^{-1} \wh \Sigma_{XY} \wh{\Sigma}_Y^{-1}.
\end{equation}
Since $X$ and $Y$ are assumed to be Gaussian and thus have finite second moments, the Law of Large Numbers implies that: 
\begin{equation}\label{eq}
        \widehat{\Sigma}_X \xrightarrow{a.s.} \Sigma_X, \qquad 
\widehat{\Sigma}_Y \xrightarrow{a.s.} \Sigma_Y \qquad \text{and } \qquad \widehat{\Sigma}_{XY}   \xrightarrow{a.s.} \Sigma_{XY}. 
\end{equation}  
Since the matrix inversion \( A \mapsto A^{-1} = \frac{1}{\det(A)} \operatorname{adj}(A) \) 
is a continuous function on the set of positive definite matrices \citep{stewart1969continuity}, due to the continuity of the determinant and adjugate functions, and since, for sufficiently large \( n \), the sample covariance matrices \( \widehat{\Sigma}_X \) and \( \widehat{\Sigma}_Y \) are almost surely positive definite, it follows that
\[
\widehat{\Sigma}_X^{-1} \overset{a.s.}{\longrightarrow} \Sigma_X^{-1}, \qquad
\widehat{\Sigma}_Y^{-1} \overset{a.s.}{\longrightarrow} \Sigma_Y^{-1}.
\]
Therefore, by the continuous mapping theorem, the matrix $\wh{\Sigma}_X^{-1} \wh \Sigma_{XY} \wh{\Sigma}_Y^{-1}$ converges entry-wise almost surely to ${\Sigma}_X^{-1}  \Sigma_{XY} {\Sigma}_Y^{-1}$ .
Under the canonical pair model (Equation \ref{reparam}), this implies that $\lim_{n \to \infty} \wh B =\lim_{n \to \infty}\wh {\Sigma}_X^{-1}  \wh \Sigma_{XY} \wh {\Sigma}_Y^{-1}=U^\star \Lambda^\star V^{\star \top} = B^\star.$ Hence, $\wh B$ is  a consistent estimator for $B^\star.$    
\end{proof}

\subsection{Proof of Theorem~\ref{theorem:consistency_low_dim}}

\begin{proof}
 By theorem~\ref{theorem_consistency_product}, we know that $\wh{B}$ is a consistent estimator of $B^\star.$ Since, additionally, the matrix square root $ A \mapsto A^{\frac12}$
is a continuous function on the set of positive definite matrices, by the continuous mapping theorem, $\wh \Sigma_X^{\frac12}\widehat{B}\wh \Sigma_Y^{\frac12}$ is a consistent estimator of $\Sigma_X^{\frac12} B^\star 
\Sigma_Y^{\frac12}.$ 
 Equation~\ref{eq:consistency_low_dim0} can be proved by application of known properties of the singular subspaces in perturbation theory (e.g. Corollary~\ref{lemma:singvectorConsistency} in Appendix~\ref{app:technical_lemmas}). A second application of the continuous mapping theorem, coupled with the fact that $\wh \Sigma_X^{-\frac12}$ and $\wh \Sigma_Y^{-\frac12}$ converge almost surely to $ \Sigma_X^{-\frac12}$ and $ \Sigma_Y^{-\frac12}$ (see proof of Theorem~\ref{theorem_consistency_product}) establishes the consistency result of Equation~\ref{eq:consistency_low_dim}.
\end{proof}

\section{Proofs of the consistency of ECCAR in high dimensions}
\subsection{The sparse Setting}
\subsubsection{Theorem \ref{theorem1}}

\textit{
     Consider the parameter space $F(s_u, s_v, p, q, r; M )$ for the covariance matrix $\Sigma$ (Equation~\ref{reparam} and conditions here therein), and let $\Delta = \widehat{B}-B^\star$, where $\widehat{B}$ is the estimate obtained in step 1 of Algorithm~\ref{alg:eccar}.
     Assume \(
         n \geq c s_u s_v \log(p + q) \)  for some sufficiently large constant~$c$.          
         There exist constants $a, b, C$ depending on $M$ and $c$ such that  if ${\rho \geq a\sqrt{\log(p + q)/n}} $,  then with probability at least 
         $1 - \exp(-b s_u\log(ep/s_u)) - \exp(- b s_v \log(eq/s_v )) - (p + q)^{-b}$, we have: 
      \begin{align*}
             \|\Delta\|_F &\le  C  \rho \sqrt{s_u s_v}.  
     \end{align*}
     In particular, if $\rho$ is of order $\sqrt{\log(p + q)/n} $, we have:
     \begin{align*}
             \|\Delta\|_F &\lesssim   \sqrt{\frac{s_us_v \log (p + q)}{n}} ,
     \end{align*}
     and $\wh{B}$ has sparse entries:
     $$\|\wh{B}\|_0 \lesssim s_us_v.$$
}

\begin{proof}[Proof of Theorem \ref{theorem1}] We first prove the bound on $\|\Delta\|_F$. The proof consists here of three main steps.\\
 \paragraph{Step 1. Showing that $\|\Delta_{(S_uS_v)^c}\|_1 \le 3\| \Delta_{S_u S_v}\|_1$:}   
 By the basic inequality, we have: 
\begin{align*}
\frac{1}{2}\|\frac{1}{n}X\wh{B}Y^\top  - I_n\|_F^2 + \rho \|\wh{B}\|_1 &\le \frac{1}{2}\|\frac{1}{n} X B^\star Y^\top  - I_n \|_F^2 + \rho \|B^\star\|_1
    \\ \frac{1}{2} \|\frac1nX\wh{B}Y^\top \|_F^2 -  \frac1n\tr(X\wh{B}Y^\top ) &\le \frac{1}{2}\|\frac1nX {B}^\star Y^\top \|_F^2 - \frac1n \tr(X {B}^\star Y^\top )  \\
    &\hspace{2cm} + \rho \|B^\star\|_1
    - \rho \|\wh{B}\|_1 
    \\ \frac{1}{2}( \|\frac1nX(B^\star+ \Delta)Y^\top \|_F^2 - \|\frac1nX B^\star Y^\top \|_F^2) &\le  \tr(\Delta^\top  \wh{\Sigma}_{XY}) + \rho \|B^\star\|_1 - \rho \|\wh{B}\|_1 
    \\  \frac{1}{2}\|\frac1nX \Delta Y^\top  \|_F^2 +  \frac{1}{n^2}\tr(YB^{\star\top} X^\top X\Delta Y^\top ) &\le  \tr(\Delta^\top  \wh{\Sigma}_{XY}) + \rho \|B^\star\|_1 - \rho \|\wh{B}\|_1 
    \\  \frac{1}{2}\|X \Delta Y^\top  / n\|_F^2 +  \tr(\wh{\Sigma}_{Y} {B}^{\star \top}  \wh{\Sigma}_{X} \Delta ) &\le  \tr(\Delta^\top  \wh{\Sigma}_{XY}) + \rho \|B^\star\|_1 - \rho \|\wh{B}\|_1 
    \\ \frac{1}{2} \|X \Delta Y^\top  / n\|_F^2 &\le  \langle  \wh{\Sigma}_{XY} -  \wh{\Sigma}_{X} B^\star \wh{\Sigma}_{Y}, \Delta  \rangle  + \rho \|B^\star\|_1 - \rho \|\wh{B}\|_1 
        \\  \frac{1}{2}\|X \Delta Y^\top  / n\|_F^2 &\le  \langle  \wh{\Sigma}_{XY} -  \wh{\Sigma}_{X} B^\star \wh{\Sigma}_{Y}, \Delta  \rangle  + \rho \|B^\star_{S_uS_v}\|_1 \\&\hspace{2cm} - \rho \|\wh{B}_{S_uS_v}\|_1  - \rho \|\wh{B}_{(S_uS_v)^c}\|_1
        \\  \frac{1}{2}\|X \Delta Y^\top  / n\|_F^2 &\le  \langle  \wh{\Sigma}_{XY} -  \wh{\Sigma}_{X}  B^\star  \wh{\Sigma}_{Y}, \Delta  \rangle \\
        &\hspace{2cm}  + \rho \| \Delta_{S_u S_v}\|_1  - \rho \|\Delta_{(S_uS_v)^c}\|_1 
        \\ \frac{1}{2} \|X \Delta Y^\top  / n\|_F^2 &\le  \| \wh{\Sigma}_{XY} -  \wh{\Sigma}_{X}  B^\star  \wh{\Sigma}_{Y}\|_\infty \|\Delta\|_1 \\
        &\hspace{2cm} + \rho \| \Delta_{S_u S_v}\|_1  - \rho \|\Delta_{(S_uS_v)^c}\|_1.
\end{align*} 
By our choice of $\rho$ and Lemma  \ref{lemma1} we have
\begin{align*}
    0 &\le  \frac{\rho}{2} \|\Delta\|_1  + \rho \| \Delta_{S_u S_v}\|_1  - \rho \|\Delta_{(S_uS_v)^c}\|_1 \\
        0 &\le  \frac{3}{2}  \| \Delta_{S_u S_v}\|_1  - \frac{1}{2}  \|\Delta_{(S_uS_v)^c}\|_1. \end{align*}
Therefore we have $$
    \|\Delta_{(S_uS_v)^c}\|_1 \le 3\| \Delta_{S_u S_v}\|_1.  $$
 \paragraph{Step 2. Showing that $ \|X \Delta Y^\top  / n\|_F^2  \lesssim   \rho \sqrt{s_u s_v} \|\Delta_{S_u S_v} \|_F$:}  This is directly implied by the conclusion of \textbf{Step 1}:
\begin{align*}
   \frac{1}{2} \|X \Delta Y^\top  / n\|_F^2 &\le   \frac{\rho}{2} \|\Delta\|_1  + \rho \| \Delta_{S_u S_v}\|_1 \\
    &= \frac{\rho}{2} (\| \Delta_{S_u S_v}\|_1  +  \|\Delta_{(S_uS_v)^c}\|_1 ) + \rho \| \Delta_{S_u S_v}\|_1 \\
    &\le \frac{\rho}{2} (\| \Delta_{S_u S_v}\|_1  +  3 \|\Delta_{S_uS_v}\|_1 ) + \rho \| \Delta_{S_u S_v}\|_1\\
    &= 3 \rho \| \Delta_{S_u S_v}\|_1 \\
    &\le 3 \rho \sqrt{s_u s_v} \|\Delta_{S_u S_v} \|_F.
\end{align*}
 \paragraph{Step 3. Showing that $ \|X \Delta Y^\top  / n\|_F  \gtrsim   \|\Delta_{S_u S_v} \|_F$: } Now let us look at $\|X \Delta Y^\top  / n\|_F^2 =tr(\Delta^\top  \wh{\Sigma}_{X} \Delta \wh{\Sigma}_{Y}) = \|\wh{\Sigma}_{X}^{1/2} \Delta \wh{\Sigma}_{Y}^{1/2}\|_F^2$. Let the index set $J_1 = \{(i_k, j_k)\}_{k = 1}^\top $ in $(S_u \times S_v)^c$
correspond to the entries with the largest absolute values in $\Delta$. and we define the set $\tilde{J} = (S_u \times S_v) \cup J_1$. Now we partition $\tilde{J}^c$
into disjoint subsets $J_2, ..., J_K$ of size $t$ (with $|J_K| \le t$),
such that $J_k$ is the set of (double) indices corresponding to the entries of $t$ largest absolute
values in $\Delta$ outside $\tilde{J} \cup \bigcup_{j = 2}^{k - 1}J_j$ . By  the triangle inequality,
\begin{align*}
    \|\wh{\Sigma}_{X}^{1/2} \Delta \wh{\Sigma}_{Y}^{1/2}\|_F
    & \geq \|\wh{\Sigma}_{X}^{1/2} \Delta_{\tilde{J}} \wh{\Sigma}_{Y}^{1/2}\|_F - \sum_{k = 2}^K \|\wh{\Sigma}_{X}^{1/2} \Delta_{J_i} \wh{\Sigma}_{Y}^{1/2}\|_F \\
    & \geq \sqrt{\phi_{min}^{\wh{\Sigma}_{X}} (s_u + t) \phi_{min}^{\wh{\Sigma}_{Y}} (s_v + t) } \|\Delta_{\tilde{J}} \|_F - \sqrt{\phi_{max}^{\wh{\Sigma}_{X}} ( t) \phi_{max}^{\wh{\Sigma}_{Y}} ( t) } \sum_{k = 2}^K \|\Delta_{J_k} \|_F. 
\end{align*}
Also we have:
\begin{align*}
     \sum_{k = 2}^K \|\Delta_{J_k} \|_F &\le \sqrt{t}  \sum_{k = 2}^K \|\Delta_{J_k} \|_\infty
     \le \frac{1}{\sqrt{t}}\sum_{k = 1}^{K - 1} \|\Delta_{J_k} \|_1 \\
     &\le \frac{1}{\sqrt{t}} \|\Delta_{(S_uS_v)^c}\|_1 \\
     &\le \frac{3}{\sqrt{t}}\|\Delta_{S_uS_v}\|_1 \\
     &\le \frac{3 \sqrt{s_us_v}}{\sqrt{t}}\|\Delta_{\tilde{J}}\|_F.
\end{align*}
Therefore 
$$ \|\wh{\Sigma}_{X}^{1/2} \Delta \wh{\Sigma}_{Y}^{1/2}\|_F \geq  \left( \sqrt{\phi_{min}^{\wh{\Sigma}_{X}} (s_u + t) \phi_{min}^{\wh{\Sigma}_{Y}} (s_v + t) }  - \frac{3 \sqrt{s_us_v}}{\sqrt{t}}\sqrt{\phi_{max}^{\wh{\Sigma}_{X}} ( t) \phi_{max}^{\wh{\Sigma}_{Y}} ( t) } \right) \|\Delta_{\tilde{J}} \|_F. $$
Take $t = c_1s_u s_v$ for some sufficiently large constant $c_1 > 1$. Then with high probability, 
$$ \left( \sqrt{\phi_{min}^{\wh{\Sigma}_{X}} (s_u + t) \phi_{min}^{\wh{\Sigma}_{Y}} (s_v + t) }  - \frac{3 \sqrt{s_us_v}}{\sqrt{t}}\sqrt{\phi_{max}^{\wh{\Sigma}_{X}} ( t) \phi_{max}^{\wh{\Sigma}_{Y}} ( t) } \right) \geq \kappa,$$
where $\kappa$ is some positive constant that depends only on $M$. To see this, note by Lemma \ref{lemma2}, $\delta_u(s_u + t)$ and $\delta_v(s_v + t)$ are bounded by $\sqrt{\frac{2c_1 s_us_v  \log(p)}{n}}$ and $\sqrt{\frac{2c_1 s_u s_v \log(q)}{n}}$, which are small by assumption. The same arguments hold for $\delta_u(t)$ and $\delta_v(t)$. Therefore taking $c_1$ to be large leads to the lower bound $\kappa$.

Combining this result with the upper bound on the right-hand side, we have 
$$ \kappa^2 \|\Delta_{\tilde{J}}\|_F^2 \le \|\wh{\Sigma}_{X}^{1/2} \Delta \wh{\Sigma}_{Y}^{1/2}\|_F^2 \le 3 \rho \sqrt{s_u s_v} \|\Delta_{S_u S_v} \|_F \le 3 \rho \sqrt{s_u s_v} \|\Delta_{\tilde{J}} \|_F.$$
Therefore $$\|\Delta_{\tilde{J}}\|_F \le \frac{3 \rho \sqrt{s_u s_v}}{\kappa^2}. $$
On the other hand, $$ \|\Delta_{\tilde{J}^c}\|_F \le  \sum_{k = 2}^K \|\Delta_{J_k} \|_F \le  \frac{3 \sqrt{s_us_v}}{\sqrt{t}}\|\Delta_{\tilde{J}}\|_F =   \frac{3 \sqrt{s_us_v}}{\sqrt{c_1 s_u s_v }}\|\Delta_{\tilde{J}}\|_F .$$
Combining the results from steps 1 through 3, the bound on $\|\Delta\|_F$ in Equation~\ref{eq:bound_D} follows.

Then we prove the support of $\wh{B}$ is sparse. By the KKT conditions, we have
$$\left(\wh{\Sigma}_{X} \wh{B} \wh{\Sigma}_{Y} - \wh{\Sigma}_{XY} \right)_{ij} = -\rho \  \frac{\partial |B_{ij}|}{\partial B}.$$
Therefore
$$\left| \left(\wh{\Sigma}_{X} \Delta \wh{\Sigma}_{Y}\right)_{ij} \right| \geq \rho - \|\wh{\Sigma}_{X} B^\star\wh{\Sigma}_{Y} - \wh{\Sigma}_{XY} \|_\infty \geq \frac{\rho}{2} $$
if $\wh{B}_{ij} \not=0$, where the last inequality follows from Lemma \ref{lemma1}. 

Then we prove by contradiction. Assume that $J \subset \mathrm{supp}_e(\wh B)$ such that $|J| =  \lceil k s_u s_v \rceil $ where $k$ is a constant. Let $J_X = \{x : (x, y) \in J\}$, $J_Y = \{y: (x, y) \in J \}$ and $T = |J|$. Then
\begin{align*}
    \frac{\rho^2}{4} T
    &\le \sum_{(i, j) \in J}   \left(\wh{\Sigma}_{X} \Delta \wh{\Sigma}_{Y}\right)_{ij} ^2 \\
    &\le \| \left(\wh{\Sigma}_{X} \Delta \wh{\Sigma}_{Y}\right)_J \|_F^2  \\
        &= \| \frac{1}{n^2} \left(X^T X \Delta Y^\top Y \right)_J \|_F^2  \\
               &\le \| \frac{1}{n^2} X_{J_X}^\top \left( X \Delta Y^T \right) Y_{J_Y} \|_F^2  \\
                       &= \mathrm{Tr}\left(  \frac{1}{n} X_{J_X} X_{J_X}^\top \left( \frac{X \Delta Y^T}{n} \right) \frac{1}{n}Y_{J_Y} Y_{J_Y}^\top \left( \frac{X \Delta Y^T}{n} \right)^\top   \right) \\
               &\lesssim \left(M + \sqrt{\frac{T\log(p + q)}{n}} \right)^2  \|\frac{1}{n}(X \Delta Y^\top)\|_F^2 ,\end{align*}
               where the last step is implied by Lemma~\ref{lemma2}. Moreover, in Step 2 of the proof, we showed $$ \|\frac{1}{n}(X \Delta Y^\top)\|_F^2 \le 6 \rho \|\Delta_{S_uS_v}\|_1.$$ Also,
$$  \|\Delta_{S_uS_v}\|_1 \le \sqrt{s_us_v} \|\Delta \|_F $$
Therefore we know $$T \lesssim \left(M + \sqrt{\frac{T\log(p + q)}{n}} \right)^2  \frac{\|\Delta\|_F}{\rho} \sqrt{s_us_v} \lesssim \left(M^2 +  {\frac{T\log(p + q)}{n}} \right) s_us_v,$$
Recall that $T  = \lceil k s_us_v\rceil $. This shows a contradiction when the inequality cannot be satisfied (when $k$ is sufficiently large).

\end{proof}

\subsubsection{Theorem \ref{thm:support}}

\textit{Consider the parameter space $\mathcal{F}(s_u, s_v, p, q, r; M )$ for the covariance matrix $\Sigma$ (Equation~\ref{reparam} and conditions here therein). Assume 
$n > \max\{s_u, s_v\}$   and 
     \begin{equation}
       { \frac{\| \wh{\Sigma}_{XY} -  \wh{\Sigma}_X B^\star\wh{\Sigma}_Y\|_\infty}{\rho} +  \frac{2(\| \wh{\Sigma}_{XY} -  \wh{\Sigma}_{X} B^\star\wh{\Sigma}_{Y}\|_\infty + \rho )}{\rho}\frac{\wh{\tau}(X, Y) \sqrt{s_u s_v} }{\sigma_{\min}( (\wh{\Sigma}_{X})_{S_uS_u}) \sigma_{\min}( (\wh{\Sigma}_{Y})_{S_vS_v}) }  < 1},
     \end{equation}
     where 
     ${S_u = \operatorname{supp}(U^\star) \mbox{ and } S_v = \operatorname{supp}(V^\star)}$ and
     \begin{align*}\wh{\tau}(X, Y ) = \max\Big\{ &\big\|(\wh{\Sigma}_X)_{S_u^cS_u}    \big\|_{2, \infty} \big \|(\wh{\Sigma}_Y)_{S_v^cS_v}    \big\|_{2, \infty} ,\\
     &\big\|(\wh{\Sigma}_X)_{S_u^cS_u}    \big\|_{2, \infty}\big\| (\wh{\Sigma}_Y)_{S_vS_v}\big\|_{op}, \\
     &\big\|(\wh{\Sigma}_X)_{S_uS_u} \big\|_{op} \big\|(\wh{\Sigma}_Y)_{S_v^cS_v}   \big\|_{2, \infty}\Big\}.
     \end{align*}  Then the minimizer of  \ref{eqn:lasso} is unique and satisfies  
      \begin{align*}
            \text{supp}_e(\wh{B}) \subset S_u \times S_v.  
     \end{align*}
}

\begin{proof}[Proof of Theorem \ref{thm:support}]
 We first prove the existence of solution with support $S_u \times S_v$.   Using strong duality, we
can write the objective in a equivalent min–max form:
\begin{align*}
    \min_{B \in \mathbb{R}^{p \times q }} \frac{1}{2}\|\frac{1}{n}XBY^\top - I_n\|_F^2 + \rho \|B \|_1 
    &\Longleftrightarrow  \min_{B \in \mathbb{R}^{p \times q }} \max_{Z \in \mathbb{B}_{p, q }} \frac{1}{2}\|\frac{1}{n}XBY^\top - I_n\|_F^2 + \rho \langle B, Z \rangle 
    \\& \Longleftrightarrow   \max_{Z \in \mathbb{B}_{p, q }} \min_{B \in \mathbb{R}^{p \times q }} \frac{1}{2}\|\frac{1}{n}XBY^\top - I_n\|_F^2 + \rho \langle B, Z \rangle
\end{align*}
where $$\mathbb{B}_{p, q } := \{Z \in \mathbb{R}^{p \times q}, \|Z\|_{\infty} \le 1 \}.$$
Then from the KKT condition, a pair $(\wh{B}, \wh{Z})$ is an optimal solution if and only if
\begin{align}
        Z_{ij} &= \text{sign} (\wh{B}_{ij}) \ \ \ \wh{B}_{ij} \not= 0  \label{kkt1}\\ 
    Z_{ij} &\in [-1, 1] \ \ \ \wh{B}_{ij} = 0  \label{kkt2} \\
        \widehat{B}&= \argmin_{B \in \R^{p \times q}}  \   \frac{1}{2}\|\frac{1}{n}XBY^\top - I_n\|_F^2 + \rho \langle B, Z \rangle .  \label{kkt3}
\end{align}
Let $S = S_u \times S_v$ be the support of the true solution. Consider the constrained solution $\tilde{B}$ that solves the following problem
\begin{equation}
     \min_{B \in \mathbb{R}^{p \times q }, \mathrm{supp}_e(B) \subset S} \frac{1}{2}\|\frac{1}{n}XBY^\top - I_n\|_F^2 + \rho \|B \|_1.  
\end{equation}
One can easily show that this is equivalent to first solving
$$     \min_{B \in \mathbb{R}^{s_u \times s_v }} \frac{1}{2}\|\frac{1}{n} X_{S_u}BY_{Y_v}^\top - I_n\|_F^2 + \rho \|B \|_1, $$ then setting entries to be zero.  By Lemma \ref{lem:constrained}, we have $$\|\tilde{B} - B^\star\|_F = \|\tilde{\Delta}\|_F \le \frac{2(\| \wh{\Sigma}_{XY} -  \wh{\Sigma}_{X} B^\star\wh{\Sigma}_{Y}\|_\infty + \rho )}{\sigma_{\min}( (\wh{\Sigma}_{X})_{S_uS_u}) \sigma_{\min}( (\wh{\Sigma}_{Y})_{S_vS_v}) } \sqrt{s_u s_v}.$$ 

Define a modified primal dual pair $(\wh{B}, \wh{Z})$ as follows
\begin{equation}
    \begin{split}
        \wh{B} &= \tilde{B} \\
        \wh{Z}_{S_uS_v} &= \tilde{Z}_{S_uS_v} \\
        \wh{Z}_{S_uS_v^c} &= \frac{1}{\rho} \left( (\wh{\Sigma}_{XY})_{S_u S_v^c} -  (\wh{\Sigma}_X)_{S_u S_u} \tilde{B}_{S_uS_v} (\wh{\Sigma}_Y)_{S_v S_v^c} \right) \\
        \wh{Z}_{S_u^cS_v} &=  \frac{1}{\rho} \left( (\wh{\Sigma}_{XY})_{S_u^c S_v} -  (\wh{\Sigma}_X)_{S_u^c S_u} \tilde{B}_{S_uS_v} (\wh{\Sigma}_Y)_{S_v S_v} \right) \\ 
        \wh{Z}_{S_u^cS_v^c} &= \frac{1}{\rho }\left( (\wh{\Sigma}_{XY})_{S_u^c S_v^c} -  (\wh{\Sigma}_X)_{S_u^c S_u} \tilde{B}_{S_uS_v} (\wh{\Sigma}_Y)_{S_v S_v^c} \right) . .\\ 
    \end{split}
\end{equation}
We shall show that both $\wh{B} \text{ and } \wh{Z}$ satisfy the KKT conditions \ref{kkt1}, \ref{kkt2}, \ref{kkt3}. Note that conditions \ref{kkt1} and \ref{kkt3} are automatically satisfied by the construction. So it suffices to verify condition \ref{kkt2}.

Note that for $(i, j) \in S_u^c \times S_v$, we have:
\begin{align*}
  \rho  |\wh{Z}_{ij}|&= | (\wh{\Sigma}_{XY})_{ij} -  (\wh{\Sigma}_X)_{i, S_u} \tilde{B}_{S_uS_v} (\wh{\Sigma}_Y)_{S_v, j} |
   \\&= | (\wh{\Sigma}_{XY})_{ij} -  (\wh{\Sigma}_X)_{i, S_u} B^\star_{S_uS_v} (\wh{\Sigma}_Y)_{S_v, j} +  (\wh{\Sigma}_X)_{i, S_u} B^\star_{S_uS_v} (\wh{\Sigma}_Y)_{S_v, j} \\
   &\hspace{1cm}-  (\wh{\Sigma}_X)_{i, S_u} \tilde{B}_{S_uS_v} (\wh{\Sigma}_Y)_{S_v, j} | 
   \\&\le \| \wh{\Sigma}_{XY} -  \wh{\Sigma}_X B^\star\wh{\Sigma}_Y\|_\infty + | (\wh{\Sigma}_X)_{i, S_u} \tilde{\Delta}_{S_uS_v} (\wh{\Sigma}_Y)_{S_v, j} | \\
   &= \| \wh{\Sigma}_{XY} -  \wh{\Sigma}_X B^\star\wh{\Sigma}_Y\|_\infty + | (\wh{\Sigma}_X)_{i, S_u} \tilde{\Delta}_{S_uS_v} (\wh{\Sigma}_Y)_{S_v, j} | 
   \\& \le \| \wh{\Sigma}_{XY} -  \wh{\Sigma}_X B^\star\wh{\Sigma}_Y\|_\infty + \| (\wh{\Sigma}_X)_{i, S_u} \|_F \|\tilde{\Delta}_{S_uS_v} \|_F \|(\wh{\Sigma}_Y)_{S_v, j} \|_{op}
      \\& \le \| \wh{\Sigma}_{XY} -  \wh{\Sigma}_X B^\star\wh{\Sigma}_Y\|_\infty + \| (\wh{\Sigma}_X)_{i, S_u}\|_F  \|(\wh{\Sigma}_Y)_{S_v, j} \|_{op} \|\tilde{\Delta}_{S_uS_v} \|_F  \\
      &\le \| \wh{\Sigma}_{XY} -  \wh{\Sigma}_X B^\star\wh{\Sigma}_Y\|_\infty +  \|(\wh{\Sigma}_X)_{S_u^cS_u}    \|_{2, \infty}\|(\wh{\Sigma}_Y)_{S_vS_v}\|_{op} \|\tilde{\Delta}_{S_uS_v} \|_F. \\ 
\end{align*}
Therefore
$$|\wh{Z}_{ij}| \le \frac{\| \wh{\Sigma}_{XY} -  \wh{\Sigma}_X B^\star\wh{\Sigma}_Y\|_\infty}{\rho} +  \|(\wh{\Sigma}_X)_{S_u^cS_u}    \|_{2, \infty}\|(\wh{\Sigma}_Y)_{S_vS_v}\|_{op} \frac{\|\tilde{\Delta}_{S_uS_v} \|_F}{\rho } .$$
Similarly, for $\wh{Z}_{S_uS_v^c}$, we have
$$|\wh{Z}_{ij}| \le \frac{\| \wh{\Sigma}_{XY} -  \wh{\Sigma}_X B^\star\wh{\Sigma}_Y\|_\infty}{\rho} +  \|(\wh{\Sigma}_Y)_{S_v^cS_v}    \|_{2, \infty}\|(\wh{\Sigma}_X)_{S_uS_u}\|_{op} \frac{\|\tilde{\Delta}_{S_uS_v} \|_F}{\rho }. $$
And, for $\wh{Z}_{S_u^cS_v^c}$, we have
\begin{align*}
   |\wh{Z}_{ij}|   &\le  \frac{\| \wh{\Sigma}_{XY} -  \wh{\Sigma}_X B^\star\wh{\Sigma}_Y\|_\infty}{\rho} +  \|(\wh{\Sigma}_Y)_{S_v^cS_v}    \|_{2, \infty}\|(\wh{\Sigma}_X)_{S_u^cS_u}    \|_{2, \infty}\frac{\|\tilde{\Delta}_{S_uS_v} \|_F}{\rho }.
\end{align*}
Define \begin{align*}\wh{\tau}(X, Y ) = \max\{ &\|(\wh{\Sigma}_X)_{S_u^cS_u}    \|_{2, \infty}  \|(\wh{\Sigma}_Y)_{S_v^cS_v}    \|_{2, \infty} , \\
&\|(\wh{\Sigma}_X)_{S_u^cS_u}    \|_{2, \infty}\| (\wh{\Sigma}_Y)_{S_vS_v}\|_{op},\\
&\|(\wh{\Sigma}_X)_{S_uS_u} \|_{op} \|(\wh{\Sigma}_Y)_{S_v^cS_v}    \|_{2, \infty}\} .\end{align*}
Then by Lemma \ref{lem:constrained}, the KKT condition is satisfied when
$$ \frac{\| \wh{\Sigma}_{XY} -  \wh{\Sigma}_X B^\star\wh{\Sigma}_Y\|_\infty}{\rho} +  \frac{2(\| \wh{\Sigma}_{XY} -  \wh{\Sigma}_{X} B^\star\wh{\Sigma}_{Y}\|_\infty + \rho )}{\rho}\frac{\wh{\tau}(X, Y)\sqrt{s_u s_v} }{\sigma_{\min}( (\wh{\Sigma}_{X})_{S_uS_u}) \sigma_{\min}( (\wh{\Sigma}_{Y})_{S_vS_v}) }  \le 1.$$\\

We then prove the uniqueness of the solution $(\wh{B}, \wh{Z})$. Let $(\wh{B}', \wh{Z}')$ be another pair of solutions. Let $F(B) = \frac{1}{2}\|XB Y^\top /n - I_n\|_F^2$. Then we must have:
\begin{align*}
    F(\wh{B}) + \rho \langle \wh{B}, \wh{Z}\rangle &=  F(\wh{B}') + \rho \langle \wh{B}', \wh{Z}'\rangle\\
        F(\wh{B}) + \rho \langle \wh{B} - \wh{B}', \wh{Z}\rangle &=  F(\wh{B}') + \rho \langle \wh{B}', \wh{Z}' - \wh{Z}\rangle.
\end{align*}
By the KKT conditions, we know $\rho \wh{Z} = - \nabla F(\wh{B})$. Therefore 
$$        F(\wh{B}) - F(\wh{B}')  - \langle \wh{B} - \wh{B}', \nabla F(\wh{B})\rangle =  \rho \langle \wh{B}', \wh{Z}' - \wh{Z}\rangle  = \rho \|\wh{B}'\|_1 -  \rho \langle \wh{B}', \wh{Z}\rangle .$$
By convexity of $F$, the left-hand side is non-positive. This implies $\|\wh{B}'\|_1 \le   \langle \wh{B}', \wh{Z}\rangle$. But we also have $ \langle \wh{B}', \wh{Z}\rangle \le \|\wh{B}'\|_1 $ by the min-max formulation. This suggests $\|\wh{B}'\|_1 =   \langle \wh{B}', \wh{Z}\rangle$. On the other hand, this inequality only occurs when $\wh{B}'_{i, j} = 0$ if $(i, j) \notin S_u \times S_v$. This implies $\wh{B}'$ also has support inside $S_u \times S_v$. The proof is finished by the strict convexity of the objective inside $S_u \times S_v$.
\end{proof}

\subsubsection{Corollary \ref{corollary}}
\textit{Consider the parameter space $\mathcal{F}(s_u, s_v, p, q, r; M )$ for the covariance matrix $\Sigma$ (Equation~\ref{reparam} and conditions therein), and assume
       \begin{equation}
        \frac{2 {\tau}(X, Y)\sqrt{s_u s_v} }{\sigma_{\min}\big( ({\Sigma}_{X})_{S_uS_u}\big) \sigma_{\min}\big(({\Sigma}_{Y})_{S_vS_v}\big) }  < 1 - \alpha,
     \end{equation}
     for some $\alpha \in (0, 1]$,
     where \begin{align*}
     {\tau}(X, Y ) = \max \Big\{ &\big\|({\Sigma}_X)_{S_u^cS_u} \big \|_{2, \infty}  \big\|({\Sigma}_Y)_{S_v^cS_v}    \big\|_{2, \infty} , \\
     &\big\|({\Sigma}_X)_{S_u^cS_u}    \big\|_{2, \infty}\big\| ({\Sigma}_Y)_{S_vS_v}\big\|_{op},  \\
     &\big\|({\Sigma}_X)_{S_uS_u} \big\|_{op} \big\|({\Sigma}_Y)_{S_v^cS_v}    \big\|_{2, \infty}\Big\}. 
     \end{align*}     
There exist some constants $a, b, c$ depending on $M$ and $\alpha$ such that  if 
   $n \geq c (s_u + s_v )s_us_v \log(p + q)$
  and $\rho \geq a\sqrt{\log(p + q)/n} $, then the minimizer of  \ref{eqn:lasso} is unique and satisfies 
      \begin{align*}
            \text{supp}_e(\wh{B}) \subset S_u \times S_v
     \end{align*} 
      with probability at least $1 - \exp(-bs _u\log(ep/s_u)) - \exp(- bs_v \log(eq/s_v )) - (p + q)^{-b}$.
}

\begin{proof}[Proof of Corollary \ref{corollary}]
By Lemma \ref{lemma1}, there exist constants $a$ and $b$ such that with probability at least $1 - (p+q)^{-b}$,
$$\|\wh{\Sigma}_{XY} - \wh{\Sigma}_X 
B^\star \wh{\Sigma}_Y\|_{\infty} \le a\sqrt{\log(p + q) / n}.$$
Also by element-wise bound on covariance matrix estimation, we have $$\|(\wh{\Sigma}_X)_{S_u^c S_u }\|_{2, \infty} \le \|({\Sigma}_X)_{S_u^c S_u }\|_{2, \infty} + a\sqrt{s_u}\sqrt{\log(p + q) / n},$$
$$\|(\wh{\Sigma}_Y)_{S_v^c S_v }\|_{2, \infty} \le \|({\Sigma}_Y)_{S_v^c S_v }\|_{2, \infty} + a\sqrt{s_v}\sqrt{\log(p + q) / n}.$$
On the other hand, by Lemma \ref{lemma2}, there exist constants $a'$ and $b'$ such that with probability at least $1 - \exp(-b's_u \log(ep/s_u)) -  \exp(-b' s_v \log(eq/s_v))$, we have 
$$\sigma_{\max}({(\wh\Sigma}_{X})_{S_uS_u}) \le \sigma_{\max}(({\Sigma}_{X})_{S_uS_u}) + a'\sqrt{s_u \log(ep/s_u)/n},$$
$$\sigma_{\min}({(\wh\Sigma}_{X})_{S_uS_u}) \geq \sigma_{\min}(({\Sigma}_{X})_{S_uS_u}) - a'\sqrt{s_u \log(ep/s_u)/n},$$
$$\sigma_{\max}({(\wh\Sigma}_{Y})_{S_vS_v}) \le \sigma_{\max}(({\Sigma}_{Y})_{S_vS_v}) + a'\sqrt{s_v \log(eq/s_v)/n},$$
$$\sigma_{\min}({(\wh\Sigma}_{Y})_{S_vS_v}) \geq \sigma_{\min}(({\Sigma}_{Y})_{S_vS_v}) - a'\sqrt{s_v \log(eq/s_v)/n},$$
Then as a consequence of Theorem \ref{thm:support}, the proof is done by choosing $n \geq c_n (s_u + s_v)s_us_v \log(p + q)$ and $\rho \geq c_\rho \sqrt{\log(p + q)/n}$ for sufficiently large constants $c_n$ and $c_\rho$
\end{proof}
\subsubsection{Theorem \ref{theorem2}}
  
 \textit{Suppose that the assumptions of Theorem \ref{theorem1} are satisfied. Assume $n \geq c s_u s_v \log(p + q)/\lambda^{\star 2}_r $ for some sufficiently large constant $c$.  
     There exist constants $a_1, a_2, b, C_1, C_2$ depending on $M$ and $c$ such that if $\rho\in \Big[a_1 \sqrt{\frac{\log(p + q)}{n}}, a_2 \sqrt{\frac{\log(p + q)}{n}}\Big]$,   then with probability at least $1 - \exp(-br) - \exp(-b (s_u + \log(ep/s_u)) )- \exp(- b (s_v + \log(eq/s_v ))) - (p + q)^{-b} $, we have 
     $$\|\wh \Lambda - \Lambda^\star\|_F \le C_1 \sqrt{ \frac{s_u s_v \log(p + q)}{n} }.$$}     And 
     $$ {\max\Big\{\min_{W \in \mathcal{O}_{r}} \| \widehat{U} - U^\star W \|_F, \min_{\tilde W \in  \mathcal{O}_{r}} \| \widehat{V} - V^\star \tilde{W} \|_F \Big\}   \le  C_2 \frac{1}{\lambda^{\star}_r } \sqrt{ \frac{s_u s_v \log(p + q)}{n} }.}$$

\begin{proof}[Proof of Theorem \ref{theorem2}] 
The proof proceeds in four steps.  
(1) Applying truncated SVD to the underlying true matrix $B^\star$ and normalizing with the population covariance matrix yields $U^\star$.  
(2) Replacing the population covariance matrix with the sample covariance matrix introduces only a small error.  
(3) Applying truncated SVD to $\widehat{B}$ likewise incurs only a small additional error (4) Showing the estimation of canonical correlation is accurate.
\paragraph{Step 1: Normalized SVD of $B^\star$ with respect to the population covariance.}

Let
$
B^\star = U_B^\star \Lambda_B^\star V_B^{\star \top},
\quad
U^\star = U_l^\star \Lambda_l^\star V_l^{\star \top}
$
be the singular value decompositions of $B^\star$ and $U^\star$, respectively.
Since $\operatorname{rank}(B^\star)=\operatorname{rank}(U^\star)=r$ and
$B^\star = U^\star \Lambda^\star V^{\star \top}$, the left singular vectors
must span the same subspace. Therefore,
$
U_B^\star = U_l^\star W_l
$
for some orthogonal matrix $W_l \in \mathbb{R}^{r \times r}$ satisfying
$W_l^\top W_l = W_l W_l^\top = I_r$.

Using the SVD of $U^\star$, we obtain
$
U_B^\star
= U_l^\star W_l
= U^\star V_l^\star (\Lambda_l^\star)^{-1} W_l .
$
Consequently,
\begin{align*}
\bigl(U_B^{\star \top} \Sigma_X U_B^\star\bigr)^{-1/2}
&=
\Bigl(
W_l^\top (\Lambda_l^\star)^{-1}
V_l^{\star \top}
U^{\star \top} \Sigma_X U^\star
V_l^\star (\Lambda_l^\star)^{-1}
W_l
\Bigr)^{-1/2} \\
&= W_l^\top \Lambda_l^\star W_l,
\end{align*}
where the equality follows from the normalization
$U^{\star \top} \Sigma_X U^\star = I_r$.

It follows that
$
U_B^\star
\bigl(U_B^{\star \top} \Sigma_X U_B^\star\bigr)^{-1/2}
=
U^\star V_l^\star (\Lambda_l^\star)^{-1}
W_l W_l^\top \Lambda_l^\star W_l
=
U^\star V_l^{\star } W_l .
$
Define $W_X^\star := V_l^{\star }  W_l$. Since both $V_l^\star$ and $W_l$ are
orthogonal, $W_X^\star$ is orthogonal. Therefore,
$
U_B^\star
\bigl(U_B^{\star \top} \Sigma_X U_B^\star\bigr)^{-1/2}
=
U^\star W_X^\star ,
$
where $W_X^\star \in \mathbb{R}^{r \times r}$ is orthogonal. Applying the same argument to the right singular vectors yields
$
V_B^\star
\bigl(V_B^{\star \top} \Sigma_Y V_B^\star\bigr)^{-1/2}
=
V^\star W_Y^\star ,
$
where $W_Y^\star \in \mathbb{R}^{r \times r}$ is orthogonal.
\paragraph{Step 2: Normalizing the SVD of $B^\star$ with respect to the sample covariance.}
Define
\begin{align*}
\tilde{U} &= U_B^\star\bigl(U_B^{\star \top}\,\widehat{\Sigma}_X\,U_B^\star\bigr)^{-1/2}, \\
\tilde{V} &= V_B^\star\bigl(V_B^{\star \top}\,\widehat{\Sigma}_Y\,V_B^\star\bigr)^{-1/2}, \\
\tilde{\Lambda} &=
\bigl(U_B^{\star \top}\,\widehat{\Sigma}_X\,U_B^\star\bigr)^{1/2}
\,\Lambda_B^\star\,
\bigl(V_B^{\star \top}\,\widehat{\Sigma}_Y\,V_B^\star\bigr)^{1/2}.
\end{align*}
Then
$$
B^\star
=
U_B^\star \Lambda_B^\star V_B^{\star \top}
=
\tilde{U}\,\tilde{\Lambda}\,\tilde{V}^{\top}.
$$

Let
$$
Z_X := X U_B^\star \sim \mathcal{N}_r\!\left(0,\,
U_B^{\star \top}\Sigma_X U_B^\star\right),
\qquad
Z_Y := Y V_B^\star \sim \mathcal{N}_r\!\left(0,\,
V_B^{\star \top}\Sigma_Y V_B^\star\right).
$$
These are $r$-dimensional random vectors whose covariance matrices have bounded operator norm. By Lemma~\ref{lemma:covariance}, with probability at least
$1 - 2 e^{-r}$,
$$
\max\!\left\{
\bigl\|U_B^{\star \top}\Sigma_X U_B^\star
-
U_B^{\star \top}\widehat{\Sigma}_X U_B^\star\bigr\|_{\mathrm{op}},
\,
\bigl\|V_B^{\star \top}\Sigma_Y V_B^\star
-
V_B^{\star \top}\widehat{\Sigma}_Y V_B^\star\bigr\|_{\mathrm{op}}
\right\}
\le 
C_1 \sqrt{\frac{r}{n}},
$$
where $C_1$ depends only on $M$.

Moreover, by Lemma~\ref{lemma:root_difference}, for any positive definite matrices
$P$ and $Q$,
$$
\|P^{1/2} - Q^{1/2}\|_{\mathrm{op}}
\le
\frac{1}{
\sqrt{\sigma_{\min}(P)} + \sqrt{\sigma_{\min}(Q)}
}
\,\|P - Q\|_{\mathrm{op}}.
$$
Consequently,
$$
\|P^{-1/2} - Q^{-1/2}\|_{\mathrm{op}}
=
\|P^{-1/2}(Q^{1/2} - P^{1/2})Q^{-1/2}\|_{\mathrm{op}}
\le
\|P^{-1/2}\|_{\mathrm{op}}
\|Q^{-1/2}\|_{\mathrm{op}}
\|P^{1/2} - Q^{1/2}\|_{\mathrm{op}}.
$$

Since the eigenvalues of $\Sigma_X$ and $\Sigma_Y$ lie in $[1/M,\,M]$,
the same holds for
$U_B^{\star \top}\Sigma_X U_B^\star$
and
$V_B^{\star \top}\Sigma_Y V_B^\star$.
If $n \ge c r$ for some sufficiently large constant $c$,
then with high probability the eigenvalues of
$U_B^{\star \top}\widehat{\Sigma}_X U_B^\star$
and
$V_B^{\star \top}\widehat{\Sigma}_Y V_B^\star$
lie in $[1/(2M),\,2M]$. Applying the above bounds with
$P = U_B^{\star \top}\widehat{\Sigma}_X U_B^\star$
and
$Q = U_B^{\star \top}\Sigma_X U_B^\star$
(and analogously for $V$), there exists a constant $C_2$
depending on $M$ and $c$ such that
\begin{align*}
\|\tilde{U} - U^\star W_X^\star\|_F
&=
\Bigl\|
U_B^\star
\Bigl(
(U_B^{\star \top}\widehat{\Sigma}_X U_B^\star)^{-1/2}
-
(U_B^{\star \top}\Sigma_X U_B^\star)^{-1/2}
\Bigr)
\Bigr\|_F \\
&\le
\|U_B^\star\|_F
\|(U_B^{\star \top}\widehat{\Sigma}_X U_B^\star)^{-1/2}
-
(U_B^{\star \top}\Sigma_X U_B^\star)^{-1/2}\|_{\mathrm{op}} \\
&\le
C_2\,\frac{r}{\sqrt{n}} .
\end{align*}
Similarly,
\begin{align*}
\|\tilde{V} - V^\star W_Y^\star\|_F
&\le
C_2\,\frac{r}{\sqrt{n}} .
\end{align*}
\paragraph{Step 3: Normalizing the SVD of $\widehat{B}$ with respect to the sample covariance.}

Let
$$
\widehat{B}_r = \widehat{U}_B \widehat{\Lambda}_B \widehat{V}_B^\top
$$
be the rank-$r$ truncated SVD of $\widehat{B}$. Note the singular values of $B^\star$ has the same order of $\lambda_r^\star$ since $B^\star = \Sigma_X^{-1/2} (\Sigma_{X}^{1/2}B^\star\Sigma_{Y}^{1/2}) \Sigma_Y^{-1/2} $. By Wedin's theorem, as a consequence of Theorem \ref{theorem1}, if
$
n \ge c\,\frac{s_u s_v \log(p+q)}{\lambda_r^{\star 2}}
$
for a sufficiently large constant $c$, then 
\begin{align*}
\min_{W \in O_{r}}
\|\widehat{U}_B - U_B^\star W\|_F
&\lesssim
\frac{1}{\lambda_r^\star / M - \sqrt{\frac{s_u s_v \log(p+q)}{n}}}
\sqrt{\frac{s_u s_v \log(p+q)}{n}}
\\
&\lesssim
\frac{1}{\lambda_r^\star}
\sqrt{\frac{s_u s_v \log(p+q)}{n}},
\\
\min_{W \in O_{r }}
\|\widehat{V}_B - V_B^\star W\|_F
&\lesssim
\frac{1}{\lambda_r^\star}
\sqrt{\frac{s_u s_v \log(p+q)}{n}} .
\end{align*}

Define the support sets
$$
\widehat{S}_u :=
\Bigl\{ i : \|(\widehat{U}_B)_{i\cdot}\| \neq 0 \Bigr\}
\cup \operatorname{Supp}(U^\star),
\qquad
\widehat{S}_v :=
\Bigl\{ i : \|(\widehat{V}_B)_{i\cdot}\| \neq 0 \Bigr\}
\cup \operatorname{Supp}(V^\star).
$$
By Theorem~\ref{theorem1},
$
\max\{|\widehat{S}_u|, |\widehat{S}_v|\}
\lesssim  s_u s_v
$. 
 It follows that there exists $W_U \in O_r$ such that
\begin{align*}
\|
\widehat{U}_B^\top \widehat{\Sigma}_X \widehat{U}_B
-
W_U^\top U_B^{\star \top} \widehat{\Sigma}_X U_B^\star W_U
\|_F
&\le
\|
\widehat{U}_B^\top \widehat{\Sigma}_X (\widehat{U}_B - U_B^\star W_U)
\|_F
+
\|
(\widehat{U}_B - U_B^\star W_U)^\top
\widehat{\Sigma}_X U_B^\star
\|_F
\\
&\le
2\,\|
(\widehat{\Sigma}_X)_{\widehat{S}_u \widehat{S}_u}
(\widehat{U}_B - U_B^\star W_U)
\|_F,
\end{align*}
where $$ 
\|\widehat{U}_B - U_B^\star W_U \|_F
\lesssim
\frac{1}{\lambda_r^\star}
\sqrt{\frac{s_u s_v \log(p+q)}{n}} . $$
Similarly there exists $W_V \in O_r$ such that
$$
\|
\widehat{V}_B^\top \widehat{\Sigma}_Y \widehat{V}_B
-
W_V^\top V_B^{\star \top} \widehat{\Sigma}_Y V_B^\star
W_V \|_F
\le
2\,
\|
(\widehat{\Sigma}_Y)_{\widehat{S}_v \widehat{S}_v}
(\widehat{V}_B - V_B^\star W_V)
\|_F .
$$
By Lemma~\ref{lemma2}, if
$
n \ge c\, s_u s_v \log(p+q)
$
for a sufficiently large constant $c$, then with probability at least
$
1
-
\exp\!\big(-b (s_u s_v  \log(ep/(s_u s_v)))\big)
-
\exp\!\big(-b (s_u s_v  \log(eq/(s_u s_v)))\big),
$
we have
\begin{align*}
 &\quad  \max\!\left\{
\|
\widehat{U}_B^\top \widehat{\Sigma}_X \widehat{U}_B 
-
W_U^\top U_B^{\star \top} \widehat{\Sigma}_X U_B^\star W_U
\|_F,\;
\|
\widehat{V}_B^\top \widehat{\Sigma}_Y \widehat{V}_B
-
W_V^\top V_B^{\star \top} \widehat{\Sigma}_Y V_B^\star
W_V \|_F
\right\}
\\& \lesssim
\frac{1}{\lambda_r^\star}
\sqrt{\frac{s_u s_v \log(p+q)}{n}}  
\end{align*}

Applying Lemma~\ref{lemma:root_difference} again yields
\begin{align*}
\|
\widehat{U}
-
U_B^\star
(U_B^{\star \top} \widehat{\Sigma}_X U_B^\star)^{-1/2}
 W_U \|_F
&=
\|
\widehat{U}_B
(\widehat{U}_B^\top \widehat{\Sigma}_X \widehat{U}_B)^{-1/2}
-
U_B^\star
(U_B^{\star \top} \widehat{\Sigma}_X U_B^\star)^{-1/2}
W_U \|_F
\\
&\le
\|\widehat{U}_B\|_{\mathrm{op}}
\|
(\widehat{U}_B^\top \widehat{\Sigma}_X \widehat{U}_B)^{-1/2}
-
W_U^\top (U_B^{\star \top} \widehat{\Sigma}_X U_B^\star)^{-1/2} W_U
\|_F
\\
&\quad
+
\|
(U_B^{\star \top} \widehat{\Sigma}_X U_B^\star)^{-1/2}
\|_{\mathrm{op}}
\|
\widehat{U}_B - U_B^\star W_U
\|_F
\\
&\le
C_3\,
\frac{1}{\lambda_r^\star}
\sqrt{\frac{s_u s_v \log(p+q)}{n}}
\end{align*}
for some constant $C_3$.

Recall from Step~2 that
$$
\|
U_B^\star
(U_B^{\star \top} \widehat{\Sigma}_X U_B^\star)^{-1/2}
-
U^\star W_X^\star
\|_F
\le
C_2\,\frac{r}{\sqrt{n}} .
$$
Combining the above bounds yields
$$
\|
\widehat{U} - U^\star W_X^\star W_U
\|_F
\le
(C_2 + C_3)\,\frac{1}{\lambda_r^\star}
\sqrt{\frac{s_u s_v \log(p+q)}{n}}
$$
The same argument applies to $V^\star$.
\paragraph{Step 4: Estimation of Canonical Correlation.} 
Let \begin{align*}
    \bar{U} &= U^\star(U^{\star \top}  \wh{\Sigma}_{X} U^{\star})^{-1/2}, \\
    \bar{V} &= V^{\star} (V^{\star \top}  \wh{\Sigma}_{Y} V^\star)^{-1/2}, \\
    \bar{\Lambda} &=  (U^{\star \top}  \wh{\Sigma}_{X} U^{\star})^{1/2} \Lambda^\star (V^{\star \top}  \wh{\Sigma}_{Y} V^\star)^{1/2} .
\end{align*}
We first show that $\wh{\Sigma}_{X}^{1/2} \wh{B} \wh{\Sigma}_{Y}^{1/2}$ is close to $\wh{\Sigma}_{X}^{1/2} \bar{U} \Lambda^\star \bar{V}^\top  \wh{\Sigma}_{Y}^{1/2}$. Note
\begin{align*}
    \| \wh{\Sigma}_{X}^{1/2} \wh{B} \wh{\Sigma}_{Y}^{1/2}  - \wh{\Sigma}_{X}^{1/2} \bar{U} \Lambda^\star \bar{V}^\top  \wh{\Sigma}_{Y}^{1/2}  \|_{F} 
    &\le       \| \wh{\Sigma}_{X}^{1/2} \Delta \wh{\Sigma}_{Y}^{1/2} \|_{F}  + \| \wh{\Sigma}_{X}^{1/2} B^\star\wh{\Sigma}_{Y}^{1/2} -\wh{\Sigma}_{X}^{1/2} \bar{U} \Lambda^\star \bar{V}^\top  \wh{\Sigma}_{Y}^{1/2}  \|_{F} \\
        &=       \| \wh{\Sigma}_{X}^{1/2} \Delta \wh{\Sigma}_{Y}^{1/2} \|_{F}  + \| \wh{\Sigma}_{X}^{1/2} \bar{U}  (\Lambda^\star - \bar{\Lambda})\bar{V}^\top  \wh{\Sigma}_{Y}^{1/2}   \|_{F} \\
        &\le  \| \wh{\Sigma}_{X}^{1/2} \Delta \wh{\Sigma}_{Y}^{1/2} \|_{F}  + \| \wh{\Sigma}_{X}^{1/2} \bar{U} \|_{op} \| (\Lambda^\star - \bar{\Lambda}) \|_{op} \| \bar{V}^\top  \wh{\Sigma}_{Y}^{1/2}   \|_{F} \\
        &\le  \| \wh{\Sigma}_{X}^{1/2} \Delta \wh{\Sigma}_{Y}^{1/2} \|_{F}  + \sqrt{r} \| \Lambda^\star - \bar{\Lambda} \|_{op} .
\end{align*}
Following the same logic as in the proof of Theorem $\ref{theorem1}$, there exists a constant $\kappa$ such that with probability at least $1 - \exp(-b (s_u \log(ep/s_u) ) ) - \exp(- b ( s_v  \log(eq/s_v )) ) - (p + q)^{-b} $, we have 
$$ \| \wh{\Sigma}_{X}^{1/2} \Delta \wh{\Sigma}_{Y}^{1/2} \|_{F} \le  \kappa \|\Delta\|_F \lesssim \sqrt{\frac{s_us_v \log(p + q)}{n}},$$
where the last inequality is a consequence of Theorem 
\ref{theorem1}. By Lemma \ref{lemma3} , with high probability we know 
\begin{align*}
     \| \Lambda^\star - \bar{\Lambda} \|_{op} &\le \|(I -  (U^{\star \top}  \wh{\Sigma}_{X} U^{\star})^{1/2} ) \Lambda^\star\|_{op} +  \|(U^{\star \top}  \wh{\Sigma}_{X} U^{\star})^{1/2} \Lambda^\star  (I -  (V^{\star \top}  \wh{\Sigma}_{Y} V^\star)^{1/2} )\|_{op}  \\
     &\lesssim
      \sqrt{\frac{s_u + \log(ep/s_u)}{n}} + \left(1 +  \sqrt{\frac{s_u + \log(ep/s_u)}{n}} \right)  \sqrt{\frac{s_v + \log(eq/s_v)}{n}} \\
     &\lesssim  \left( \sqrt{\frac{s_u + \log(ep/s_u)}{n}}  + \sqrt{\frac{s_v + \log(eq/s_v)}{n}} \right).
\end{align*}
Therefore
\begin{align*}
    \| \wh{\Sigma}_{X}^{1/2} \wh{B} \wh{\Sigma}_{Y}^{1/2}  - \wh{\Sigma}_{X}^{1/2} \bar{U} \Lambda^\star \bar{V}^\top  \wh{\Sigma}_{Y}^{1/2}  \|_{F}  
    & \lesssim \sqrt{\frac{s_u s_v \log(p + q)}{n}}\\
    &+ \sqrt{r}\left( \sqrt{\frac{s_u + \log(ep/s_u)}{n}}  + \sqrt{\frac{s_v + \log(eq/s_v)}{n}} \right)\\
    & \lesssim \sqrt{\frac{s_u s_v \log(p + q)}{n}}. 
\end{align*}
By Mirsky’s inequality, we have $$\|\wh \Lambda - \Lambda^\star\|_F \lesssim \sqrt{\frac{s_u s_v \log(p + q)}{n}}.$$

\end{proof}

\subsubsection{Proof: determining an appropriate rank for CCA}\label{app:rank}
\begin{proof}[Proof of Corollary \ref{corollary:rank}]
{Note the results for the singular values of $\wh \Sigma_{X}^{1/2} \wh B  \wh \Sigma_{Y}^{1/2}$ is directly implied from Theorem \ref{theorem2}.

Next we show the results for singular values of $\wh B$.  On the event in Theorem \ref{theorem1}, we know 
$$\|\wh B - B^\star\|_{op} = \|\Delta\|_{op} \le \|\Delta\|_F \lesssim \sqrt{\frac{s_us_v\log(p + q)}{n}}$$ In addition, we know
$$d_r(B^\star) = d_r\left( \Sigma_{X}^{-1/2} (\Sigma_{X}^{1/2} B^\star \Sigma_{Y}^{1/2}) \Sigma_{Y}^{-1/2}  \right) \geq \frac{\lambda_r^{\star}}{M} , \quad d_{r + 1}(B^\star) = 0$$ Then the claimed bound on $d_r(\wh B) $ and $d_{r + 1}(\wh B)$ is proved by Mirsky’s inequality.
}
\end{proof}

\paragraph{Interpretation of the two screeplots.}
The spectrum of $\widehat B$ provides an alternative diagnostic for
rank selection. Its singular values should not, however, be interpreted
as estimates of the canonical correlations. Indeed, the singular values
of $B^{\star}$ depend on the coordinate-wise scaling of the two data
views. In contrast, the nonzero singular values of
$\Sigma_X^{1/2}B^{\star}\Sigma_Y^{1/2}$ are exactly the canonical
correlations. Nevertheless, under the bounded-eigenvalue assumptions on
$\Sigma_X$ and $\Sigma_Y$, the nonzero singular values of $B^{\star}$ are uniformly
comparable to the canonical correlations. This justifies the screeplot of the
untruncated estimator $\widehat B$ as a rank-selection diagnostic,
particularly when the variables have first been standardized. The
covariance-adjusted spectrum remains preferable when the numerical
values of the singular values are to be interpreted on the canonical-
correlation scale.

\paragraph{Permutation-based rank selection using $\widehat B$.}
The permutation procedure described above can be applied directly to
the singular values of $\widehat B$. Specifically, randomly permute the
rows of one data view, refit the untruncated estimator, and record
\[
    \widehat d_j^{(b)}
    :=\sigma_j\!\left(\widehat B^{(b)}\right),
    \qquad b=1,\ldots,N_{\mathrm{perm}}.
\]
The rank may then be selected as the number of observed singular values
$\widehat d_j$ that exceed a permutation reference level, such as the
pointwise $95\%$ quantile of
$\{\widehat d_j^{(b)}\}_{b=1}^{N_{\mathrm{perm}}}$.
\end{appendix}

\bibliographystyle{apalike}
\bibliography{Bibliography}

\end{document}